\documentclass{article} 
\usepackage{iclr2027_conference,times}

\usepackage{amsmath,amsfonts,bm}

\def\eqref#1{equation~\ref{#1}}

\def\1{\bm{1}}

\DeclareMathAlphabet{\mathsfit}{\encodingdefault}{\sfdefault}{m}{sl}
\SetMathAlphabet{\mathsfit}{bold}{\encodingdefault}{\sfdefault}{bx}{n}

\usepackage[utf8]{inputenc} 
\usepackage[T1]{fontenc}    
\usepackage[colorlinks, urlcolor=mydarkblue, linkcolor=mydarkblue, citecolor=gray]{hyperref}      
\usepackage{url}            
\usepackage{booktabs}       
\usepackage{amsfonts}       
\usepackage{nicefrac}       
\usepackage{microtype}      
\usepackage{xcolor}         
\usepackage{amssymb}
\usepackage{pifont}

\usepackage{amsmath}
\usepackage{graphicx}
\usepackage{algorithm}
\usepackage{algpseudocode}
\usepackage{pifont}
\usepackage{mathtools}
\usepackage{adjustbox}
\usepackage{enumitem}
\usepackage{wrapfig}
\usepackage[table]{colortbl}
\usepackage{multirow}
\usepackage{makecell}
\usepackage[T1]{fontenc}
\usepackage[utf8]{inputenc}
\usepackage{babel}
\usepackage[font=small]{caption}
\usepackage{tikz, tcolorbox}
\usepackage{pgfplots}
\usepackage{subcaption}
\usepackage{amsmath}
\usepackage{caption}
\usepackage{pgfplotstable}
\usepackage[linewidth=0.5pt]{mdframed}
\usepackage{tikz}
\usepackage{pgfplots}
\usetikzlibrary{pgfplots.groupplots} 
\usetikzlibrary{pgfplots.groupplots,intersections} 
\pgfplotsset{compat=1.18}
\usepackage{pgfplotsthemetol}  
\usetikzlibrary{fillbetween}
\usepackage{amsthm}
\usepackage{circledsteps}
\usepackage{algorithm}
\usepackage{algpseudocode}
\usepackage[table]{xcolor}
\usepackage{array}

\newcolumntype{M}{>{\columncolor{gray!12}\bfseries}c}

\usepackage{minitoc}

\definecolor{mydarkblue}{rgb}{0,0.08,0.45}
\newtheorem{definition}{Definition}
\newtheorem{theorem}{Theorem}
\newtheorem{corollary}{Corollary}
\usepackage{tcolorbox}
\tcbuselibrary{skins, breakable}
\definecolor{implicationborder}{RGB}{210,210,210}
\definecolor{implicationbg}{RGB}{252,252,252}

\newtcolorbox{implicationbox}{
  enhanced,
  breakable,
  colback=implicationbg,
  colframe=implicationborder,
  boxrule=0.6pt,
  left=10pt, right=10pt, top=7pt, bottom=7pt,
  arc=5pt,
}

\title{Do Multi-Agent LLMs Explore Each Other?}

\author{Hyeong Kyu Choi$^1$, Jiatong Li$^1$, Wendi Li$^1$, Xin Eric Wang$^2$, Sharon Li$^1$ \\[0.3em]
  $^1$University of Wisconsin-Madison, $^2$University of California, Santa Barbara
   \\
}

\iclrfinalcopy 
\begin{document}
\doparttoc 
\faketableofcontents 

\maketitle

\begin{abstract}
Exploration is essential for reliable autonomy in multi-agent systems, yet it remains unclear whether large language model~(LLM) agents can explore effectively when interacting with one another. 
We show that modern LLM agents often fail to do so, exhibiting myopic and polarized interaction patterns that lead to suboptimal coordination and increased regret. 
We formalize this challenge as the Multi-Agent Exploration problem, modeling it as a partially observable stochastic game~(POSG), in which agents must probe peers to infer their capabilities and identify effective interaction strategies.
To address this, we introduce Multi-Agent Contextual Exploration~(MACE), a lightweight framework that explicitly promotes exploration through structured peer selection. 
Across both contextual and parametric diversity settings, exploration behavior and downstream task performance substantially improve when MACE is applied.
We further provide theoretical insight that MACE achieves a sublinear expected-regret bound.
Overall, our results highlight a fundamental limitation of current LLM agents and underscore the importance of explicitly guided exploration for reliable multi-agent autonomy.
Code will be released for reproducible research.
\end{abstract}

\section{Introduction}
Large language models are increasingly deployed not only as isolated assistants, but as autonomous agents embedded in multi-agent systems: communicating, delegating, and making decisions in a decentralized fashion to accomplish complex tasks~\citep{li2023camel,wu2024autogen,qian2024chatdev,hong2023metagpt,gao2024agentscope,du2024improving}. This paradigm is rapidly expanding: open-world platforms now instantiate heterogeneous agent populations that interact autonomously to discuss problems, divide labor, and collectively reason toward shared goals~\citep{piao2025agentsociety,park2023generative,jiang2026humans,zhang2026agents}. 
Such systems implicitly rely on the assumption that agents are capable of autonomous decision-making, and that meaningful collective behavior can emerge from agent-to-agent interactions.

But what is required for agents to act autonomously in a reliable manner? 
A large body of literature in intrinsic motivation and reinforcement learning has demonstrated that \emph{exploration} is foundational to autonomous behavior~\citep{pathak2017curiosity,burda2019large}. Exploration is not merely a mechanism for improving task performance, but the core process by which agents self-generate goals, discover novel states, and acquire reusable strategies without external guidance~\citep{chentanez2004intrinsically,baker2019emergent,forestier2022intrinsically}. 
The importance of exploration is further amplified in multi-agent settings, where
the environment is shaped not only by environmental dynamics, but also by the behaviors and capabilities of other agents, which are often heterogeneous and can only be revealed through direct interaction. 
Thus, in such settings, 
agents must proactively explore peers to identify effective collaborators, uncover complementary information, and adapt as the interaction evolves.

Yet despite the centrality of exploration to reliable autonomy, we find that current LLM agents fail to explore effectively, even in the simplest possible settings. 
We begin with a simple autonomous multi-agent setting---a controlled two-armed bandit experiment, where an LLM agent must repeatedly choose between two peers with unknown success rates and must infer the better one through exploration. 
Rather than accumulating evidence and converging to the superior peer, Section~\ref{sec:toy} reveals that LLM agents exhibit premature commitment, locking onto one peer within the first few rounds and persisting regardless of correctness. 
Critically, this failure is not confined to smaller or less capable models: even capable frontier models like GPT-4~\citep{achiam2023gpt} and GPT-5~\citep{openai2025gpt5} exhibit the failure mode, suggesting that insufficient exploration is a fundamental and structural limitation of current LLM agents rather than a capacity deficit. 

To characterize this rigorously and capture real-world complexities, we formalize the \textit{\textbf{multi-agent exploration problem}}, in which agents differ in capabilities or possess distinct contextual information. {Multi-agent exploration} thus requires actively probing other agents strategically and identifying effective interaction strategies. Unlike standard bandit settings where arms are stationary~\citep{auer2002finite}, peers here are themselves adaptive agents whose responses evolve
as they accumulate their own interaction experience. Moreover, agents operate in a decentralized manner, making decisions based only on partial and dynamically evolving information shaped by ongoing interactions, and exploration failures compound across agents and rounds simultaneously.


To address this, we introduce \textbf{Multi-Agent Contextual Exploration (MACE)}, a lightweight solution for the multi-agent exploration problem, promoting exploration through explicit algorithmic guidance. 
Since solving the multi-agent exploration problem exactly is intractable, the key idea behind MACE is to decompose the joint problem into independent sequential decision-making problems under contextual bandits, where the non-stationarity of multi-agent interactions is encoded into a novel structural feature representation.  
Specifically, 
MACE encodes the \textit{relational structure} of each potential interaction into a contextual representation that reflects the utility of querying a given peer in a given context. This allows MACE to distinguish between a peer that is globally under-tested and one that is under-tested in the specific relational context of the current task, providing a finer-grained form of exploration. 

Empirically, MACE substantially improves existing baselines across heterogeneous agent settings spanning contextual and capability diversity. Interestingly, in-context exploration---where agents are explicitly prompted to balance exploration and exploitation using the same information available to MACE---often underperforms even random peer selection, confirming that the failure cannot be remedied through prompting alone and that algorithmic structure is necessary. MACE's gains persist into an exploitation phase with frozen parameters and transfer to unseen benchmarks, indicating that the learned interaction strategies capture generalizable structure rather than task-specific artifacts. 
Theoretically, we analyze peer selection under a realizable linear-bandit model.
We show that MACE achieves $O(\sqrt{T}\log T)$ expected cumulative regret.
We also construct a stationary instance on which empirical-mean greedy selection incurs
$\Omega(T)$ expected regret through premature commitment.
Together, these results establish an asymptotically linear expected-regret separation on the constructed instance, illustrating how explicit exploration can prevent persistent peer-selection errors.


We summarize our contributions as follows:
\begin{itemize}[leftmargin=*]
    \item[$\circ$]  
    We {formalize the Multi-Agent Exploration problem}, motivated by the observation that modern LLM agents often exhibit limited exploration in multi-agent environments. This is fundamental yet underexplored for reliable autonomy in open-world multi-agent settings.  
    \item[$\circ$] We introduce Multi-Agent Contextual Exploration (MACE), a lightweight and tractable solution to address the problem via contextual bandit-driven peer selection with explicit exploration incentives.
    \item[$\circ$] We {show both empirically and theoretically that explicit exploration significantly improves task performance}, and how explicit exploration can prevent persistent peer-selection errors. Our evaluation spans seven models across four major LLM families, involving three benchmarks.
\end{itemize}
\section{Do LLMs Explore in Multi-Agent Environments?}
\label{sec:toy}

Exploration is a fundamental prerequisite for reliable autonomy of a multi-agent system.
An agent must not only act on current beliefs, but also actively probe its environment to discover better strategies and collaborators. 
In multi-agent LLM systems, this challenge of exploration manifests as deciding which peers to trust, under uncertainty about their capabilities. 
Despite their strong abilities, it remains unclear whether LLM agents can autonomously balance exploration and exploitation when interacting with other agents. 
To investigate this, we study a simplified delegation scenario centered on a single agent, isolating the core decision problem of selecting peers with unknown reliability. 

\noindent\textbf{Setup.} Consider an LLM agent that must repeatedly delegate tasks to one of two peer LLMs, knowing nothing about their abilities upfront. Each delegation reveals a small piece of evidence (\emph{i.e.}, whether the peer succeeds or fails), and the agent must decide how to use that evidence. The agent can keep testing the peer it knows less about, or commit to the one that has performed well so far. This is the classic exploration-exploitation dilemma in its purest form, and it is precisely the kind of judgment that reliable multi-agent coordination demands.

To study whether LLM agents can exercise this judgment, we design a two-armed bandit setting where an agent must repeatedly delegate arithmetic questions to one of its two peers, A or B. Each peer has a fixed but unknown probability of answering correctly~($p_\text{A}$ and $p_\text{B}$ respectively). 
At every round, the agent observes the historical performance and the number of times it delegated to each peer, and is prompted to select an agent. Ideally, the  agent should balance exploration (trying less-tested peers) and exploitation (selecting the empirically better one). 
Specifically, we test settings with probability $(p_\text{A}, p_\text{B}) = (0.6, 0.5)$, where one peer is better on average, but with a noisy performance gap. This setting requires the agent to perform sufficient exploration before converging to optimal exploitation. Additional experiments with varying $p_\text{A}$ values are provided in Appendix~\ref{apdx:full_hist}.

\noindent\textbf{Behavioral reference.} We evaluate three LLM agents on this task: Qwen2.5-7B-Instruct~\citep{yang2024qwen2}, GPT-4~\citep{achiam2023gpt} and GPT-5~\citep{openai2025gpt5}.
In all cases, peer selection is performed purely through in-context information: at each round, the agent is presented with running history and prompted to select a peer. 
To anchor our interpretation of their behavior, we include a classical exploration algorithm as {behavioral reference}: Upper Confidence Bound (UCB)~\citep{auer2002finite}. Unlike the LLM agents, UCB imposes explicit mathematical structure on the selection decision that provably incentivizes exploration. More specifically, UCB maintains an optimistic estimate of each peer's true success rate, and selects the peer with the highest such estimate. This optimism-under-uncertainty principle naturally drives exploration toward under-tested peers while transitioning to exploitation as evidence accumulates. We include UCB as a concrete illustration of what algorithmically guaranteed exploration looks like in this setting, making the absence of exploration in LLM agents directly visible. See Appendix~\ref{apdx:toy_setup} for further descriptions on the UCB algorithm and other experimental details.



\begin{wrapfigure}{r}{0.5\textwidth}
  \vspace{-5mm}
  \centering
  \includegraphics[width=0.45\textwidth]{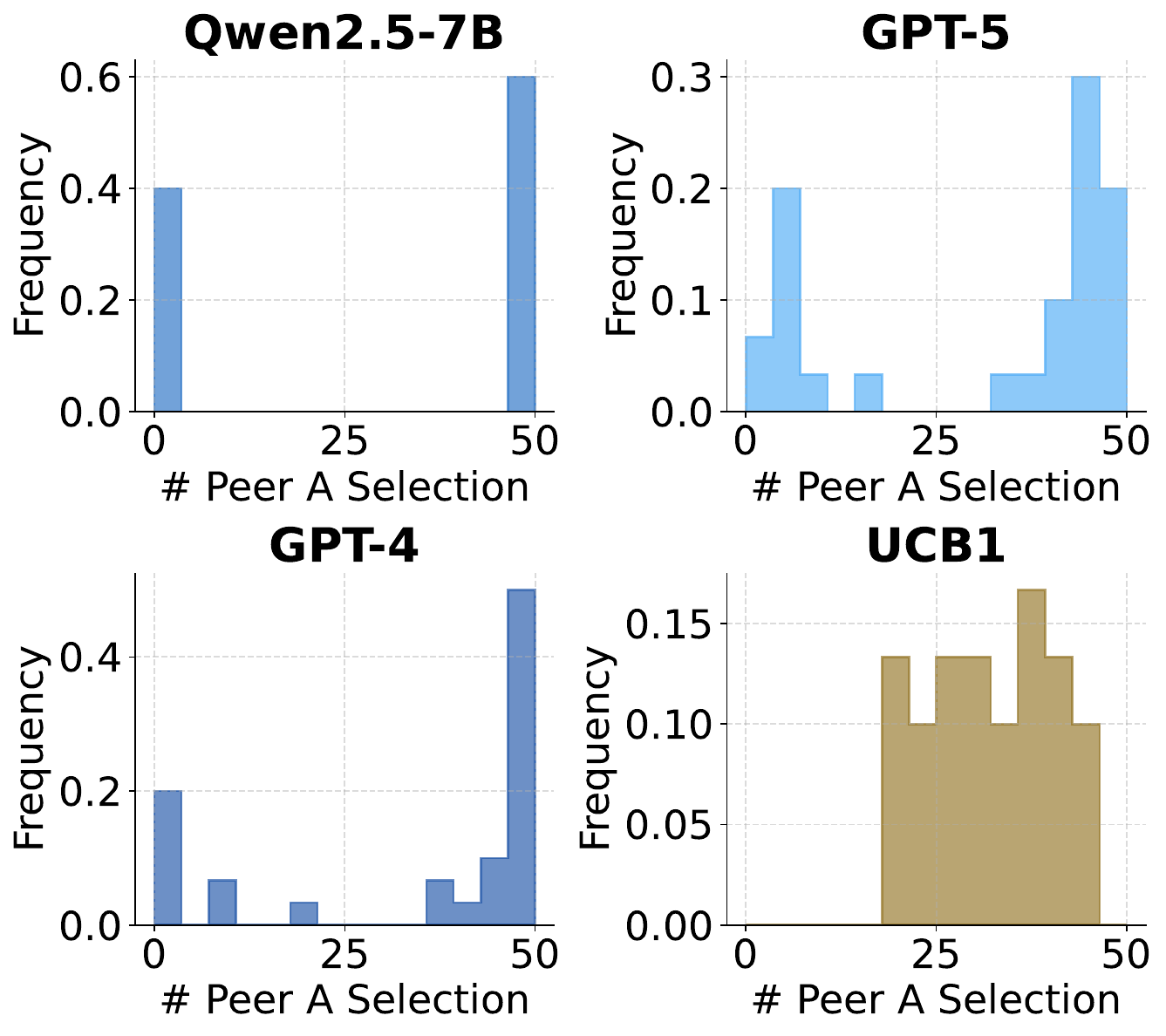}
  \vspace{-2mm}
  \caption{Histogram of the number of peer A selection. }
  \label{fig:toy_exp}
  \vspace{-7mm}
\end{wrapfigure}



\noindent\textbf{{LLM agents exhibit insufficient exploration.}}
Figure~\ref{fig:toy_exp} shows the distribution of peer A selection frequency across 30 independent runs, where each run consists of $T=50$ delegation rounds. 
Optimal behavior corresponds to predominantly selecting peer A while still occasionally exploring peer B.
 We observe that the behavioral reference UCB~(in brown) exhibits such desired pattern: the distribution is  concentrated toward higher selection counts, indicating effective exploration followed by consistent exploitation of the better peer.

In stark contrast, all LLM agents display highly polarized bimodal distributions, with mass concentrated near both extremes~(close to 0 or 50 selections out of 50 rounds).
This pattern reveals a characteristic failure mode: rather than gradually accumulating evidence and converging to a reliable preference, LLM agents commit prematurely to one peer within the first few rounds and maintain that commitment for the remainder of the task.
Crucially, this early commitment is not always correct---the mass near 0 indicates frequent runs where the agent locked onto the inferior peer and never reconsidered. Overall, LLM agents do not exhibit an innate capability to explore and reliably identify the better option.
More detailed analyses are in Appendix~\ref{apdx:full_hist}.

\section{The Multi-Agent Exploration Problem: A Formalization}
\label{sec:body}
\vspace{-4mm}
The observations in Section~\ref{sec:toy} reveal that modern LLM agents fail to exhibit sufficient exploration in its simplest form.
In realistic multi-agent systems, however, this challenge is further amplified.
Agents operate in a decentralized manner, making decisions based only on partial and dynamically evolving information shaped by ongoing interactions.  The space of possible interaction protocols is vast, the capabilities and context of peers are heterogeneous, and the consequences of poor exploration compound across agents and rounds simultaneously. 
To capture these complexities, we formalize the problem of \emph{\textbf{multi-agent exploration}}.

\subsection{Problem Formalization}

\begin{wrapfigure}{r}{0.52\textwidth}
  \vspace{-9mm}
  \centering
    \includegraphics[width=0.52\textwidth]{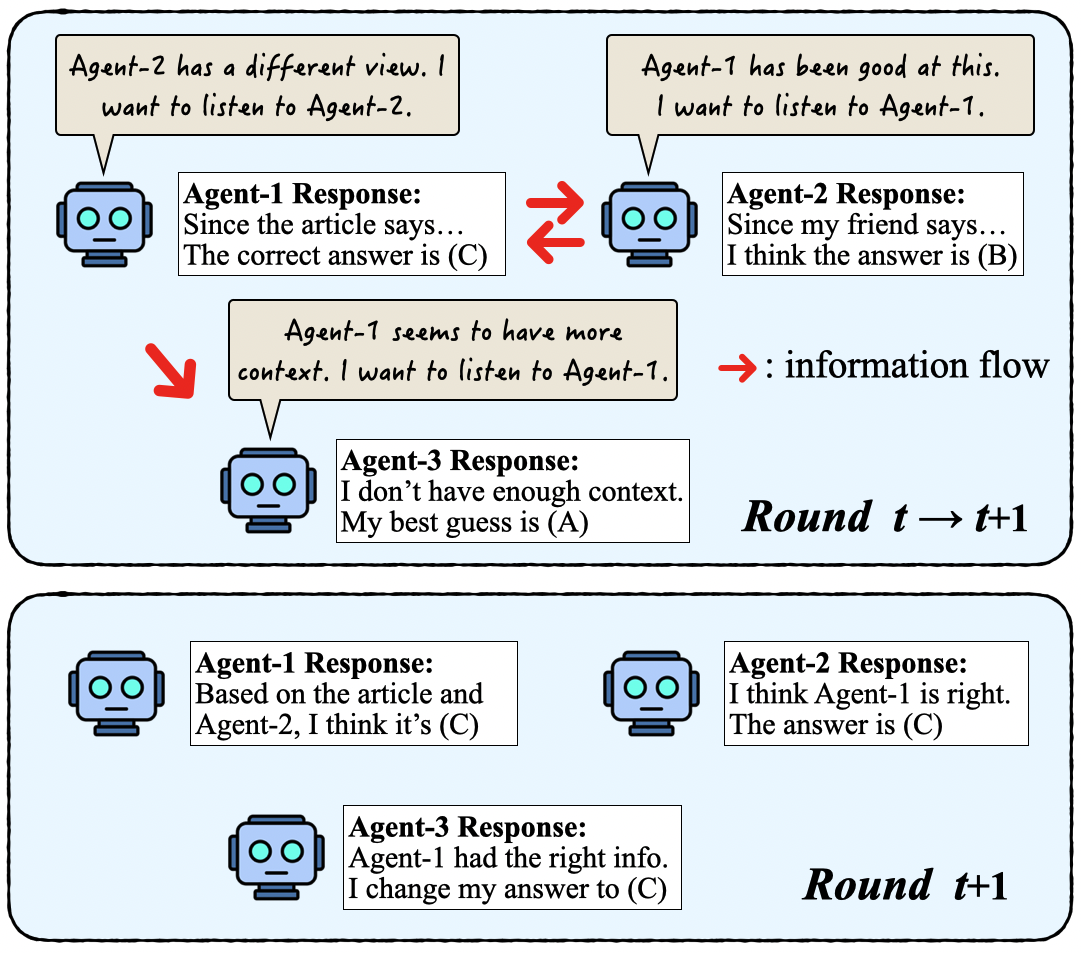}
  \caption{\textbf{The Multi-Agent Exploration problem}. At round $t$, each agent needs to select a peer to interact with, and update its response in the subsequent interaction round.}
  \label{fig:protocol}
  \vspace{-6mm}
\end{wrapfigure}

We consider a system of $\mathcal{N}=\{1,...,N\}$ agents interacting over $T$ timesteps  to solve a sequence of tasks. 
Each agent may possess distinct capabilities, which may arise from differences in model family or available context. And these differences are latent: no agent has direct access to the capabilities of its peers, and they become partially observable through interactions. 
Each agent $i$ observes its selected peer's response, accumulated over interaction rounds.
{Multi-agent exploration} thus arises as a problem of actively probing other agents to infer their capabilities from partial observation history and identify effective interaction strategies. 


Formally, we model this process as a Partially Observable Stochastic Game (POSG), defined by the tuple
$
\mathcal{G} = \langle \mathcal{S}, \mathcal{A}, \mathcal{O}, \mathcal{P}, \mathcal{Q}, \mathcal{R} \rangle.
$
Here, latent state $s_t \in \mathcal{S}$ encodes all task-relevant information together with the unknown capabilities of every agent in the system. This encapsulates everything that governs what an effective interaction strategy would look like, yet is not fully observable to  anindividual agent. At each round $t \in \{1, \ldots, T\}$, agent $i \in \mathcal{N}$ selects action $a_{i,t} \in \mathcal{N}$, designating the peer to interact with. The joint action $\mathbf{a}_t = (a_{1,t}, \ldots, a_{N,t}) \in \mathcal{A} = \mathcal{N}^N$ captures the system-level interaction structure at round $t$. The transition function $\mathcal{P} : \mathcal{S} \times \mathcal{A} \rightarrow \Delta(\mathcal{S})$ governs how information propagates through the system as a consequence of $\mathbf{a}_t$, inducing a dynamically evolving interaction graph whose topology is itself determined by the agents' decisions.
The high-level architecture of the multi-agent exploration problem is illustrated in Figure~\ref{fig:protocol}.

Upon taking action $a_{i,t}$, agent $i$ receives an observation $o_{i,t} \in \mathcal{O}_i$ drawn from the observation function $\mathcal{Q} : \mathcal{S} \times \mathcal{A} \rightarrow \Delta(\mathcal{O})$, where $\mathcal{O} = \prod_{i=1}^{N} \mathcal{O}_i$. 

The observation $o_{i,t}$ consists of the queried peer's response together with the agent's accumulated interaction history. 
Crucially, agents do not have access to the full state $\mathcal{S}$, and must instead act based on partial observations accumulated over time from its selected peers.
That is, before selecting a peer, each agent observes its own current response, local interaction statistics, and summary features computed from all agents' previous-round responses. 
Only after selection it receives the
selected peer's full response.
Finally, the reward $r_{i,t} = R_i(s_t, \mathbf{a}_t)$ for agent $i$, drawn from $\mathcal{R} = \{R_i\}_{i=1}^{N}$ with $R_i : \mathcal{S} \times \mathcal{A} \rightarrow \mathbb{R}$, reflects task performance (e.g., the correctness of the answer, the gain in performance from the round, or the utility of intermediate responses).


\subsection{Regret}

Each agent $i \in \mathcal{N}$ should ideally  minimize the cumulative regret over $T$ rounds with respect to the peer with the highest conditional expected reward at each round:
\begin{equation}
\label{eq:regret}
\mathrm{Regret}_i
:=
\sum_{t=1}^T
\Big(
\max_{a \in [N]} \mu_{i,a}^{(t)}
-
\mu_{i,a_{i,t}}^{(t)}
\Big),
\end{equation}
where $\mu_{i,a_{i,t}}^{(t)}$ is the expected reward for agent $i$ selecting peer $a_{i,t}$ at round $t$.
To characterize the practical effect of insufficient exploration, we additionally consider the expected cumulative reward:
$
\mathbb{E}\left[\sum_{t=1}^T r_{i,t}\right],
$
where $r_{i,t}$ denotes the reward obtained at round $t$. 
A central challenge in this setting is the exploration-exploitation trade-off under partial observability and agent heterogeneity. 
Effective policies must balance:
(i) \emph{exploration}, \emph{i.e.}, querying diverse peers to estimate their capabilities and reduce uncertainty and regret, and 
(ii) \emph{exploitation}, \emph{i.e.}, selecting peers believed to yield high rewards.
In the next section, we introduce exploration-driven strategies that address this challenge.


\section{Multi-Agent Contextual Exploration~(MACE)}
\label{sec:max}
\vspace{-2mm}
While our formalization captures the full structure of the multi-agent exploration problem, solving the finite-horizon POSG problem is NEXP-hard. The core difficulty is that an agent must estimate peer capabilities from sparse, noisy interaction history, while the peers themselves are adaptive (\emph{i.e.}, their responses evolve as they accumulate their own interaction experience).
To address the challenge, we introduce \textbf {Multi-Agent  Contextual Exploration} (\textbf{MACE)}, a tractable and lightweight solution that explicitly promotes exploration in multi-agent interactions by decomposing the joint problem into independent per-agent decisions. Rather than requiring complex coordination protocols, MACE treats peer selection as a learning problem: each agent adaptively chooses which peer to query based on past observations and contextual signals, with an explicit mechanism to avoid premature commitment.

\textbf{Contextual bandit formulation.}  
MACE models multi-agent interaction as a contextual multi-armed bandit problem, where each agent independently learns which peer to query at each round. 
For agent $i$, selecting a peer $a \in \mathcal{N}$ at round $t$ corresponds to pulling an arm, and the expected reward is modeled as a linear function:
$
    \mu_{i,a,t} \approx \boldsymbol{\mathbf{x}}_{i,a,t}^\top \boldsymbol{\theta}_{i,a},
$
where $\boldsymbol{\mathbf{x}}_{i,a,t} \in \mathbb{R}^d$ is a context feature vector encoding the interaction between agent $i$ and peer $a$ at timestep $t$, and $\boldsymbol{\theta}_{i,a} \in \mathbb{R}^d$ is the corresponding weight vector. 
This reduction to the contextual bandit problem is non-trivial: unlike standard bandit settings where arms are stationary~\citep{auer2002finite}, the peers here are themselves adaptive agents. 

Before each interaction round, a feature-extraction
module computes summary features from the agents’ previous-round responses. Each selector
receives these features for all candidate peers, together with its own response and local interaction
statistics. After selecting a peer, the selector receives that peer’s full response and uses it to revise
its answer. Thus, access to full peer responses is restricted, while the selection policy has access to
population-wide response summaries. Both MACE and In-Context Exploration receive the same
pre-selection features.

\textbf{Relational feature design.} While scalar performance summaries are simple, they may fail to capture the non-stationary evolution of agent capabilities over time. Therefore, instead of relying solely on scalar summaries, we propose constructing these features to encode the relational structure of the interaction between agent $i$ and peer $a$.
(i) \emph{Response diversity} captures the degree to which peer $a$'s response diverges from agent $i$'s current answer, encoding the potential information gain from the interaction; 
(ii) \emph{Peer distinctiveness} measures how distant peer $a$ is to others within the agent network; 
(iii) \emph{Historical performance} provides an exploitation signal via the peer's empirical success rate across prior rounds; and 
(iv) the \emph{Interaction round} encodes the interaction round within a sample, allowing the agent to modulate exploration intensity over time.
We provide detailed descriptions of the features in Appendix~\ref{apdx:features}, and discuss the learned feature weights in Appendix~\ref{apdx:feature_importance}.

\textbf{Exploration via optimism.} For each agent $i$ and candidate peer $a \in \mathcal{N}$, we maintain a design matrix $\mathbf{A}_{i,a} \in \mathbb{R}^{d \times d}$ and reward vector $\mathbf{b}_{i,a}
\in \mathbb{R}^d$, initialized as
$
    \mathbf{A}_{i,a} = \lambda \mathbf{I}$ and 
    $\mathbf{b}_{i,a} = \mathbf{0}$,
where $\lambda>0$ is a regularization parameter.
These statistics yield a ridge-regression estimate of the expected reward model weights:
\begin{equation}
    \hat{\boldsymbol{\theta}}_{i,a} = \mathbf{A}_{i,a}^{-1} \mathbf{b}_{i,a}.
    \label{eq:maxmab_weight}
\end{equation}
At interaction round $t$, given contextual features $\mathbf{x}_{i,a,t}$ describing the potential interaction between agent $i$ and peer $a$, agent $i$ selects a peer according to the LinUCB~\citep{li2010contextual} rule:
\begin{equation}
    a_{i,t} = \arg\max_{a \in [N]} \left[
        \hat{\boldsymbol{\theta}}_{i,a}^\top \boldsymbol{\mathbf{x}}_{i,a,t}
        + \alpha \sqrt{\boldsymbol{\mathbf{x}}_{i,a,t}^\top \mathbf{A}_{i,a}^{-1}
        \boldsymbol{\mathbf{x}}_{i,a,t}}
    \right],
    \label{eq:ucb}
\end{equation}
where the first term estimates the expected reward of querying peer $a$, and the second term is an uncertainty bonus that encourages exploration, and $\alpha>0$ is the exploration coefficient~(see Appendix~\ref{apdx:alpha_sensitivity} for sensitivity analysis). Critically, because the uncertainty bonus operates over the
relational feature space rather than raw selection counts, the agent can
distinguish between a peer that is under-tested globally and one that is
under-tested in the specific relational context of the current task. This provides a
finer-grained form of exploration than scalar UCB affords.

\textbf{Sufficient statistic updates.} After selecting peer $a_{i,t}$, agent $i$ receives reward $r_{i,t}$, defined as the average of the performance metric and the improvement of its response after incorporating information from the selected peer. 
The sufficient statistics are then updated as
\begin{equation}
\label{eq:param_update}
\mathbf{A}_{i,a'}
\leftarrow
\mathbf{A}_{i,a'}
+
\boldsymbol{\mathbf{x}}_{i,a',t}\boldsymbol{\mathbf{x}}_{i,a',t}^{\top} \;\;,\;\;
\mathbf{b}_{i,a'}
\leftarrow
\mathbf{b}_{i,a'}
+
r_{i,t}\boldsymbol{\mathbf{x}}_{i,a',t} \qquad (a':=a_{i,t}),
\end{equation}
after which the parameter estimate $\hat{\boldsymbol{\theta}}_{i,a}$ is recomputed via \eqref{eq:maxmab_weight}.
The reward is computed based on both the agent's performance and the magnitude of improvement.
Detailed reward function description is in Appendix~\ref{apdx:reward}, and the algorithm pseudocode is provided in Appendix~\ref{apdx:algorithm}.

\vspace{-3mm}
\section{Experiments}
\vspace{-3mm}
\subsection{Agent Heterogeneity Setup}
\vspace{-3mm}
To evaluate exploration in heterogeneous multi-agent environments, we consider two representative sources of heterogeneity commonly encountered in realistic agent ecosystems: \textit{contextual diversity} and \textit{parametric diversity}. These settings capture complementary forms of uncertainty that agents must navigate when selecting collaborators and communication strategies.

\textbf{Exploration under contextual diversity.}
This setting models environments in which agents have access to different subsets of the global context, such as distinct retrieved documents, memories, or tools some of which may be incomplete, irrelevant, or misleading. 
Each agent is unaware of the contextual information available to its peers, and must therefore explore which collaborators have the necessary information to solve a task through interaction.
To simulate this setting, we adopt the distractor configuration of HotpotQA~\citep{yang2018hotpotqa}, which contains 10 disjoint contexts: 2 relevant evidence passages and 8 irrelevant distractors. 
Each evidence passage is then distributed across 10 \texttt{Qwen2.5-7B-Instruct}~\citep{yang2024qwen2} agents.
For each data sample, we let the agents engage in $R=5$ rounds of discussions, where each round corresponds to a single timestep.

\textbf{Exploration under parametric diversity.}
This setting models environments in which agents differ in intrinsic reasoning ability due to heterogeneous model families or parameter scales, as in realistic open-agent ecosystems. 
To simulate this setting, we construct a multi-agent system composed of heterogeneous LLMs, comprising 4 model types:
\{\texttt{GPT-5}~\citep{openai2025gpt5}, \texttt{Qwen2.5-7B-Instruct}~\citep{yang2024qwen2}, \texttt{Llama3.1-8B-Instruct}~\citep{grattafiori2024llama}, and \texttt{Mistral-7B-v0.3}~\citep{jiang2023mistral}\}. 
For each sample, agents interact for $R=3$ rounds on two downstream tasks, Math500~\citep{lightman2023lets} and GPQA~\citep{rein2024gpqa}.
Note that we use fewer interaction rounds in this setup, as it involves only four agents.


\subsection{Experimental Details}
\vspace{-2mm}
\textbf{Baselines.}
We compare MACE against three baselines: (1) \textit{In-Context Exploration}: agents
make peer selection decisions autonomously using only in-context information.
At each round, each agent is presented with the full interaction history,
including how many times each peer has been queried and whether their responses
were correct, and is explicitly prompted to balance exploration and exploitation.
Crucially, this baseline has access to the same information as MACE but relies
entirely on the LLM's in-context reasoning to act on it, with no algorithmic
structure imposed on the selection decision. 
(2) \textit{Random}: At each interaction step, agents uniformly sample a peer to interact with at random. (3) \textit{Pre-defined}: The connectivity between agents is pre-defined, where each agent chooses its right-side neighbor as its peer. 
This shares the spirit of most multi-agent debate architectures with fixed communication structures~\citep{du2024improving,li2024improving}.

\textbf{Evaluation.}
For contextual diversity, we evaluate answer quality and regret computation on HotpotQA distractor mode~\citep{yang2018hotpotqa} using Exact Match (EM) and F1, and restrict evaluation to 600 samples labeled as {hard}.
For parametric diversity, we evaluate final-answer accuracy on \texttt{Math500}~\citep{lightman2023lets} and \texttt{GPQA}~\citep{rein2024gpqa}. Specifically, we use Level 4-5 problems from Math500 and the Diamond split of GPQA.
Furthermore, for each setup, we evaluate performance across two phases.
During the \emph{trial-and-error phase}, all methods interact with peers and accumulate
observations. During the \textit{exploitation phase}, the linear parameters
learned by MACE are frozen and reused without further updates; In-Context
Exploration retains its accumulated interaction statistics; and Random and
Pre-defined, which maintain no learnable parameters, operate identically to
their exploration phase.
Each benchmark dataset is split evenly into two halves: the first half is used for the trial-and-error phase, and the second half is reserved for the exploitation phase. 
Additional experimental details and prompt templates are provided in Appendix~\ref{apdx:experiment_details}.



\subsection{Results and Diagnosis}

\begin{figure}[t]
    \centering
    \vspace{-10mm}
    \begin{subfigure}[t]{0.49\textwidth}
        \centering
        \includegraphics[width=\linewidth]{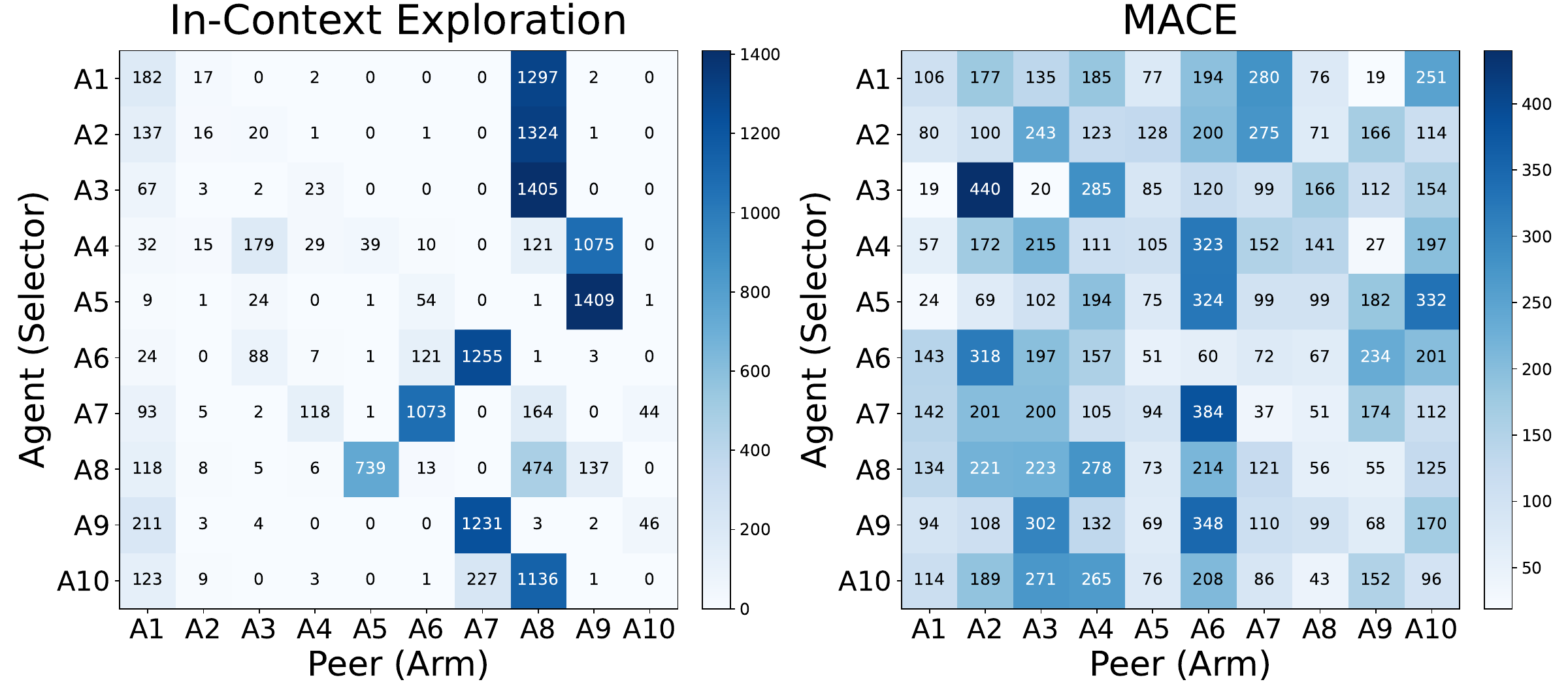}
        \caption{Contextual Diversity}
        \label{fig:exploration_f1}
    \end{subfigure}
    \hfill
    \begin{subfigure}[t]{0.49\textwidth}
        \centering
        \includegraphics[width=\linewidth]{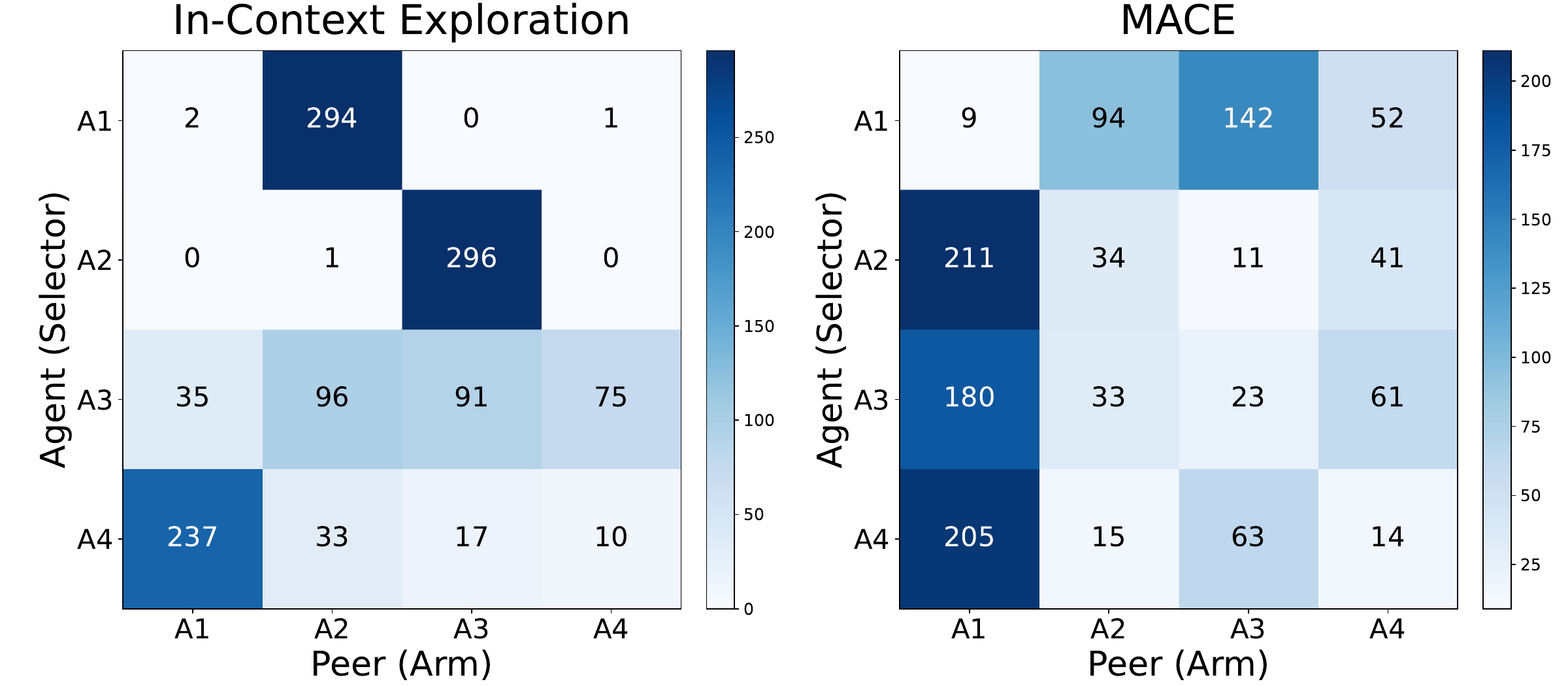}
        \caption{Parametric Diversity}
        \label{fig:exploration_f1}
    \end{subfigure}
    \vspace{-2mm}
    \caption{\textbf{Peer Selection distribution comparison of In-context Exploration and MACE}. (a) HotpotQA benchmark; All 10 agents are Qwen2.5-7B-Instruct, each with heterogeneous context. (b) GPQA benchmark; A1--A4 agents are GPT-5, Qwen2.5-7B-Instruct, Llama3.1-8B-Instruct, and Mistral-7B-v0.3, respectively.}
    \label{fig:heatmap}
\end{figure}

\textbf{\textcolor{mydarkblue}{LLM agents exhibit insufficient exploration in multi-agent environments.}}
Figure~\ref{fig:heatmap} visualizes peer-selection distributions under exploration settings with both contextual and parametric diversity. In Figure~\ref{fig:heatmap}(a), corresponding to contextual diversity, In-Context Exploration produces highly concentrated selection patterns, with agents often locking onto a single peer. This is undesirable because the agent holding the relevant context is randomized across samples and interaction rounds; consistently selecting the same peer therefore signals a clear failure to explore. By contrast, MACE induces substantially more distributed selections, suggesting better coverage of the available peers.
A different pattern appears in Figure~\ref{fig:heatmap}(b), corresponding to parametric diversity. Here, some convergence toward a generally strong agent is expected. However, In-Context Exploration still exhibits unnaturally sharp and poorly calibrated concentration, with agents collapsing onto what appears to be an arbitrary early favorite rather than exploring the full pool. For example, even GPT-5~(A1) selects Qwen-7B~(A2) in 294 out of 297 interactions, while almost never querying the other agents. 
This suggests that LLMs do not reliably identify strong collaborators through systematic exploration, but instead overcommit prematurely to early signals. 
In contrast, MACE produces more balanced selection patterns, indicating more stable and effective exploration before exploitation.


\begin{figure}[t]
    \centering
    \vspace{-3mm}
    \begin{subfigure}[t]{0.32\textwidth}
        \centering
        \includegraphics[width=\linewidth]{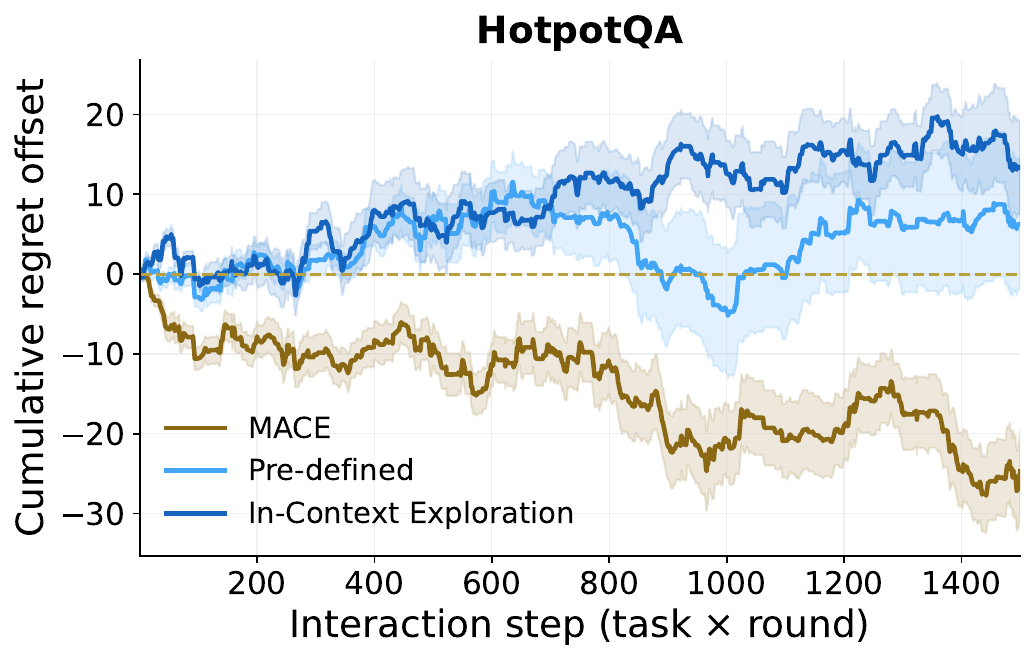}
        \label{fig:exploration_f1}
    \end{subfigure}
    \hfill
    \begin{subfigure}[t]{0.32\textwidth}
        \centering
        \includegraphics[width=\linewidth]{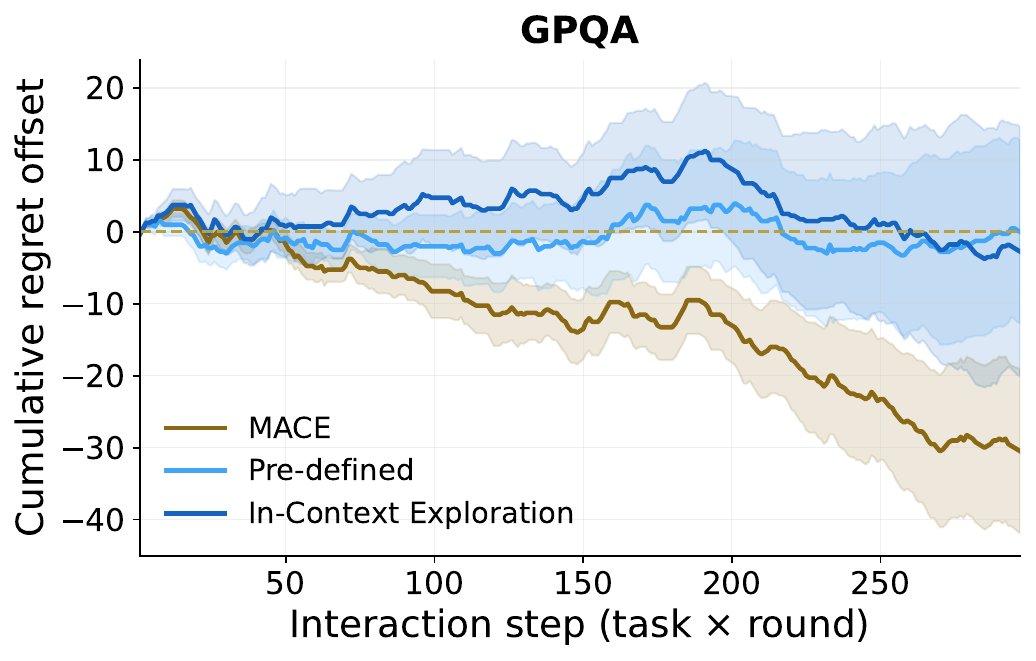}
        \caption{Trial-and-Error Phase}
        \label{fig:exploration_f1}
    \end{subfigure}
    \hfill
    \begin{subfigure}[t]{0.32\textwidth}
        \centering
        \includegraphics[width=\linewidth]{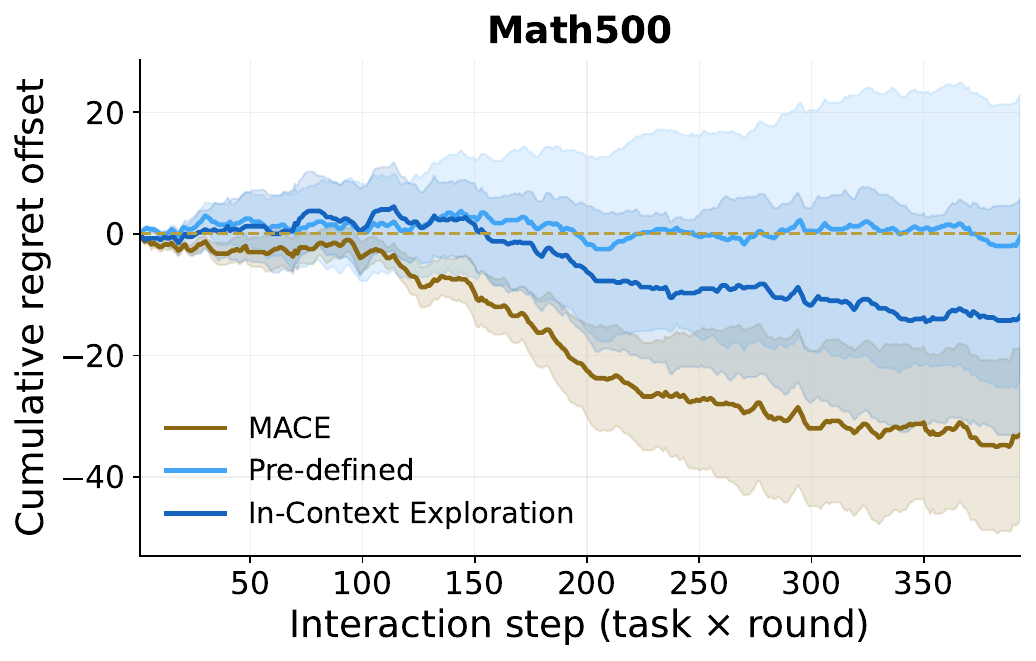}
        \label{fig:exploration_f1}
    \end{subfigure}
    \hfill
    \begin{subfigure}[t]{0.32\textwidth}
        \centering
        \includegraphics[width=\linewidth]{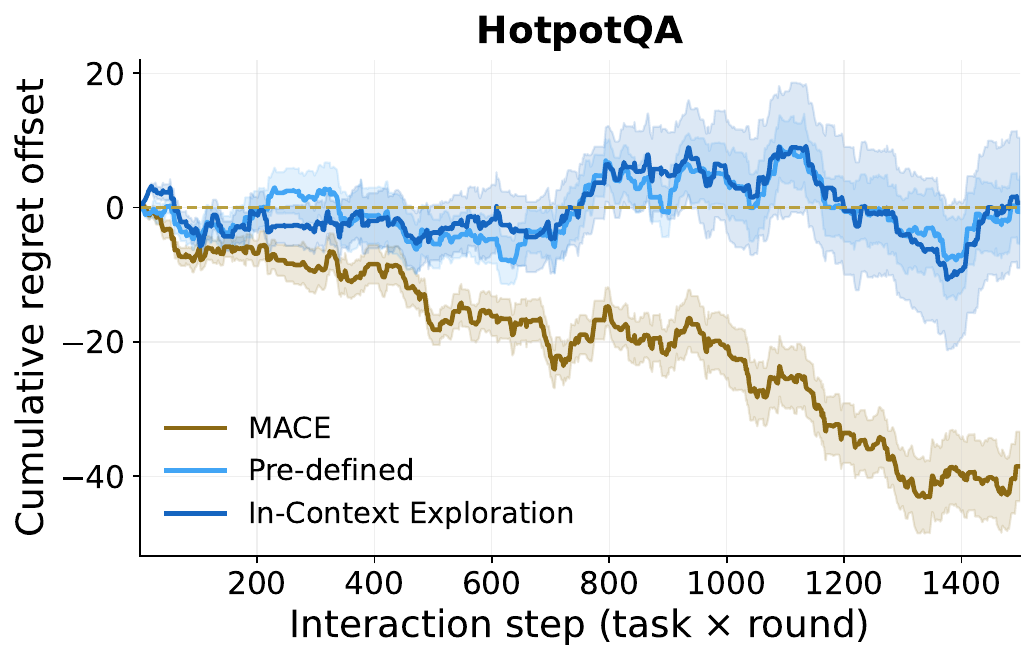}
        \label{fig:exploitation_f1}
    \end{subfigure}
    \hfill
    \begin{subfigure}[t]{0.32\textwidth}
        \centering
        \includegraphics[width=\linewidth]{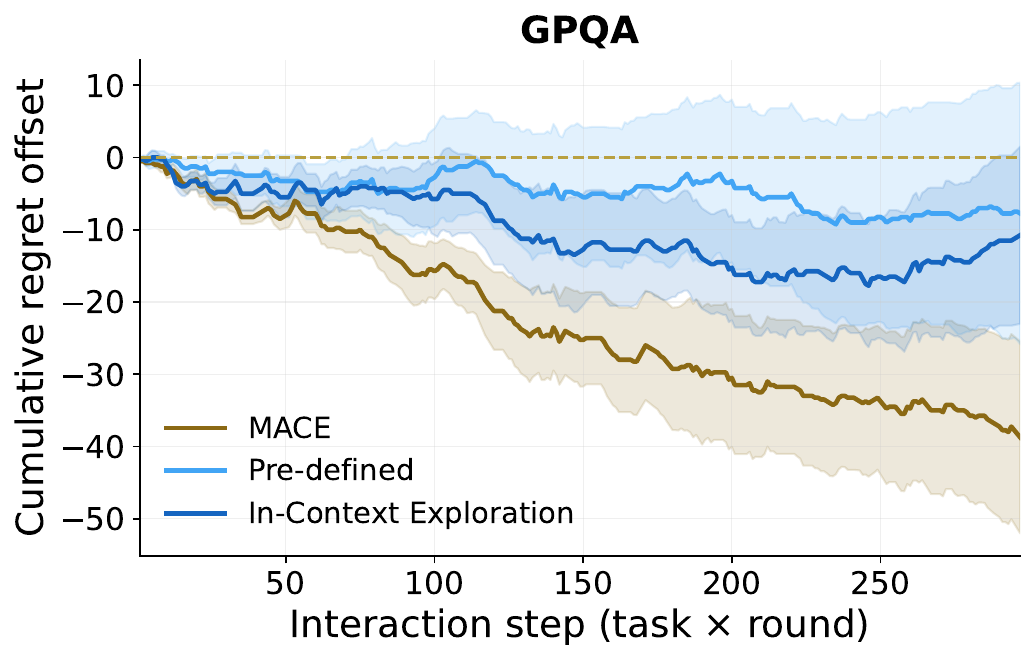}
        \caption{Exploitation Phase}
        \label{fig:exploitation_f1}
    \end{subfigure}
    \hfill
    \begin{subfigure}[t]{0.32\textwidth}
        \centering
        \includegraphics[width=\linewidth]{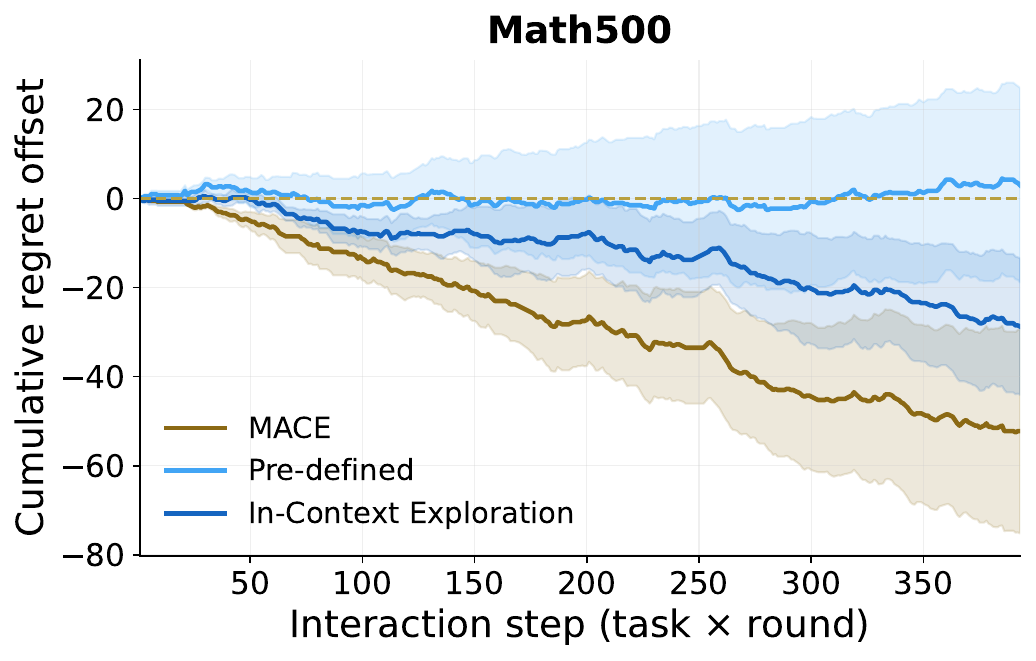}
        \label{fig:exploitation_f1}
    \end{subfigure}
    \vspace{-2mm}
    \caption{\textbf{Cumulative regret offset across tasks and phases.} The cumulative regret offset~(Baseline $-$ Random), averaged across agents, is shown over interaction steps~(lower is better). The horizontal dashed line at Offset $=0$ is the Random baseline, and the shaded areas indicate the standard error across participating agents.}
    \vspace{-4mm}
    \label{fig:explore_exploit}
\end{figure}

\textbf{\textcolor{mydarkblue}{In-context exploration may underperform random exploration, while explicit guided-exploration yields large gains.}}
In Figure~\ref{fig:explore_exploit}, we compare cumulative regret offsets across exploration strategies, defined as the difference in cumulative regret between a given strategy and the random baseline, \textit{i.e.}, $\textrm{Regret}_i^{\textrm{Strategy}} - \textrm{Regret}_i^{\textrm{Random}}$~(\eqref{eq:regret}).
The figure reveals a striking result that under \textit{contextual diversity} (top row), In-Context Exploration performs worse than Random peer selection, indicating that autonomous LLM agents not only fail to explore effectively, but can be less reliable than a naive stochastic policy. 
This suggests that current LLM agents relying on prompting alone produce systematically poor exploration behavior in multi-agent interactions. 
In contrast, MACE substantially improves performance throughout the exploration phase, demonstrating that even lightweight exploration guidance can dramatically enhance interaction quality. 
These gains persist into the exploitation phase (bottom row), where the learned policies continue to outperform all baselines after exploration is deliberately disabled. 
We observe similar trends under parametric diversity setup on both Math500 and GPQA, demonstrating that the benefits of explicitly guided exploration generalize across heterogeneous model pools and task domains. 
In Appendix~\ref{apdx:roundwise}, we also show the task performance for each interaction round.
Notably, even the strongest model in the system, \textsc{GPT-5}, benefits from the proposed exploration mechanisms~(Table~\ref{tab9}--\ref{tab16}), suggesting that insufficient exploration is a fundamental limitation of current multi-agent LLMs in general.

\begin{wrapfigure}{r}{0.4\textwidth}
  \vspace{-4mm}
  \centering
  \includegraphics[width=0.39\textwidth]{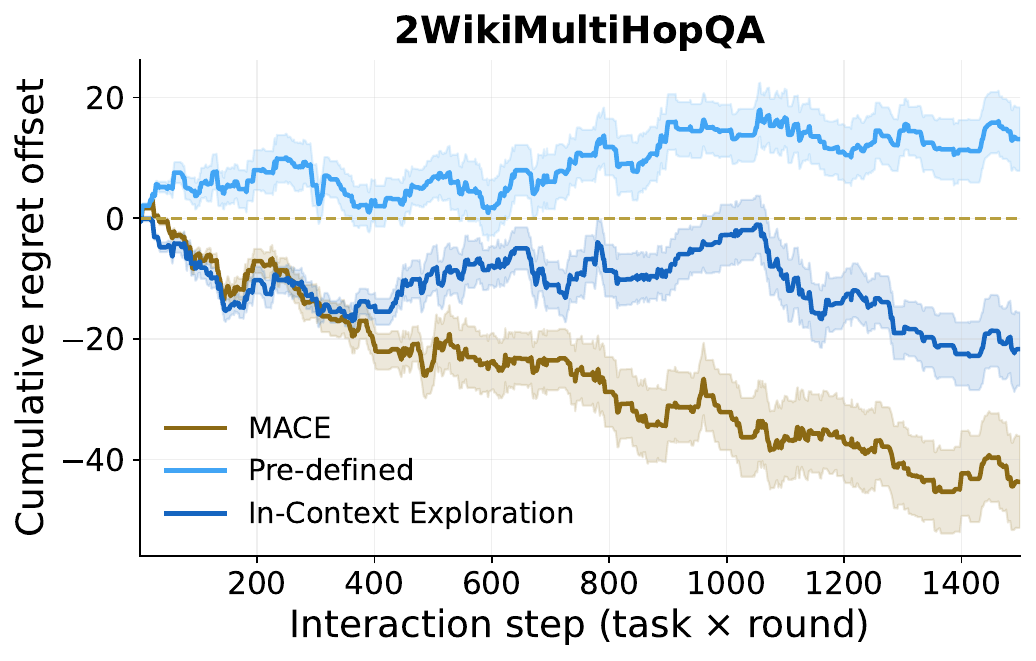}
  \caption{The parameters learned from HotpotQA can be transferred to a different benchmark, 2WikiMultiHopQA~\citep{xanh2020_2wikimultihop}.}
  \label{fig:transfer_f1}
  \vspace{-3mm}
\end{wrapfigure}

\textbf{\textcolor{mydarkblue}{Strategies learned by MACE generalize to an unseen task.}}
We evaluate transferability by applying the parameters learned on HotpotQA to a more challenging benchmark, 2WikiMultihopQA~\citep{xanh2020_2wikimultihop}~(details in Appendix~\ref{apdx:dataset_details}). 
As in the contextual diversity setup, each agent receives partial context, with only a subset having access to the information required to solve the question. 
All other baselines explore on a 300-sample subset of benchmark, while MACE operates purely in exploitation mode with frozen parameters. 
Despite never being trained on this benchmark, MACE achieves robust performance that outperforms all baselines in Figure~\ref{fig:transfer_f1}.
This indicates that the learned interaction strategies capture generalizable structure rather than task-specific artifacts.

We evaluate transferability by applying the parameters learned on HotpotQA to a more challenging benchmark, 2WikiMultihopQA~\citep{xanh2020_2wikimultihop}~(details in Appendix~\ref{apdx:dataset_details}).
As in the contextual diversity setup, each agent receives partial context, with only a subset having access to the information required to solve the question.
{\color{black}All other baselines explore on a 300-sample subset of the benchmark, while MACE operates purely in exploitation mode with both its parameters and source-derived historical reward statistics frozen; only response-based features and the interaction-round feature vary during transfer, without target-task reward feedback.}
Despite never being trained on this benchmark, MACE achieves robust performance that outperforms all baselines in Figure~\ref{fig:transfer_f1}.
This indicates that the learned interaction strategies capture generalizable structure rather than task-specific artifacts.

The following additional analyses are deferred to the Appendix: 
$\Circled{1}$ Interaction round-wise effect on task performance~(Appendix~\ref{apdx:roundwise}), 
$\Circled{2}$ Sensitivity analysis on the exploration coefficient~$\alpha$~(Appendix~\ref{apdx:alpha_sensitivity}), 
$\Circled{3}$ Experiments with a set of stronger frontier models~(Appendix~\ref{apdx:frontier}),
$\Circled{4}$ Homogeneous agent setting experiments~(Appendix~\ref{apdx:homogeneous}), 
$\Circled{5}$ Temporal-Difference algorithm extension of MACE and discussions on potential emergence of cooperative behaviors~(Appendix~\ref{apdx:mace_td}),
$\Circled{6}$ Feature selection analyses~(Appendix~\ref{apdx:feature_importance}).


\section{Theoretical Analysis}
\label{sec:theory}
\vspace{-2mm}
In this section, we provide formal guarantees on the benefit of MACE with explicit exploration under simplified conditions.
Theorem~\ref{thm:mace_regret} establishes a sublinear expected-regret bound for MACE under the stated assumptions.
Theorem~\ref{thm:greedy_regret} constructs a stationary instance on which empirical-mean greedy selection incurs linear expected regret through premature commitment.
Corollary~\ref{cor:gap} then establishes an asymptotically linear expected-regret separation on the constructed instance.
We specify assumptions for our theorems in Appendix~\ref{app:theory}, and all proofs are in Appendix~\ref{apdx:proofs}.



\begin{definition}[\textbf{Capability Diversity}]
\label{def:diversity}
Let each agent $a \in \mathcal{N}$ be associated with a latent capability vector
$\mathbf{c}_a \in \mathbb{R}^k$, encoding its strengths across reasoning domains
or knowledge areas. The {capability diversity} of the agent pool is
defined as:
\begin{equation}
    \delta := \frac{1}{N} \sum_{a=1}^{N} \|\mathbf{c}_a - \bar{\mathbf{c}}\|_2,
    \qquad \bar{\mathbf{c}} := \frac{1}{N} \sum_{a=1}^{N} \mathbf{c}_a.
    \label{eq:diversity}
\end{equation}
\end{definition}
The quantity $\delta$ measures dispersion in the agents' capability
representations.
Its relationship to peer-selection rewards depends on how capabilities affect performance on the task distribution.
In the stationary construction used below, we substitute scalar capabilities with peer reward means.
For instance, with one optimal peer and $N-1$ equally inferior peers separated by a reward gap $\Delta$, this gives $\delta=2(N-1)\Delta/N^2$.

\begin{theorem}[\textbf{Expected regret of MACE}]
\label{thm:mace_regret}
Under exact linear realizability, boundedness, and noise assumptions~(Appendix~\ref{app:theory}), let $\lambda\geq1$
and choose $\alpha_T$ as a simultaneous confidence radius
with failure probability $1/T$. For $T\geq2$, MACE satisfies
\begin{equation}
    \mathbb{E}\!\left[
        \mathrm{Regret}_i^{\mathrm{MACE}}
    \right]
    \leq
    2\alpha_T
    \sqrt{
        2TNd\log\!\left(1+\frac{T}{d\lambda}\right)
    }
    +\frac{3}{2},
    \label{eq:mace_regret}
\end{equation}
where $\alpha_T=O(\sqrt{\log T})$,
yielding an $O(\sqrt{T}\log T)$ expected-regret bound.
\end{theorem}

\begin{theorem}[\textbf{Regret of non-exploring policy}]
\label{thm:greedy_regret}
For every $N\geq2$, there exists a stationary $N$-peer
instance with capability diversity $\delta>0$ on which
a greedy agent, after querying each peer once, selects
$a_{i,t}\in\arg\max_a\widehat{\mu}_{i,a,t}$ and incurs
\begin{equation}
    \mathbb{E}\!\left[    \mathrm{Regret}_i^{\mathrm{non\text{-}exploring}}
    \right]
    \geq
    \frac{1}{8}\delta(T-1),
    \label{eq:greedy_bound}
\end{equation}
for all $T\geq N$, where $\widehat{\mu}_{i,a,t}$ is the
empirical reward mean, and scalar capabilities are identified with peer reward means.
\end{theorem}
\begin{corollary}[\textbf{Benefit of explicit exploration}]
\label{cor:gap}
On the stationary $N$-peer instance constructed in
Theorem~\ref{thm:greedy_regret}, represented as a linear
bandit with constant scalar features, MACE with the
confidence radius from Theorem~\ref{thm:mace_regret}
satisfies, for $T\geq N$,
\begin{equation*}
    \mathbb{E}\!\left[
        \mathrm{Regret}_i^{\mathrm{non\text{-}exploring}}
        -
        \mathrm{Regret}_i^{\mathrm{MACE}}
    \right]
    \geq\;
    \frac{\delta}{8}(T-1) -
    2\alpha_T
    \sqrt{2TN\log\!\left(1+\frac{T}{\lambda}\right)}
    -\frac{3}{2}.
\label{eq:exploration_gap}
\end{equation*}
For each fixed $N$, the expected regret gap is
$\Omega(T)$ as $T\to\infty$.
\end{corollary}

\vspace{-3mm}
\begin{implicationbox}
{\textbf{Implication: Explicit exploration can prevent
persistent peer-selection errors.}
For every fixed population size $N\geq2$, our construction
exhibits a stationary environment in which greedy selection
incurs linear expected regret through premature commitment.
Under the stated linear-bandit assumptions, MACE achieves
sublinear expected regret, yielding an asymptotically linear
separation. This establishes a concrete benefit of explicit
exploration.}
\end{implicationbox}

\begin{figure}[t!]
\centering
\vspace{-10mm}
\hspace*{-10mm}
\begin{tikzpicture}

\begin{axis}[
    name=plotA,
    width=0.42\textwidth,
    height=0.33\textwidth,
    xlabel={Interaction Steps $T$},
    ylabel={Expected Cumulative Regret},
    xmin=0, xmax=280,
    ymin=0, ymax=140,
    legend style={
        at={(0.03,0.97)},
        anchor=north west,
        font=\scriptsize,
        draw=none,
        fill=white,
        fill opacity=0.8,
        text opacity=1,
    },
    grid=major,
    grid style={gray!20},
    tick label style={font=\scriptsize},
    label style={font=\small},
    title style={font=\scriptsize},
    title={Schematic: fixed $N$ and $\delta$},
]

\addplot[
    red, thick, solid,
    domain=2:280, samples=100
] {0.6 * sqrt(x) * ln(x)};
\addlegendentry{MACE: $O(\sqrt{T}\log T)$}

\addplot[
    blue!80!black, thick, dashed,
    domain=2:280, samples=100
] {0.5 * (x-1)};
\addlegendentry{Greedy: $\Omega(T)$}

\draw[<->, thick, black!60]
    (axis cs:260,{0.6 * sqrt(260) * ln(260)})
    --
    (axis cs:260,{0.5 * (260-1)});

\node[
    font=\scriptsize,
    black!70,
    rotate=90
] at (axis cs:240,75)
    {Asymptotic gap: $\Omega(T)$};
\end{axis}

\node[anchor=west]
at ([xshift=0.3cm, yshift=-0.4cm]plotA.east) {
    \includegraphics[width=0.35\textwidth]
    {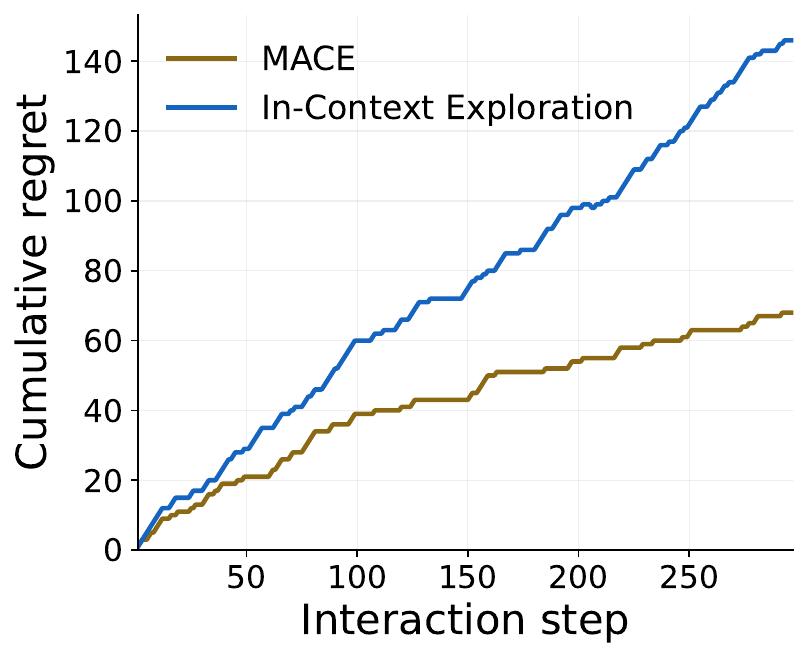}
};

\end{tikzpicture}
\vspace{-3mm}
 \caption{\textbf{Illustration of Theorem.} The asymptotic expected-regret  bounds~(left) align with the empirical behavior of Qwen2.5-7B-Instruct on the GPQA benchmark~(right).}
\vspace{-5mm}
\label{fig:corollary}
\end{figure}
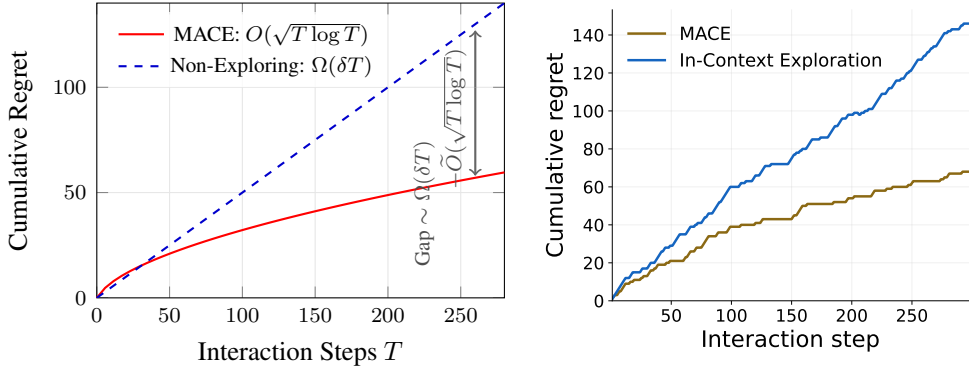

\vspace{-0.1cm}
\paragraph{Empirical illustration.}
Figure~\ref{fig:corollary} (left) schematically illustrates
the separation established by Corollary~\ref{cor:gap}.
The right panel shows empirical regret trajectories for
\texttt{Qwen2.5-7B-Instruct}. The observed advantage of
MACE is qualitatively consistent with the benefit of
explicit exploration.


\vspace{-2mm}
\section{Related Works}
\vspace{-2mm}
Recent works have built foundations for multi-agent LLM coordination through frameworks like CAMEL~\citep{li2023camel}, AutoGen~\citep{wu2024autogen}, ChatDev~\citep{qian2024chatdev}, and MetaGPT~\citep{hong2023metagpt}, and through debate protocols~\citep{du2024improving,chan2024chateval,liu2024groupdebate,liang2024encouraging,becker2025mallm}, though most rely on fixed roles or orchestration rules, with a few attempting to grant autonomy~\citep{hou2025halo,dang2025multi,wang2025megaagent,chen2023agentverse}. Separately, studies on single-agent exploration show that LLMs exhibit insufficient exploration in bandits~\citep{krishnamurthy2024can,nie2025evolve}, underperform humans in open-ended settings~\citep{pan2025large,zhang2025comparing}, and benefit from explicit algorithmic guidance~\citep{arumugam2025toward}, while related work on delegation reveals that verbalized confidence is a poor proxy for partner selection~\citep{tomavsev2026intelligent,buyl2025building,debnath2025can}. 
Our work \textit{extends these findings to the multi-agent setting, showing that exploration failures compound under agent interaction and proposing MACE to explicitly induce the exploration that current agents lack.}
A complete and extended discussion of related works is in Appendix~\ref{apdx:related_works}.


\vspace{-2mm}
\section{Conclusion}
\vspace{-2mm}
In this work, we showed that LLM agents often fail to explore effectively in multi-agent environments, leading to premature commitment, poor peer selection, and higher regret. 
We formalized this as the Multi-Agent Exploration problem and introduced Multi-Agent Contextual Exploration~(MACE), a lightweight solution that explicitly induces exploration through structured peer selection. 
Across diverse settings, MACE improves both exploration and downstream performance, with gains that transfer to unseen tasks. 
Our theory establishes that explicit exploration can prevent persistent peer-selection errors, yielding an asymptotically linear expected-regret advantage over greedy selection on a constructed stationary instance.
Overall, results imply that reliable multi-agent autonomy may require explicit exploration rather than expecting it to emerge on its own.


\section*{AI Use Statement}

In this work, we used generative AI tools for polishing the text, to generate scientific plots, and to help prove theorems.
We have not used generative AI tools for automating research and experiments.
We have reviewed all AI-assisted work. 
We manually checked the writing faithfully conveys our intended content, and if the plots correctly express values.
We take responsibility for the final content of this work, including text, claims or artifacts produced with the aid of generative AI.

\section*{Ethics Statement}

This work studies exploration and peer selection in multi-agent LLM systems. 
All experiments are conducted in simulated multi-agent environments using existing language models and publicly available question-answering benchmarks. 
The study does not involve human participants, personal data, or deployment in real-world decision-making settings. 
We use benchmark datasets and model APIs solely for research purposes and in accordance with their respective terms of use.

\section*{Reproducibility Statement}

We provide the formal problem definition, optimization objective, feature construction, reward functions, and complete pseudocode for \textsc{MACE} in the main paper and Appendix~\ref{apdx:algorithm}. Appendix~\ref{apdx:experiment_details} documents the benchmark datasets, evaluation splits, agent configurations, backbone models, decoding parameters, interaction rounds, evaluation metrics, computational resources, and complete prompt templates used in our experiments.
We additionally report results across both trial-and-error and frozen-parameter exploitation phases, together with baseline comparisons, parameter sensitivity analyses, and round-wise performance.
Upon publication, we will also release the implementation needed to reproduce the experiments, including preprocessing and evaluation scripts, configuration files, random seeds, and instructions for constructing the contextual- and parametric-diversity environments.

\bibliography{iclr2027_conference}
\bibliographystyle{iclr2027_conference}

\newpage
\appendix

\addcontentsline{toc}{section}{Appendix}
\part{Appendix} 
\parttoc 

\clearpage
\section{Appendix for the Delegation Experiment}
\label{apdx:sec2}

\subsection{Experimental Setup}
\label{apdx:toy_setup}

\paragraph{Specifications for the Delegated Tasks.}
Questions are randomly generated very simple arithmetic expressions involving addition, subtraction, and multiplication over integers in $[1,100]$.
Some example questions are: ``What is 10 + 32?", ``What is 40 - 21", or ``What is 24 $\times$ 3?".
When delegated a question, the chosen peer returns the correct answer with probability $p_i$, and an incorrect answer otherwise.

\paragraph{LLM Prompt Template.}
For all LLM-based delegation agents, we use the prompt shown below. 
At each round, the model is given the current arithmetic question together with the historical performance statistics of Peer A and Peer B, including the number of prior delegations, number of correct responses, and empirical success rate. 
The prompt explicitly reminds the model to balance exploitation~(favoring peers with stronger observed accuracy) and exploration (testing peers with fewer prior trials). 
The model is then asked to output a final delegation decision in a constrained XML-style format, either \texttt{<choice>Peer A</choice>} or \texttt{<choice>Peer B</choice>}. 

\paragraph{Runtime Setting.} We use temperature 1.2 for Qwen2.5-7B-Instruct to allow various decision trajectories, and use the default setting for GPT-4 and GPT-5.
The maximum number of tokens is set to 2048 for all settings.

\begin{figure}[h!]
    \centering
    \begin{tcolorbox}[title={Delegation Task Prompt Template}]\small
You are coordinating a team of two peers---Peer A and Peer B---who answer arithmetic questions.
Your sole responsibility is to decide which peer to delegate each question to, based on their past performance.
You cannot answer the question yourself; you must pick a peer.\\

Reason carefully: a peer with a higher historical success rate is probably more reliable, but also consider how many times you have tried each peer—a peer with very few attempts may be underexplored.

Always end your response with your choice in exactly this format:
<choice>Peer A</choice>  or  <choice>Peer B</choice>\\

Question to delegate: \texttt{<question>}\\

Peer performance history:\\
Peer A: asked $\texttt{<N>}$ times, correct $\texttt{<n>}$ times (success rate: $\texttt{<K>}$\%)\\
Peer B: asked $\texttt{<N>}$ times, correct $\texttt{<n>}$ times (success rate: $\texttt{<K>}$\%)\\

Which peer do you want to delegate this question to?

State your final answer as: <choice>Peer A</choice> or <choice>Peer B</choice>
    \end{tcolorbox} 
\end{figure}

\paragraph{UCB1 Algorithm Details.}
As a principled exploration baseline, we compare against the UCB1 algorithm~\citep{auer2002finite}. 
Let $n_i(t)$ denote the number of times peer $i \in \{A,B\}$ has been selected up to round $t$, and let $\hat{\mu}_i(t)$ be its empirical success rate. At each round, UCB1 selects the peer maximizing
\[
\hat{\mu}_i(t) + c \sqrt{\frac{\log t}{n_i(t)}},
\]
where the first term encourages exploitation of high-performing peers and the second term encourages exploration of under-sampled peers. 
In our experiments, we use an exploration coefficient of $c=1.0$. 
Each peer is selected once initially to avoid undefined confidence bonuses. This baseline provides a statistically grounded reference for adaptive exploration under uncertainty.

\subsection{Histograms for All \texorpdfstring{$\boldsymbol{p}_\text{A}$}{pA} Values}
\label{apdx:full_hist}

\begin{figure}[t!]
    \centering
    \begin{subfigure}[h!]{0.3\textwidth}
        \centering
        \includegraphics[width=\linewidth]{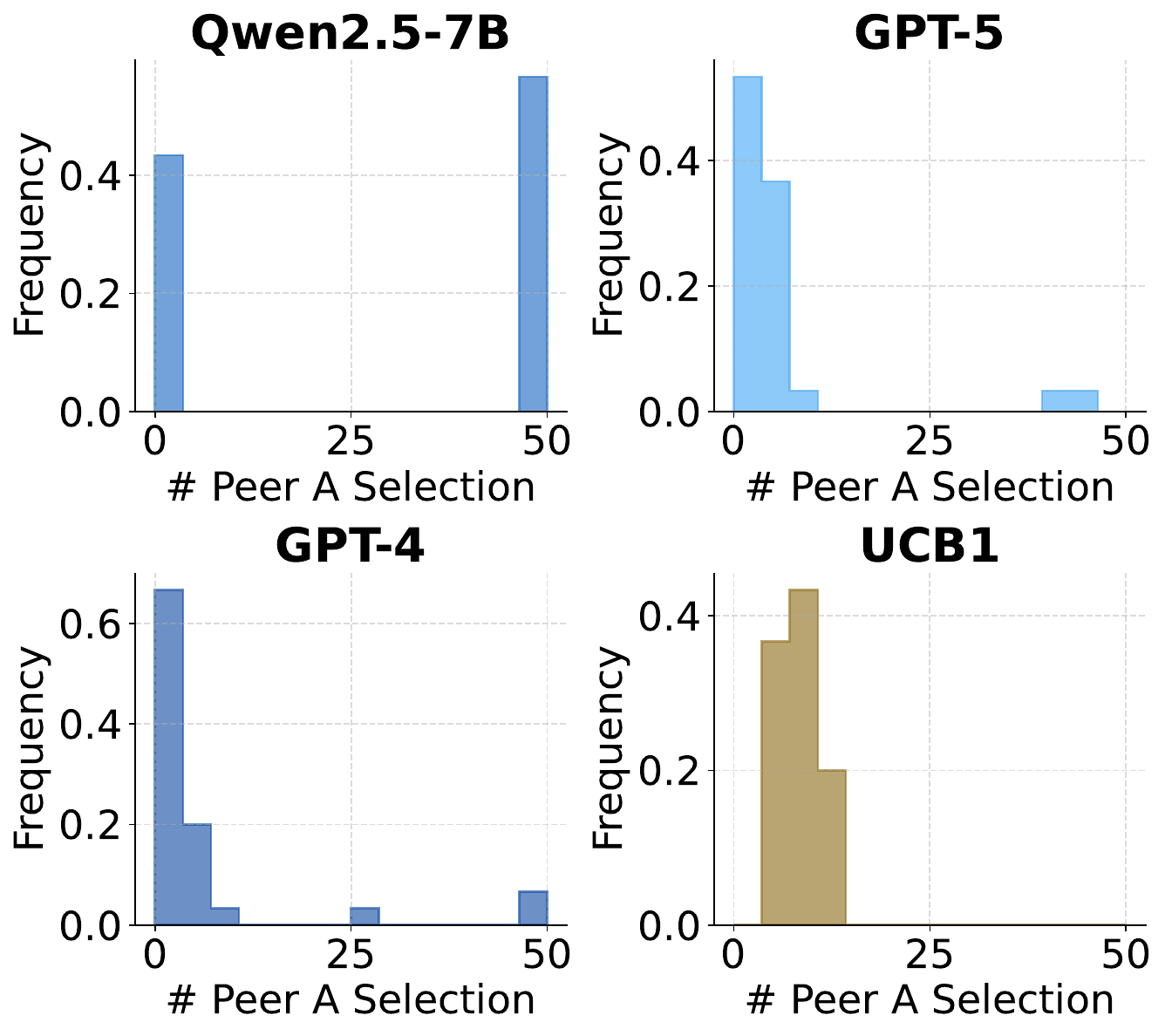}
    \end{subfigure}
    \hfill
    \begin{subfigure}[h!]{0.3\textwidth}
        \centering
        \includegraphics[width=\linewidth]{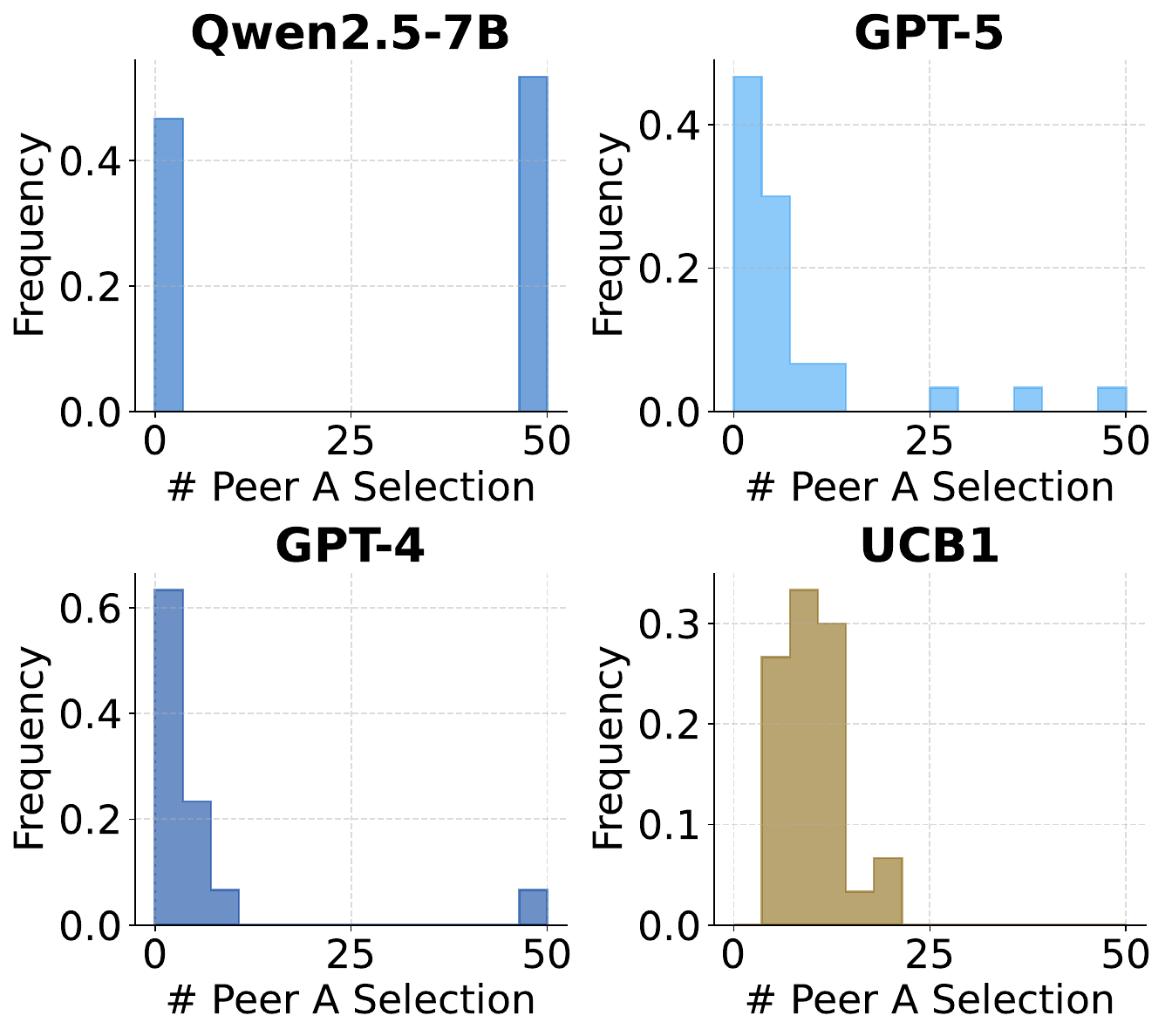}
    \end{subfigure}
    \hfill
    \begin{subfigure}[h!]{0.3\textwidth}
        \centering
        \includegraphics[width=\linewidth]{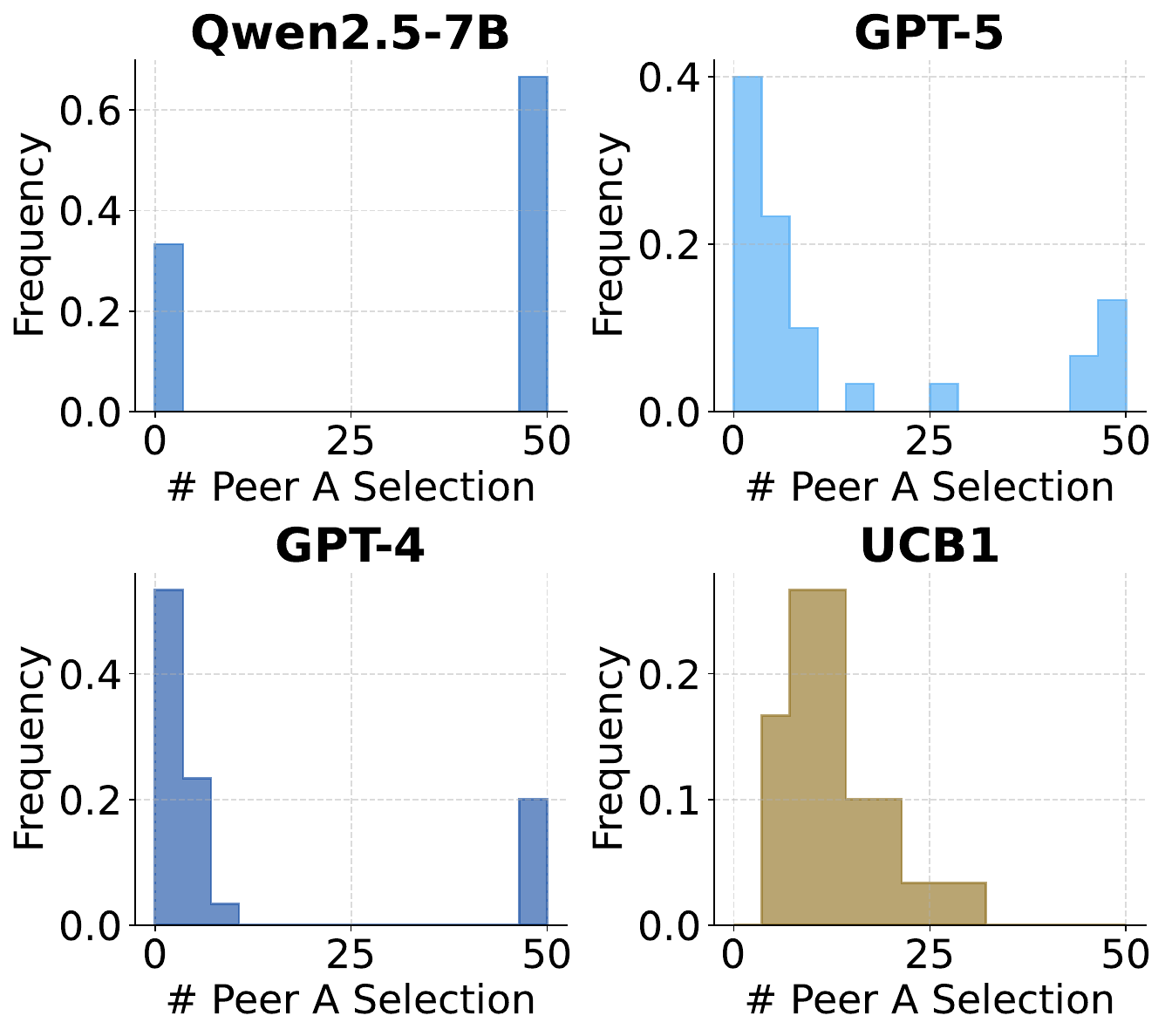}
    \end{subfigure}
    \hfill\vspace{3mm}
    \begin{subfigure}[h!]{0.3\textwidth}
        \centering
        \includegraphics[width=\linewidth]{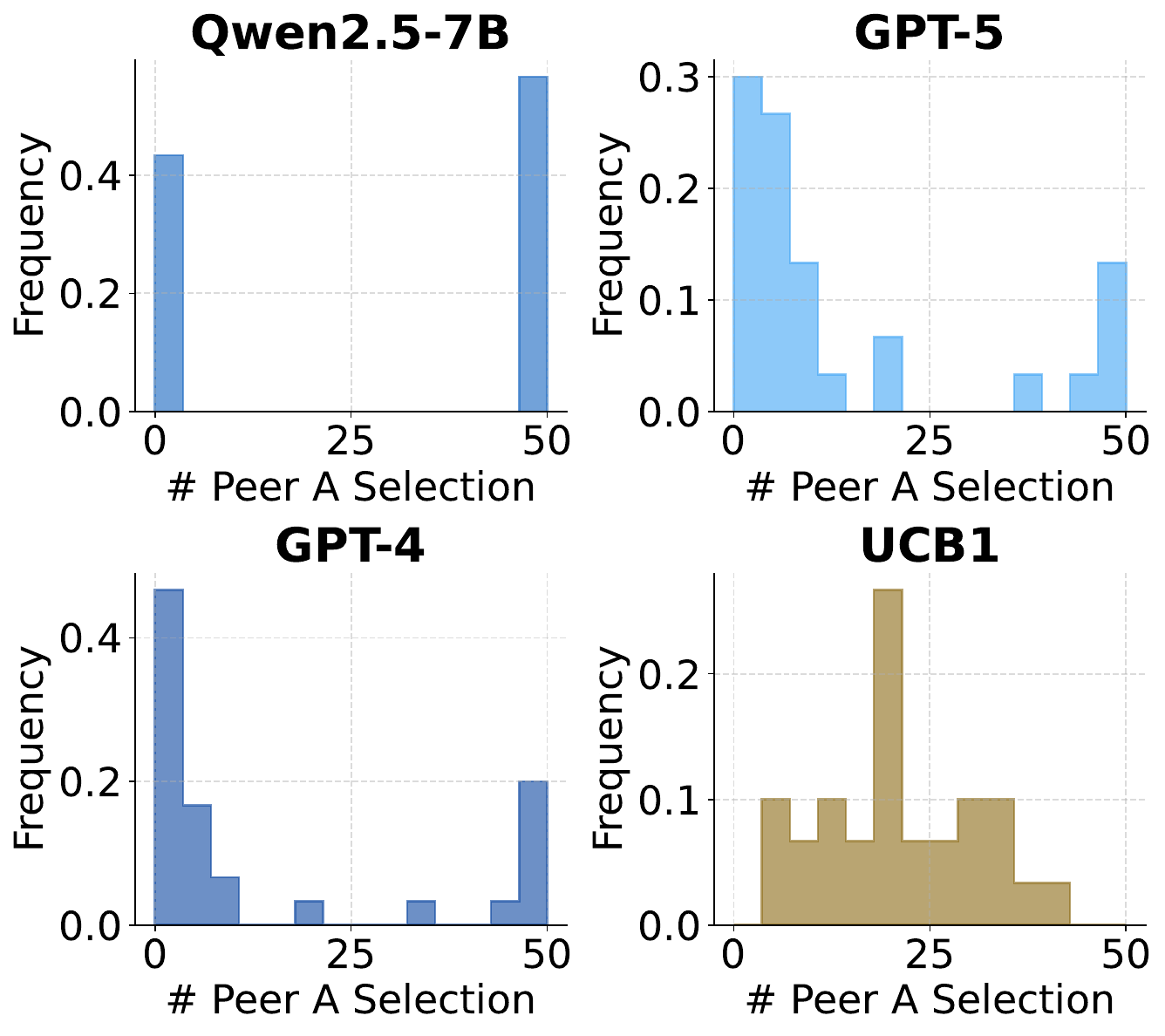}
    \end{subfigure}
    \hfill
    \begin{subfigure}[h!]{0.3\textwidth}
        \centering
        \includegraphics[width=\linewidth]{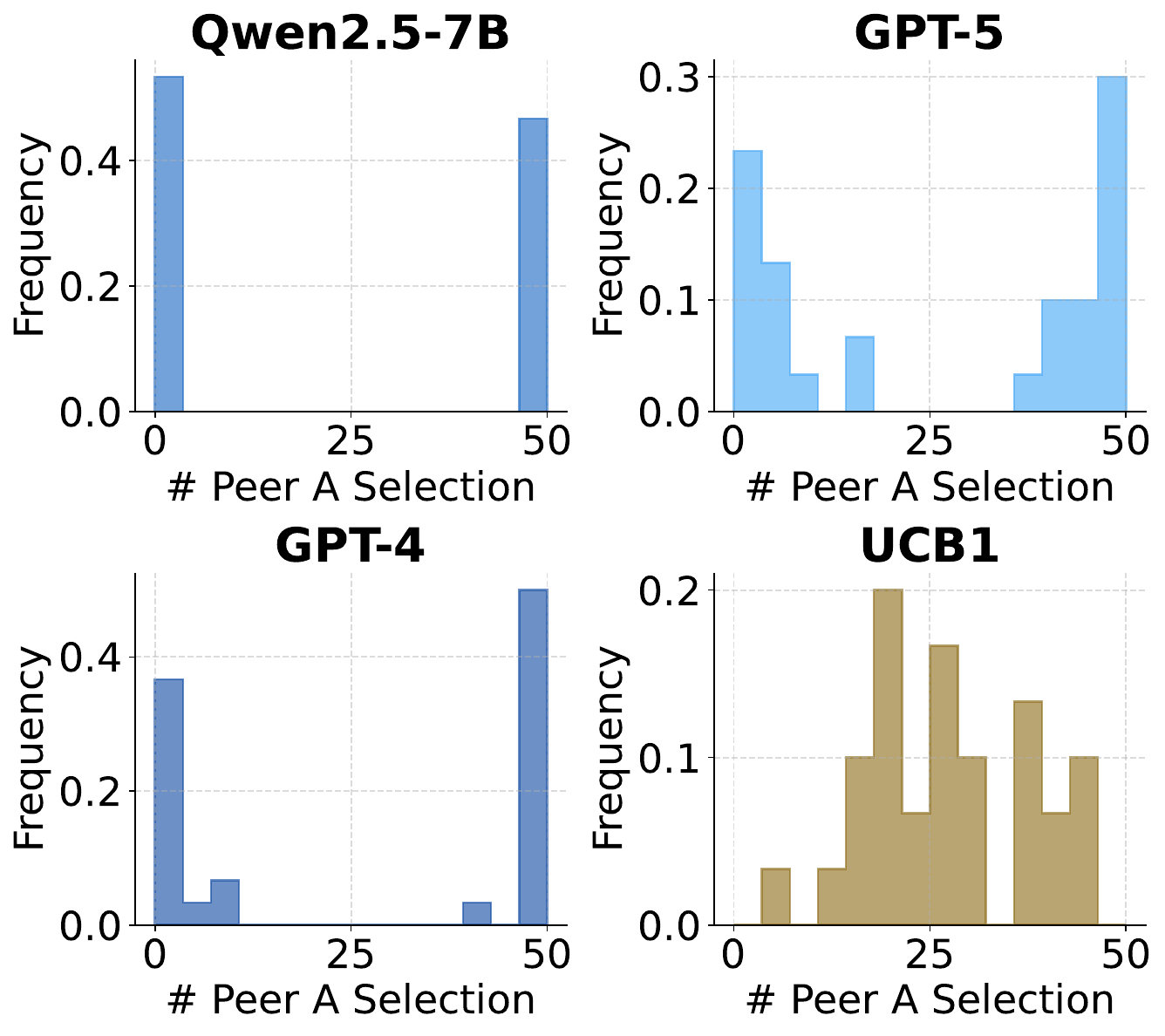}
    \end{subfigure}
    \hfill
    \begin{subfigure}[h!]{0.3\textwidth}
        \centering
        \includegraphics[width=\linewidth]{B_Figures/figs/hist_pA_0.6_grid.pdf}
    \end{subfigure}
    \hfill\vspace{3mm}
    \begin{subfigure}[h!]{0.3\textwidth}
        \centering
        \includegraphics[width=\linewidth]{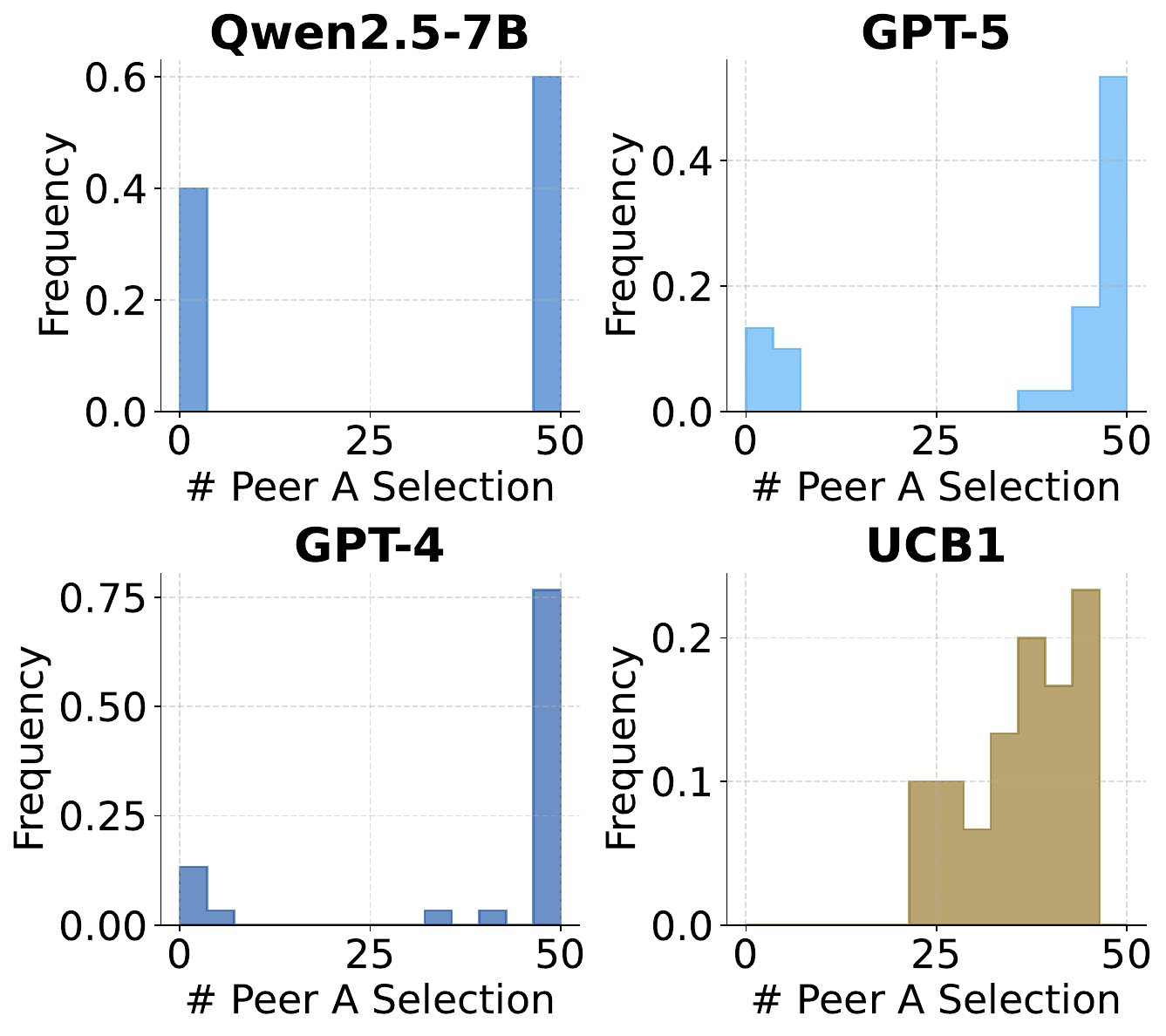}
    \end{subfigure}
    \hfill
    \begin{subfigure}[h!]{0.3\textwidth}
        \centering
        \includegraphics[width=\linewidth]{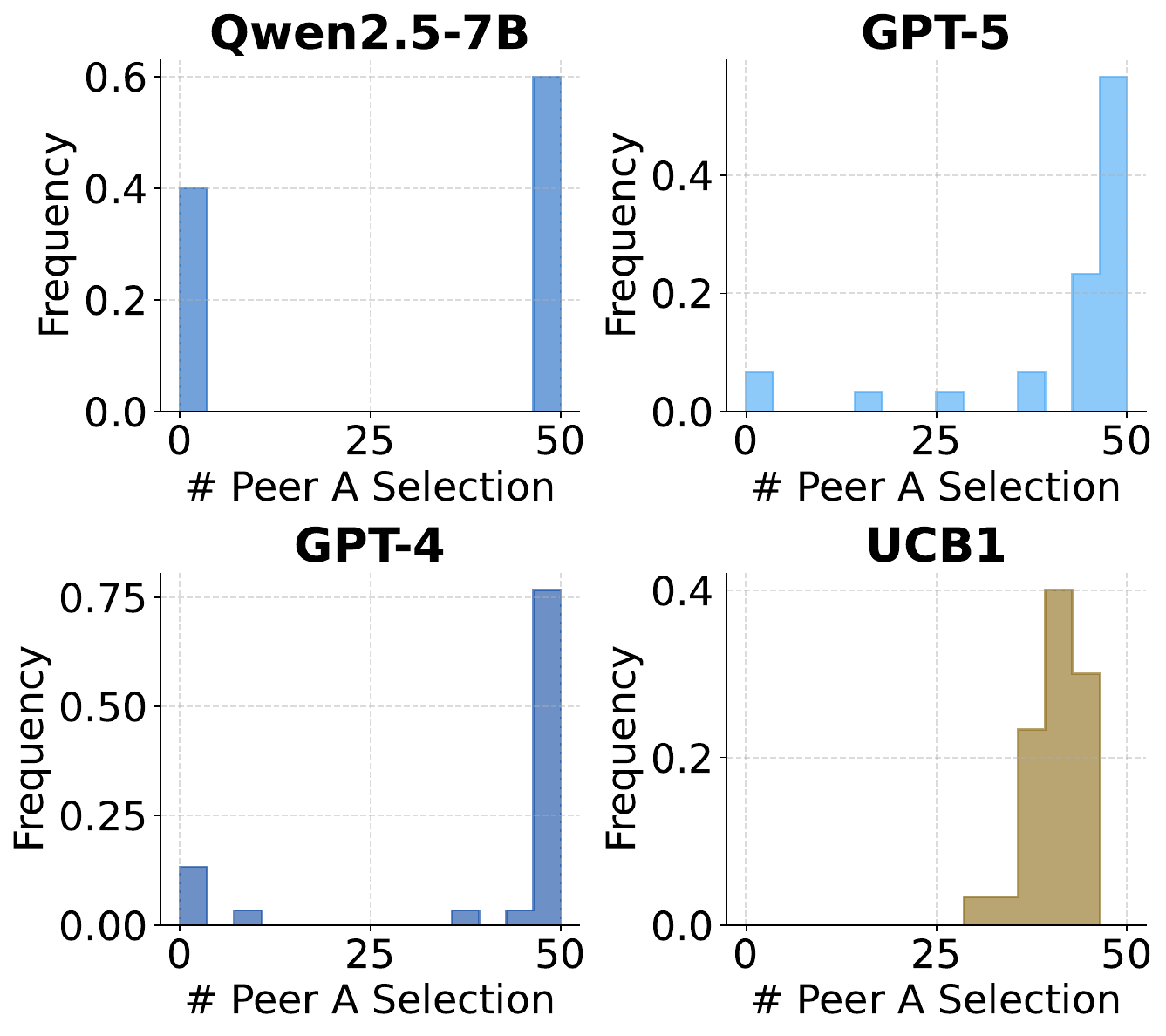}
    \end{subfigure}
    \hfill
    \begin{subfigure}[h!]{0.3\textwidth}
        \centering
        \includegraphics[width=\linewidth]{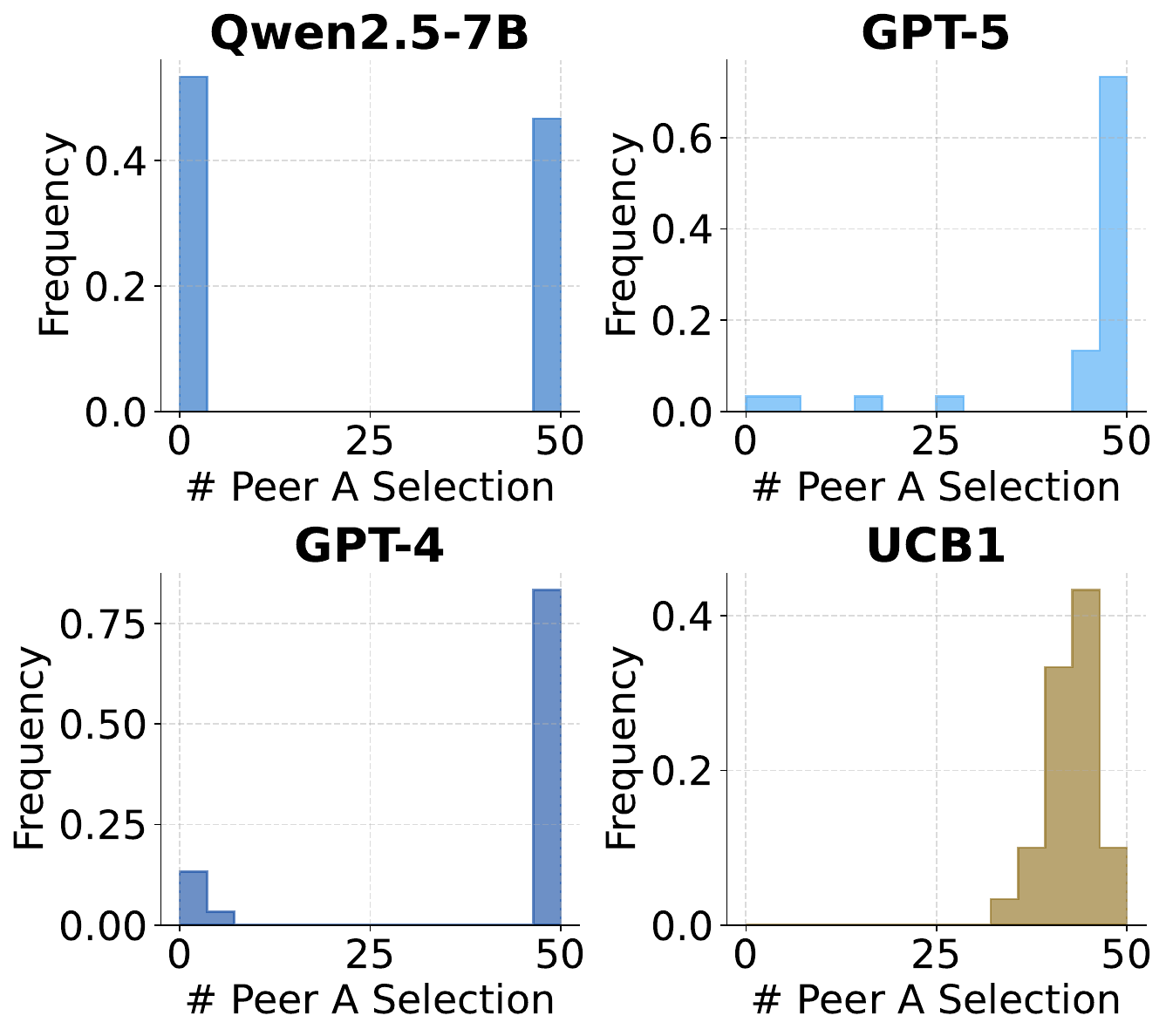}
    \end{subfigure}
    \caption{\textbf{Histogram comparison across various $p_\text{A}$ values}. From left to right and from top to bottom, $p_\text{A}=[0.1, 0.2, \ldots, 0.9]$ while $p_\text{B}$ is fixed to 0.5. }
    \label{fig:full_hist}
\end{figure}

Here, we present the complete set of histogram comparisons for the delegation experiment described in Section~\ref{sec:toy}, shown in Figure~\ref{fig:full_hist}. Across all settings, the contrast with the UCB baseline is striking. 
Whereas UCB exhibits smooth unimodal distributions centered near the optimal allocation implied by each $p_A$, all three LLMs instead display highly polarized behaviors, with selections sharply concentrated near the extremes of either $0$ or $50$. 
It is also noteworthy that the LLMs exhibit polarized behaviors even when $p_\text{A}=p_\text{B}=0.5$.
These results indicate that, rather than gradually balancing exploration and exploitation, the LLMs often commit early to one peer and repeatedly reinforce that initial choice.

Among the tested models, Qwen2.5-7B-Instruct demonstrates the most surprising failures of calibration. Even in settings where peer A is clearly inferior---for example when $p_A=0.1$ and $p_B=0.5$---the model still frequently allocates nearly all selections to peer A. Such behavior suggests severe overreaction to noisy early outcomes and an inability to recover through continued exploration. GPT-5 exhibits comparatively stronger exploratory tendencies than the other two LLMs, producing somewhat broader distributions in several settings. Nevertheless, its behavior remains far from principled exploration strategies such as UCB, and it still often collapses into near-deterministic decisions with insufficient trial-and-error search. Overall, these results reinforce that current LLM agents struggle to sustain statistically grounded exploration, even in simple delegation environments.


\begin{figure}[h!]
    \centering
    \includegraphics[width=0.7\linewidth]{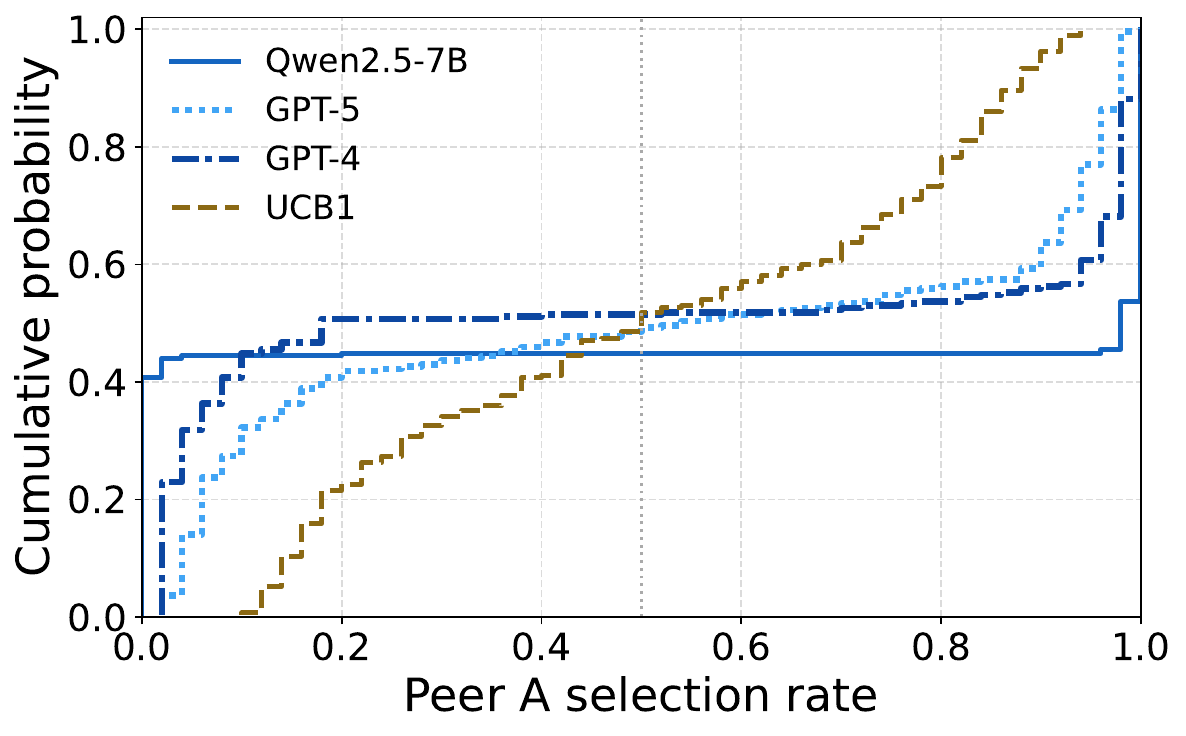}
    \caption{CDF plot of the Peer A selection ratio}
    \label{fig:sweep_cdf}
\end{figure}

\paragraph{CDF of the Peer A Selection Ratios.}

For clearer comparison, we plot in Figure~\ref{fig:sweep_cdf} the cumulative distribution function~(CDF) of the Peer A selection rate, aggregated across different $p_A$ settings $(0.1,0.2,\dots,0.9)$ and 30 independent runs. 
The CDF provides a compact summary of how frequently each method allocates its selections to Peer A across the full sweep of environments.

Consistent with the histogram results in Figure~\ref{fig:full_hist}, all LLM agents exhibit sharp jumps in the CDF near selection rates close to $0$ and close to $1$. These abrupt increases indicate that the models overwhelmingly favor near-extreme behaviors: they tend to allocate almost all selections either to Peer B or to Peer A, with relatively little mass in the intermediate region. In other words, the LLMs rarely maintain balanced exploratory strategies and instead collapse to near-deterministic commitments. By contrast, UCB displays a much smoother and more gradual increase over the entire range, with little probability mass at the extremes. This pattern reflects the adaptive exploration behavior of principled bandit algorithms, which continuously trade off exploration and exploitation rather than prematurely locking onto a single option.

\subsection{Further Analysis: Explicit Policies and Computational Support}
\label{apdx:ucb-controls}

To distinguish difficulty in spontaneously adopting an exploration strategy from difficulty in executing a prescribed policy, we further try augmenting the two-arm delegation experiment with three controls:
\begin{itemize}[leftmargin=*]
    \item \textbf{LLM + UCB Formula:} The complete UCB selection formula is provided in the prompt, and the LLM computes the scores and selects a peer.
    \item \textbf{LLM + UCB Calculator:} An external calculator computes the UCB scores, and the LLM makes the final peer selection.
    \item \textbf{Algorithmic UCB:} UCB1 directly selects the peer without LLM involvement. This control follows the same principle of algorithmically guided peer selection used by MACE.
\end{itemize}

Specifically, we evaluate Qwen-7B at multiple temperatures in a setting where peers A and B succeed with probabilities $0.6$ and $0.4$, respectively. 
We measure the \emph{inferior-peer lock-in rate}, defined as the percentage of runs in which the selector chooses peer B in a majority of the final ten rounds, and cumulative regret incurred by selecting the inferior peer.

Table~\ref{tab:ucb-controls} shows that providing the UCB formula reduces the lock-in rate from $50.0\%$ to $23.3\%$ and cumulative regret from $20.0$ to $13.5$. Providing externally computed UCB scores further reduces the observed lock-in rate to $0.0\%$, with cumulative regret closely matching Algorithmic UCB ($9.9$ versus $9.8$).
The calculator result shows that the LLM can correctly interpret and act on externally computed UCB values, suggesting that the LLM can use externally computed exploration scores to make effective selections in this setting, but providing the formula alone is insufficient for reliable execution of the policy. 
The gap between the formula and calculator conditions highlights the value of computational support for implementing an explicit exploration strategy.
More broadly, these findings motivate MACE's use of algorithmic peer selection to support systematic exploration.

\begin{table}[h!]
    \centering
    \caption{\textbf{Explicit exploration policies and computational support.}
    Comparison of peer selectors in the two-arm delegation experiment.
    Lower values indicate better performance for both metrics.}
    \label{tab:ucb-controls}
    \renewcommand{\arraystretch}{1.15}
    \begin{tabular}{lcc}
        \toprule
        \textbf{Selector}
        & \textbf{Inferior-Peer Lock-In Rate}
        & \textbf{Cumulative Regret} \\
        \midrule
        Prompt-only
        & $50.0\%$ & $20.0$ \\
        LLM + UCB Formula
        & $23.3\%$ & $13.5$ \\
        LLM + UCB Calculator
        & $\mathbf{0.0\%}$ & $9.9$ \\
        Algorithmic UCB
        & $\mathbf{0.0\%}$ & $\mathbf{9.8}$ \\
        \bottomrule
    \end{tabular}
\end{table}
\section{Appendix for Multi-Agent Contextual Exploration~(MACE)}

\subsection{Feature Design}
\label{apdx:features}

We describe the relational features used by MACE to represent each candidate interaction between an agent $i$ and a peer $a$ at round $t$. 
To reduce the complexity of learning, we do not directly use high-dimensional response embeddings as contextual features. 
Instead, we construct a compact relational feature vector that captures the pairwise relationship between agent $i$ and peer $a$, together with peer-level utility signals. 
For each possible interaction $(i,a)$, we define
\begin{equation}
    \mathbf{x}_{i,a,t}
    =
    \Big[
        d_{i,a}^{(1)},\;
        d_{i,a}^{(2)},\;
        d_{i,a}^{(3)},\;
        c_{a}^{(1)},\;
        c_{a}^{(2)},\;
        c_{a}^{(3)},\;
        \hat{p}_{a},\;
        \tau_t,\;
        1
    \Big]^\top
    \in \mathbb{R}^{9},
\end{equation}
where $d_{i,a}^{(n)}$ denotes the response diversity between agent $i$ and peer $a$ measured using $n$-gram similarity, $c_a^{(n)}$ denotes the centrality of peer $a$ in the response graph constructed from $n$-gram similarities, $\hat{p}_a$ denotes the peer's historical performance, $\tau_t$ denotes the normalized interaction round $t$, and the final constant $1$ is a bias feature. 
These features are normalized before being used by the bandit or value-based policy.
Further details are described below for each feature type.
\paragraph{Response diversity.}
The response diversity feature measures how much peer $a$'s current response differs from agent $i$'s current response. 
Let $y_i$ and $y_a$ denote the answers produced by agent $i$ and peer $a$ at round $t$, respectively. 
For each $n \in \{1,2,3\}$, we first compute the Jaccard similarity between their $n$-gram sets:
\begin{equation}
    s_{i,a}^{(n)}
    =
    \frac{
        \left|\mathcal{N}_n(y_i) \cap \mathcal{N}_n(y_a)\right|
    }{
        \left|\mathcal{N}_n(y_i) \cup \mathcal{N}_n(y_a)\right|
    },
\end{equation}
where $\mathcal{N}_n(y)$ denotes the set of $n$-grams extracted from response $y$. 
The corresponding diversity feature is defined as
\begin{equation}
    d_{i,a}^{(n)}
    =
    1 - s_{i,a}^{(n)}.
\end{equation}
Thus, higher values of $d_{i,a}^{(n)}$ indicate that peer $a$ provides a response that is lexically more distinct from agent $i$'s current answer. 
This feature captures the potential information gain from querying a peer whose answer differs from the agent's own belief.

\paragraph{Peer distinctiveness.}
In addition to pairwise diversity between agent $i$ and peer $a$, we include a network-level distinctiveness feature that measures how much peer $a$'s response is distinct compared to the overall agent population. 
For each $n \in \{1,2,3\}$, we define the divergence degree of peer $a$ as
\begin{equation}
    q_a^{(n)}
    =
    \sum_{u \in \mathcal{V},\, u \neq a}
    \left(1 - s_{a,u}^{(n)}\right),
\end{equation}
where
\begin{equation}
    s_{a,u}^{(n)}
    =
    \frac{
        \left|\mathcal{N}_n(y_a) \cap \mathcal{N}_n(y_u)\right|
    }{
        \left|\mathcal{N}_n(y_a) \cup \mathcal{N}_n(y_u)\right|
    }.
\end{equation}
We then normalize this quantity by the total divergence mass across all agents:
\begin{equation}
    c_a^{(n)}
    =
    \frac{
        q_a^{(n)}
    }{
        \sum_{p \in \mathcal{V}} q_p^{(n)}
    },
\end{equation}
with $c_a^{(n)} = 0$ when the denominator is zero. 
This feature can be interpreted as peer $a$'s share of the total response-divergence mass in the population. 

\paragraph{Historical performance.}
The historical performance feature provides an exploitation signal based on the empirical utility of peer $a$ in previous interactions. 
Let $N_a$ denote the number of previous times peer $a$ has been selected, and let $r_a^{(s)}$ denote the reward observed from the $s$-th selection of peer $a$. 
We use the plain running mean
\begin{equation}
    \hat{p}_a
    =
    \begin{cases}
    \dfrac{1}{N_a}\sum_{s=1}^{N_a} r_a^{(s)}, & N_a > 0, \\[1.2em]
    0, & N_a = 0.
    \end{cases}
\end{equation}
This feature signals the policy to exploit peers based on historical performance.

\paragraph{Interaction round.}
We include the normalized  per-sample interaction round 
\begin{equation}
    \tau
    =
    \frac{r}{R},
\end{equation}
where $r \in [0,R]$ is the current round and $R$ is the total number of interaction rounds for each sample.
This feature allows the policy to condition its decision on the stage of the interaction process for a shared task sample, enabling different behavior in earlier versus later rounds.
For the contextual diversity setting, $R=5$, and $R=3$ for the parametric diversity setting.

\paragraph{Bias feature.}
Finally, we append a constant bias feature,
\begin{equation}
    x_{\mathrm{bias}} = 1.
\end{equation}
This allows the learned policy to include an intercept term independent of the relational features.

\subsection{Reward Function}
\label{apdx:reward}

At each interaction round $t$, agent $i$ first updates its response after observing the selected peer. 
We then assign a scalar reward based on the change in a task-specific quality score. 
Let $s_i^{(t)} \in [0,1]$ denote the score of agent $i$'s response after round $t$, evaluated against the ground-truth answer. 
Depending on the task, we use token-level F1 for question answering, and the binary exact match score for multiple-choice and mathematical reasoning tasks.

A natural reward is the improvement in task score:
\begin{equation}
    r_{i,\mathrm{imp}}^{(t)}
    =
        s_i^{(t)} - s_i^{(t-1)}.
\end{equation}
This reward is positive when the agent improves after interaction, negative when its answer degrades, and zero when the score remains unchanged. 

However, using only improvement has an important limitation: an agent that is already correct and remains correct receives zero reward, even though the interaction successfully preserved a high-quality answer. 
To address this, we use a blended reward that combines improvement with the agent's absolute post-interaction score:
\begin{equation}
    r_i^{(t)}
    =
    \frac{1}{2}
    \left[
           \big(s_i^{(t)} - s_i^{(t-1)}\big)
        +
            s_i^{(t)}
    \right].
\end{equation}
The first term credits interactions that improve the agent's answer, while the second term credits interactions that lead to a high-quality final response.

\clearpage
\subsection{Algorithm}
\label{apdx:algorithm}

In Algorithm~\ref{alg:mace}, we provide the pseudo-code for Multi-Agent Contextual Exploration~(MACE).
Note that the total interaction step for the contextual bandit problem is $T = |\mathcal{Q}| \times R$, $\lambda$ is set to 1.0 across all settings.

\begin{algorithm}[h!]
\caption{Multi-Agent Contextual Exploration~(MACE)}
\label{alg:mace}
\begin{algorithmic}[1]
\Require Questions $\mathcal{Q}$, $N$ agents, $R$ rounds, $\alpha$ (exploration)
\State \textbf{Initialize} for all $i,j$: $\mathbf{A}_{ij} \leftarrow \lambda\mathbf{I}_9$,\; $\mathbf{b}_{ij} \leftarrow \mathbf{0}$,\; $n_{ij} \leftarrow 0$,\; $\bar{r}_{ij} \leftarrow 0$,\; $t \leftarrow 0$

\For{each task $q \in \mathcal{Q}$}
  \State \textbf{Round 0:} each agent $i$ generates $y_i^{(0)}$ independently given $q$

  \For{$r = 1, \ldots, R$}
    \State $t \leftarrow t + 1$
      \State \textit{// Peer selection}
      \For{each agent $i$}
          \For{each candidate peer $j$}
              \State Compute context $\boldsymbol{\mathbf{x}}_{ij}^{(t)} \in \mathbb{R}^9$
              \State $\boldsymbol{\theta}_{ij} \leftarrow \mathbf{A}_{ij}^{-1}\mathbf{b}_{ij}$
          \EndFor
          \State $a_i^* \leftarrow \arg\max_{j}\,\Bigl[\boldsymbol{\theta}_{ij}^\top \boldsymbol{\mathbf{x}}_{ij}^{(t)}
                     + \alpha\sqrt{\boldsymbol{\mathbf{x}}_{ij}^{(t)\top}
                     \mathbf{A}_{ij}^{-1}
                     \boldsymbol{\mathbf{x}}_{ij}^{(t)}}\Bigr]$
          \EndFor

          \State \textit{// Response generation}
          \For{each agent $i$}
              \State Agent $i$ reads $y_{a_i^*}^{(t-1)}$ and produces updated response $y_i^{(t)}$
          \EndFor

          \State \textit{// Reward computation and LinUCB update}
          \For{each agent $i$}
              \State $r_i \leftarrow 
    \displaystyle\frac{1}{2}
    \left[
           \big(s_i^{(t)} - s_i^{(t-1)}\big)
        +
            s_i^{(t)}
    \right]$
              \State $\mathbf{A}_{i,a_i^*} \leftarrow \mathbf{A}_{i,a_i^*} + \boldsymbol{\mathbf{x}}_{i,a_i^*}^{(t)}\boldsymbol{\mathbf{x}}_{i,a_i^*}^{(t)\top}$
              \State $\mathbf{b}_{i,a_i^*} \leftarrow \mathbf{b}_{i,a_i^*} + r_i\,\boldsymbol{\mathbf{x}}_{i,a_i^*}^{(t)}$
              \State $n_{i,a_i^*} \mathrel{+}= 1$,\quad
                     $\bar{r}_{i,a_i^*} \leftarrow \bar{r}_{i,a_i^*} + (r_i - \bar{r}_{i,a_i^*})\,/\,n_{i,a_i^*}$
          \EndFor
      \EndFor
  \EndFor
  \end{algorithmic}
  \end{algorithm}
\clearpage
\section{Appendix for the Main Experiments}
\label{apdx:experiment_details}

In this section, we provide more detailed explanations on the benchmark datasets~(Appendix~\ref{apdx:dataset_details}), backbone LLMs~(Appendix~\ref{apdx:backbone_details}), and prompt templates~(Appendix~\ref{apdx:prompt_templates}).

\subsection{Dataset Details}
\label{apdx:dataset_details}

We evaluated MACE under two representative forms of agent heterogeneity: \emph{contextual diversity} and \emph{parametric diversity}. 
These settings are designed to capture complementary sources of uncertainty in realistic multi-agent systems. 
In contextual diversity, agents share the same backbone model but observe different pieces of task-relevant context. 
In parametric diversity, agents differ in their underlying model family and scale, inducing heterogeneity in reasoning capabilities.

\paragraph{HotpotQA for the contextual diversity setting.}
For contextual diversity, we use the distractor setting of HotpotQA~\citep{yang2018hotpotqa}, a multi-hop question answering benchmark in which each example consists of multiple context passages required to answer an open-ended question. 
Each instance contains two gold evidence passages required to answer the question, together with eight distractor passages. 
We construct a multi-agent environment with $N=10$ agents, where each agent receives a distinct passage as its local context. 
Thus, only a subset of agents has access to the information necessary for solving the task, while the remaining agents observe irrelevant or potentially misleading context.
To avoid a case where a single agent consistently receives the gold evidence passage, we randomly assign the passage to each agent for every sample question.
This setting requires each agent to explore its peers in order to identify which agents possess useful evidence.
We evaluate on the first $600$ examples labeled as hard, as the full dataset takes an excessive amount of time to evaluate on, where the first 300 samples are used for the trial-and-error phase evaluation and the latter 300 samples are used for the exploitation phase where the MACE parameters are frozen to test generalizability.
The agents interact for $R=5$ rounds per example, totaling $300 \times 5 = 1500$ steps of interaction for each evaluation phase.
Answer quality is measured using Exact Match~(EM) and token-level F1 against the ground-truth answer. 

\paragraph{Math500 for the parametric diversity setting.}
For mathematical reasoning, we use Math500~\citep{lightman2023lets}, focusing on Level 4--5 problems to emphasize challenging examples that require non-trivial reasoning. 
This results in a total of 262 questions, where the first 131 samples are used for the trial-and-error phase, and the remaining 131 is evaluated for the exploitation phase.
Unlike the contextual-diversity setting, all agents receive the same problem statement, but differ in their underlying backbone models. 
This setting tests whether agents can learn which peers are more reliable for difficult mathematical reasoning problems. 
The agents interact for $R=3$ rounds per sample, totaling $131 \times 3 = 393$ steps of interaction for each evaluation phase.
We evaluate performance using a binary exact match between the extracted final answer and the ground-truth solution.

\paragraph{GPQA for the parametric diversity setting.}
We also evaluate on the Graduate-Level Google-Proof Question Answering~(GPQA) benchmark~\citep{rein2024gpqa}, a challenging multiple-choice question answering dataset designed to test expert-level scientific reasoning. 
We use the Diamond split, which contains 198 particularly difficult questions. 
Similar to other benchmarks, we use the first 99 samples for trial-and-error, and the latter 99 samples for the exploitation phase evaluation.
Also, as in Math500, all agents receive the same question, but differ in their model backbones, and the agents interact for $R=3$ rounds per sample, totaling $99 \times 3 = 297$ steps of interaction for each evaluation phase.
The evaluation metric is a binary exact match after extracting the selected answer option.

\paragraph{2WikiMultihopQA for the parameter transfer experiment.}
To evaluate cross-task transfer of learned exploration behavior, we use 2WikiMultihopQA~\citep{xanh2020_2wikimultihop}, another multi-hop question answering benchmark that requires combining evidence from multiple documents, also providing the distractor mode.
Generally, this benchmark is considered more challenging compared to HotpotQA.
As in the HotpotQA setup, we construct a contextual-diversity environment with $N=10$ agents, where each agent observes a different local context and only a subset of agents has access to the information required to answer the question correctly.
The learned MACE parameters are transferred without further training, and agents interact for $R=5$ rounds across a subset of 300 examples.
We evaluate answer quality using Exact Match~(EM) and token-level F1 against the ground-truth answer.

{\color{black}
\subsection{Empirical Regret Computation}
For each method $m$, we define its \emph{cumulative regret offset} relative to Random as
\begin{equation}
D_m(T)
=
\frac{1}{N}\sum_{i=1}^{N}\sum_{t=1}^{T}
\left(
r_{i,t}^{\mathrm{Random}}-r_{i,t}^{(m)}
\right).
\end{equation}
Here, $r_{i,t}^{(m)}$ is the realized blended reward of agent $i$ under method $m$, as defined in Appendix~B.2, and $t$ indexes interactions across task samples and rounds within the evaluation
phase. 
Each method, including Random, follows its own peer-selection and response-update trajectory on the same task sequence.

Figure~4 labels this offset as \emph{Baseline $-$ Random}, where Baseline denotes the method $m$ and the subtraction is in terms of cumulative regret. 
To see the equivalence, let $G_i(T)$ be a common reference cumulative reward for agent $i$. Subtracting Random's
cumulative regret from method $m$'s cumulative regret cancels the
reference:
\[
\left(G_i(T)-\sum_{t=1}^{T}r_{i,t}^{(m)}\right)
-
\left(G_i(T)-\sum_{t=1}^{T}r_{i,t}^{\mathrm{Random}}\right)
=
\sum_{t=1}^{T}
\left(r_{i,t}^{\mathrm{Random}}-r_{i,t}^{(m)}\right).
\]
Averaging over agents yields $D_m(T)$. 
}

\subsection{Backbone Model Details}
\label{apdx:backbone_details}

We use different backbone configurations depending on the source of heterogeneity being evaluated.

\textbf{Contextual diversity.}
In the contextual-diversity experiments, all agents use the same backbone model, \texttt{Qwen2.5-7B-Instruct}~\citep{yang2024qwen2}.
Since we instantiate ten independent Qwen agents in this setup, we use stochastic decoding with a temperature of 1.2 and nucleus sampling with \texttt{top\_p}=0.95 to encourage response diversity across agents.

\textbf{Parametric diversity.}
In the parametric-diversity experiments, we construct a heterogeneous agent pool consisting of four model types:
    \{
    \texttt{GPT-5}~\citep{openai2025gpt5},
    \texttt{Qwen2.5-7B-Instruct}~\citep{yang2024qwen2},
    \texttt{Llama3.1-8B-Instruct}~\citep{grattafiori2024llama},
    \texttt{Mistral-7B-v0.3}~\citep{jiang2023mistral}
    \},
each with (temperature, \texttt{top\_p}) values of (\texttt{default}, \texttt{default}), (1.2, 0.95), (1.0, 0.95), (1.0, 0.9), respectively.
Each model corresponds to one agent in the multi-agent system. 
This setup reflects realistic open-agent ecosystems in which agents may differ substantially in model family, scale, training data, and reasoning ability. 

\subsection{Resources}
\label{apdx:resources}
All contextual diversity experiments were done on a single NVIDIA RTX A6000 GPU, and the parametric diversity experiments were done on three NVIDIA RTX A6000 GPUs (GPT-5 used Azure API calls).

\newpage
\subsection{Prompt Templates}
\label{apdx:prompt_templates}


\textbf{Prompt templates for the Contextual Diversity setup}
\begin{figure}[h]
    \centering
    \begin{tcolorbox}[title={Prompt Template Under Contextual Diversity~(initial response generation)}]\small

Context: 

\texttt{<context passage or sentence>} \\

Question: \texttt{<question>} \\

First, briefly reason over the provided context. Then, place your final answer between <answer> and </answer> tags. Keep the answer concise -- a word, phrase, or short sentence. Example: <answer>Paris</answer>
    \end{tcolorbox} 
\end{figure}

\begin{figure}[h!]
    \centering
    \begin{tcolorbox}[title={Prompt Template Under Contextual Diversity~(interaction round response generation)}]\small

Context: 

\texttt{<context passage or sentence>} \\

Question: \texttt{<question>} \\

YOUR PREVIOUS RESPONSE:

\texttt{<agent's previous round response>} \\

RESPONSE FROM \texttt{<chosen peer's name>}:

\texttt{<chosen peer's previous round response>} \\

INSTRUCTIONS:

Review your previous response and the peer's response carefully. You may agree, disagree, or selectively incorporate the peer's reasoning. Then provide your updated, complete response to the QUESTION.\\

First, briefly reason over the provided context. Then, place your final answer between <answer> and </answer> tags. Keep the answer concise -- a word, phrase, or short sentence. Example: <answer>Paris</answer>
    \end{tcolorbox} 
\end{figure}

\newpage
\textbf{Prompt templates for the Parametric Diversity setup} 

Note that the specific instructions on how to format answers differ between the datasets: Math500 and GPQA.
\begin{figure}[h]
    \centering
    \begin{tcolorbox}[title={Prompt Template Under Parametric Diversity~(initial response generation)}]\small

Question: \texttt{<question>} \\

\textit{If Math500:}

First, briefly state your step-by-step reasoning. Then, make sure to place your final answer between <answer> and </answer> tags. Example: <answer>$\backslash$boxed\{123\}</answer>

\textit{If GPQA:}

First, briefly state your step-by-step reasoning. Then, make sure to place only your final answer label between <answer> and </answer> tags. Example: <answer>(A)</answer>
    \end{tcolorbox} 
\end{figure}

\begin{figure}[h!]
    \centering
    \begin{tcolorbox}[title={Prompt Template Under Parametric Diversity~(interaction round response generation)}]\small

Question: \texttt{<question>} \\

YOUR PREVIOUS RESPONSE:

\texttt{<agent's previous round response>} \\

RESPONSE FROM \texttt{<chosen peer's name>}:

\texttt{<chosen peer's previous round response>} \\

INSTRUCTIONS:

Review your previous response and the peer's response carefully. You may agree, disagree, or selectively incorporate the peer's reasoning. Then provide your updated, complete response to the QUESTION.\\

\textit{If Math500:}

First, briefly state your step-by-step reasoning. Then, make sure to place your final answer between <answer> and </answer> tags. Example: <answer>$\backslash$boxed\{123\}</answer>

\textit{If GPQA:}

First, briefly state your step-by-step reasoning. Then, make sure to place only your final answer label between <answer> and </answer> tags. Example: <answer>(A)</answer>
    \end{tcolorbox} 
\end{figure}

\newpage
\textbf{Prompt template for peer selection of In-Context Exploration}
\begin{figure}[h]
    \centering
    \begin{tcolorbox}[title={Prompt Template for Peer Selection~(In-Context Exploration)}]\small
QUESTION: \texttt{<question>}\\

HISTORY:

\texttt{<agent 1 name>} [chosen \texttt{<n>} times, mean reward: \texttt{<r>} | features: [Diversity (unigram): \texttt{<f1>}, Diversity (bigram): \texttt{<f2>}, Diversity (trigram): \texttt{<f3>}, Centrality (unigram): \texttt{<f4>}, Centrality (bigram): \texttt{<f5>}, Centrality (trigram): \texttt{<f6>}, Hist. Reward: \texttt{<f7>}, Norm. Round: \texttt{<f8>}, Bias: 1.0]] \\

\texttt{<agent 2 name>} [chosen \texttt{<n>} times, mean reward: \texttt{<r>} | features: [Diversity (unigram): \texttt{<f1>}, Diversity (bigram): \texttt{<f2>}, Diversity (trigram): \texttt{<f3>}, Centrality (unigram): \texttt{<f4>}, Centrality (bigram): \texttt{<f5>}, Centrality (trigram): \texttt{<f6>}, Hist. Reward: \texttt{<f7>}, Norm. Round: \texttt{<f8>}, Bias: 1.0]] 

\begin{center}
    $\vdots$
\end{center}

\texttt{<agent $N$ name>} [chosen \texttt{<n>} times, mean reward: \texttt{<r>} | features: [Diversity (unigram): \texttt{<f1>}, Diversity (bigram): \texttt{<f2>}, Diversity (trigram): \texttt{<f3>}, Centrality (unigram): \texttt{<f4>}, Centrality (bigram): \texttt{<f5>}, Centrality (trigram): \texttt{<f6>}, Hist. Reward: \texttt{<f7>}, Norm. Round: \texttt{<f8>}, Bias: 1.0]] \\

Based on the historical information, whose full response would you like to read to help refine your answer? You may also choose your own.

Reply with only the agent number (1 to \texttt{<$N$>}).
    \end{tcolorbox} 
\end{figure}

For the features used in In-Context Exploration, the feature values used for MACE are identically shown to each agent as context for selection.
\section{Further Experiments and Analyses}
\label{apdx:further_experiments}

\subsection{Round-wise Performance}
\label{apdx:roundwise}

\begin{figure}[t!]
    \centering
    \begin{subfigure}[h!]{0.32\textwidth}
        \centering
        \includegraphics[width=\linewidth]{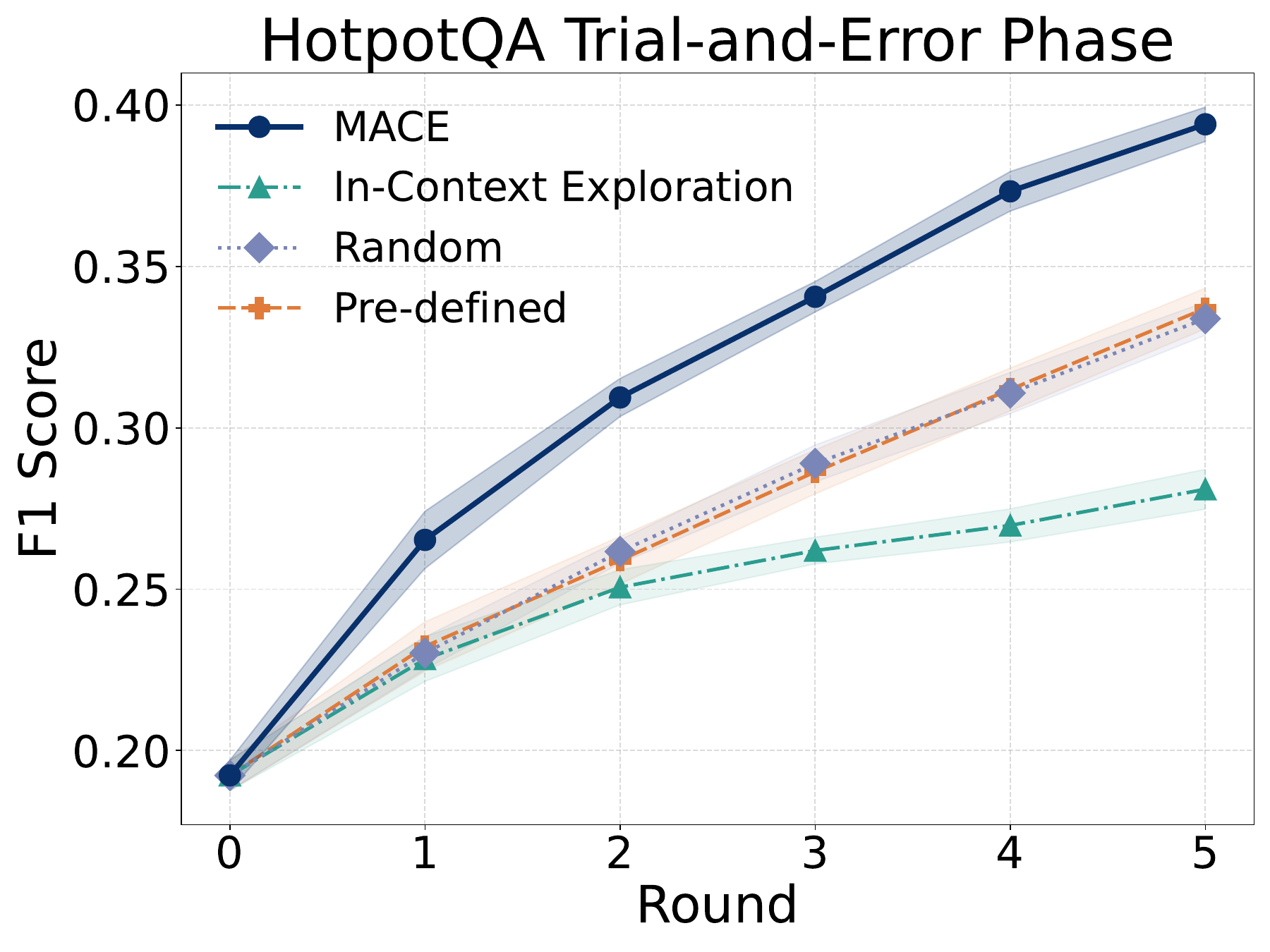}
        \label{fig:exploration_f1}
    \end{subfigure}
    \hfill
    \begin{subfigure}[h!]{0.32\textwidth}
        \centering
        \includegraphics[width=\linewidth]{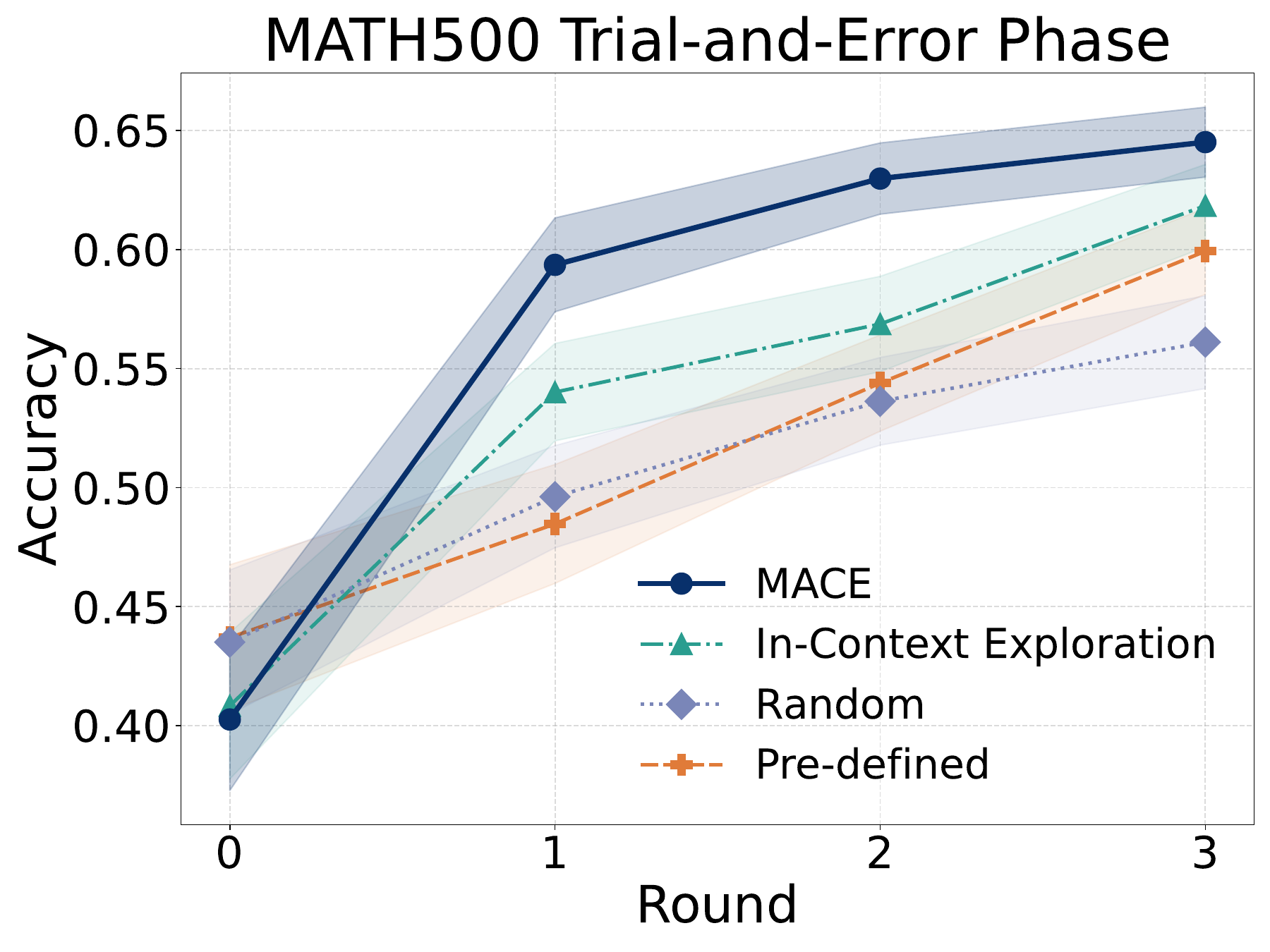}
        \label{fig:exploration_f1}
    \end{subfigure}
    \hfill
    \begin{subfigure}[h!]{0.32\textwidth}
        \centering
        \includegraphics[width=\linewidth]{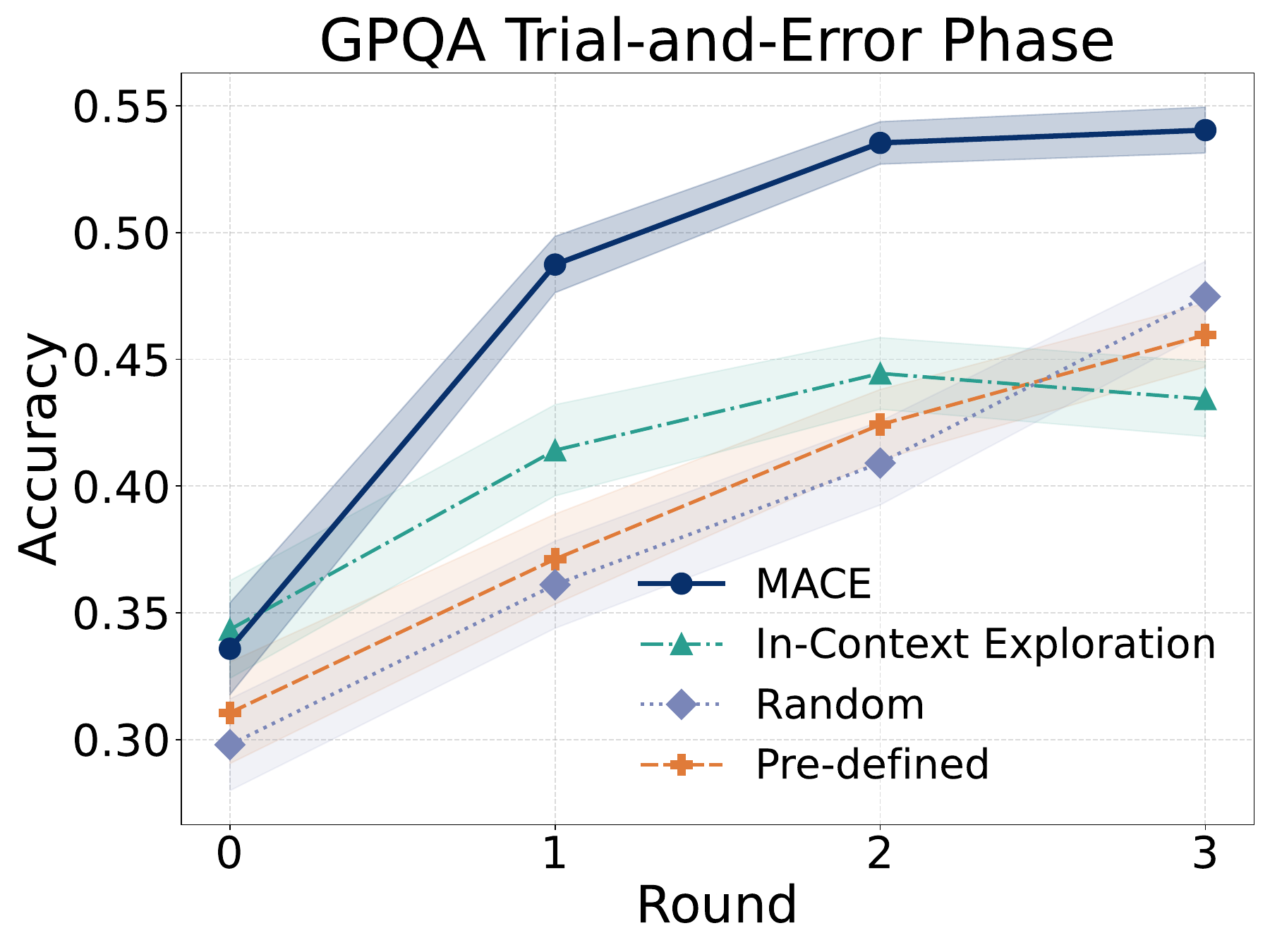}
        \label{fig:exploration_f1}
    \end{subfigure}
    \hfill
    \begin{subfigure}[h!]{0.32\textwidth}
        \centering
        \includegraphics[width=\linewidth]{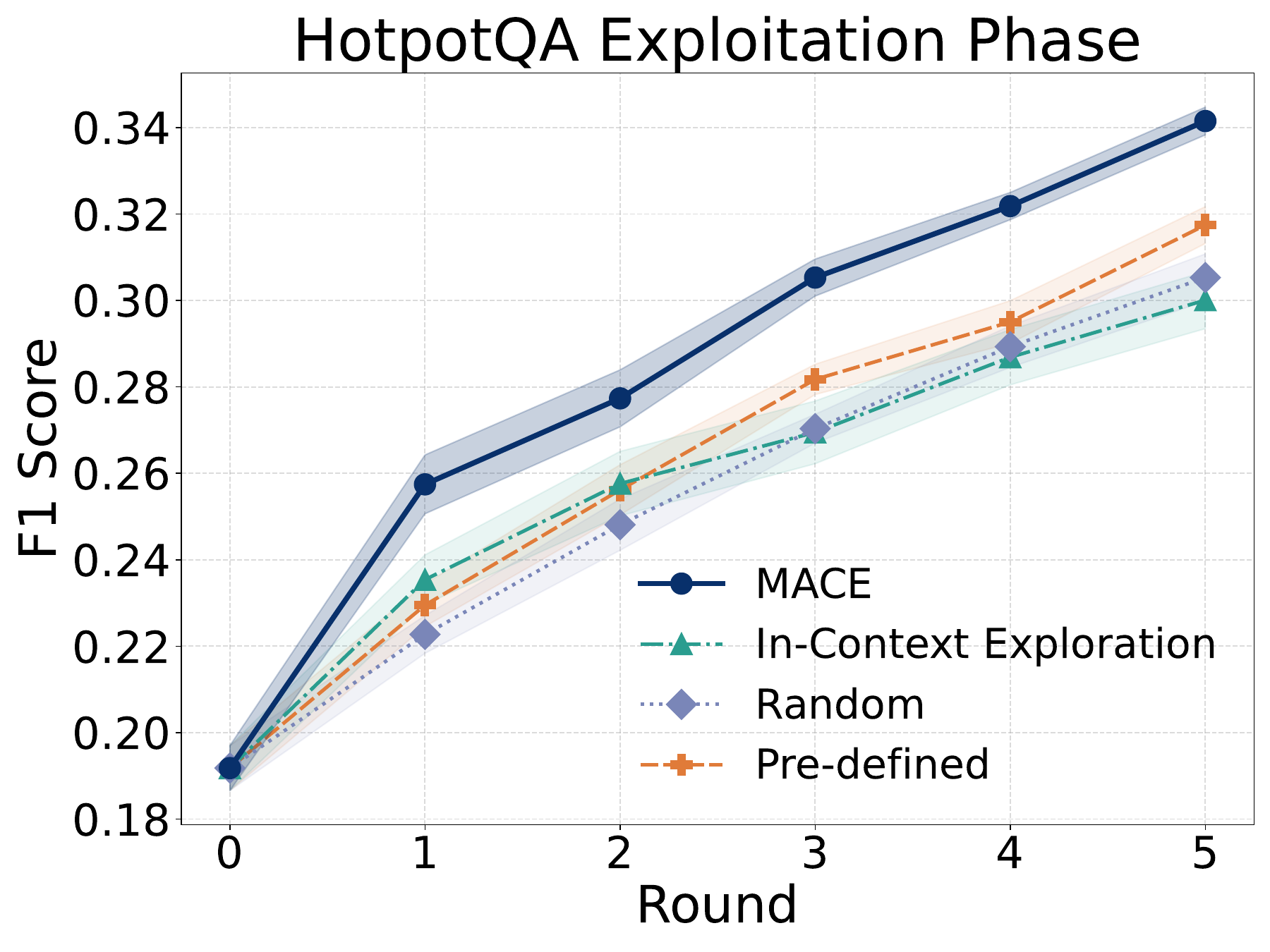}
        \label{fig:exploitation_f1}
    \end{subfigure}
    \hfill
    \begin{subfigure}[h!]{0.32\textwidth}
        \centering
        \includegraphics[width=\linewidth]{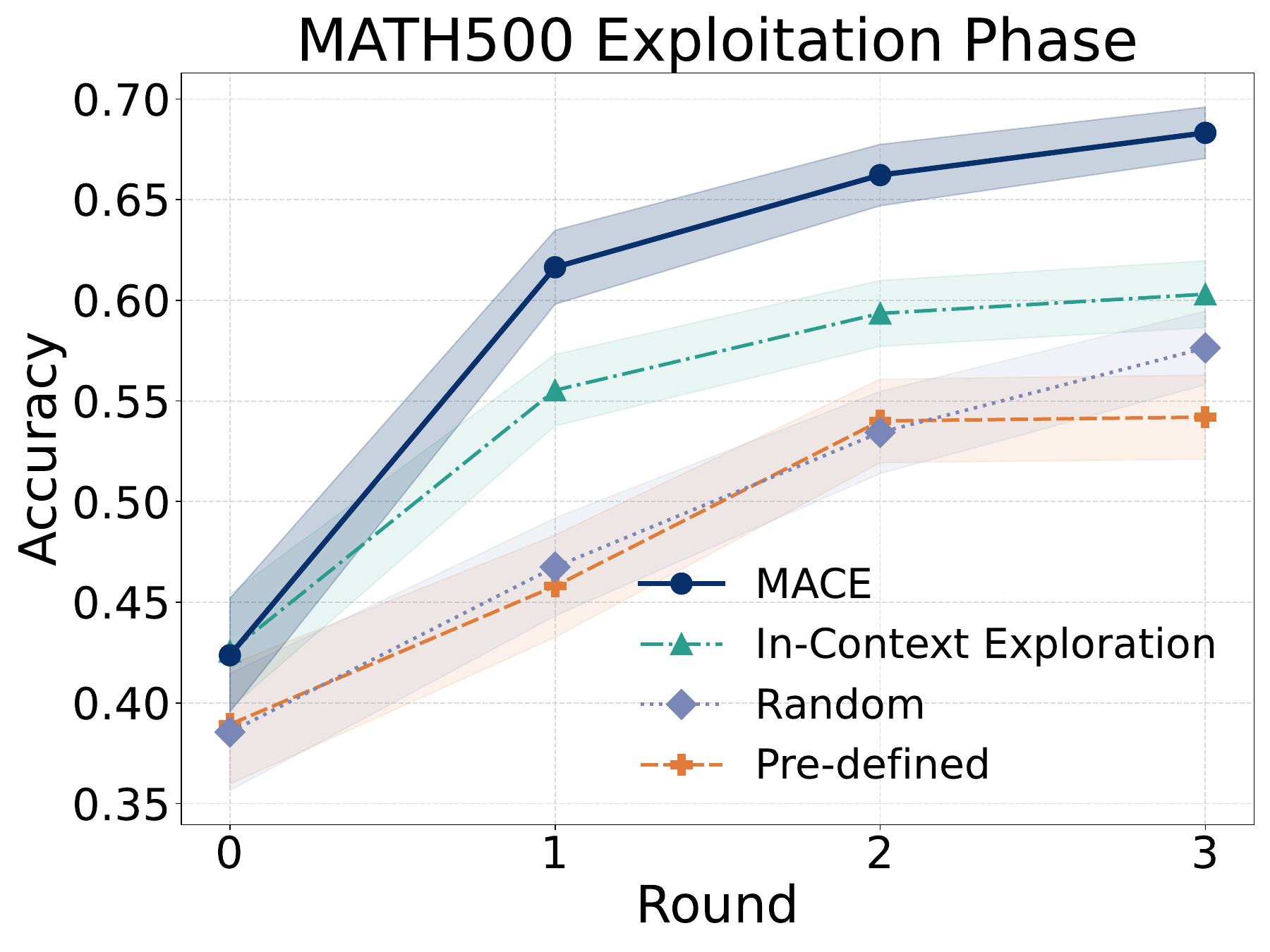}
        \label{fig:exploitation_f1}
    \end{subfigure}
    \hfill
    \begin{subfigure}[h!]{0.32\textwidth}
        \centering
        \includegraphics[width=\linewidth]{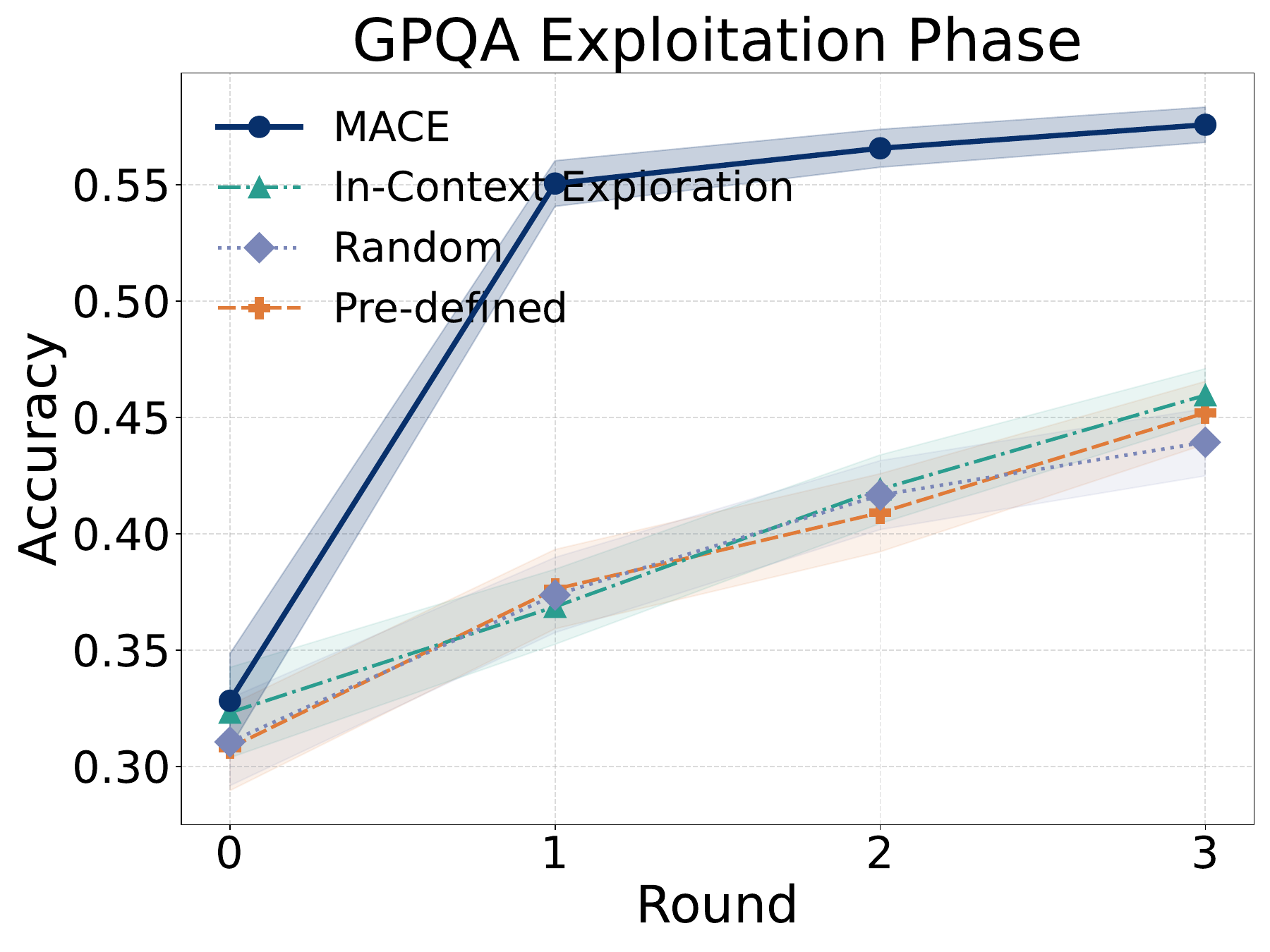}
        \label{fig:exploitation_f1}
    \end{subfigure}
    \caption{Comparison of exploration and exploitation performance for each interaction round. Shaded are the standard errors across participating agents; standard error * 0.2 are shown for Math500 and GPQA to avoid visual clutter. Also, note that the variance in the 0-th round is induced by GPT-5. }
    \label{fig:roundwise_perf}
\end{figure}

\paragraph{Round-wise task performance.}
In Figure~\ref{fig:roundwise_perf}, we show the average task performance across task samples during both the exploration and exploitation phases, evaluated for each per-sample interaction round $r \in [0,R]$.
Compared to the global step regret evaluation in Figure~\ref{fig:explore_exploit}, this analysis provides a concrete intuition of how proper exploration can significantly impact task performance.
Across all three benchmarks, MACE consistently improves as the interaction round proceeds, indicating that the learned peer-selection policy is able to convert additional communication rounds into better task performance. 
This trend is especially clear on Math500 and GPQA, where MACE quickly separates from Random, Pre-defined, and In-Context Exploration after the first interaction round and maintains the largest gains through the final round.
Generally, MACE yields steady improvements in F1 score across rounds, suggesting that explicit exploration helps agents identify peers with useful evidence under contextual diversity. 
In contrast, In-Context Exploration often improves only marginally after the initial rounds and can remain close to or below Random, reinforcing that prompting agents to explore is insufficient for reliable peer discovery. 
Importantly, the same pattern persists in the exploitation phase, where MACE parameters are frozen: the policy learned during trial-and-error continues to produce stronger round-wise gains than all baselines.
This suggests that MACE does not merely overfit to transient exploration feedback, but learns reusable interaction strategies that remain effective when deployed without further updates.

\begin{figure}[h!]
    \centering
    \includegraphics[width=0.45\linewidth]{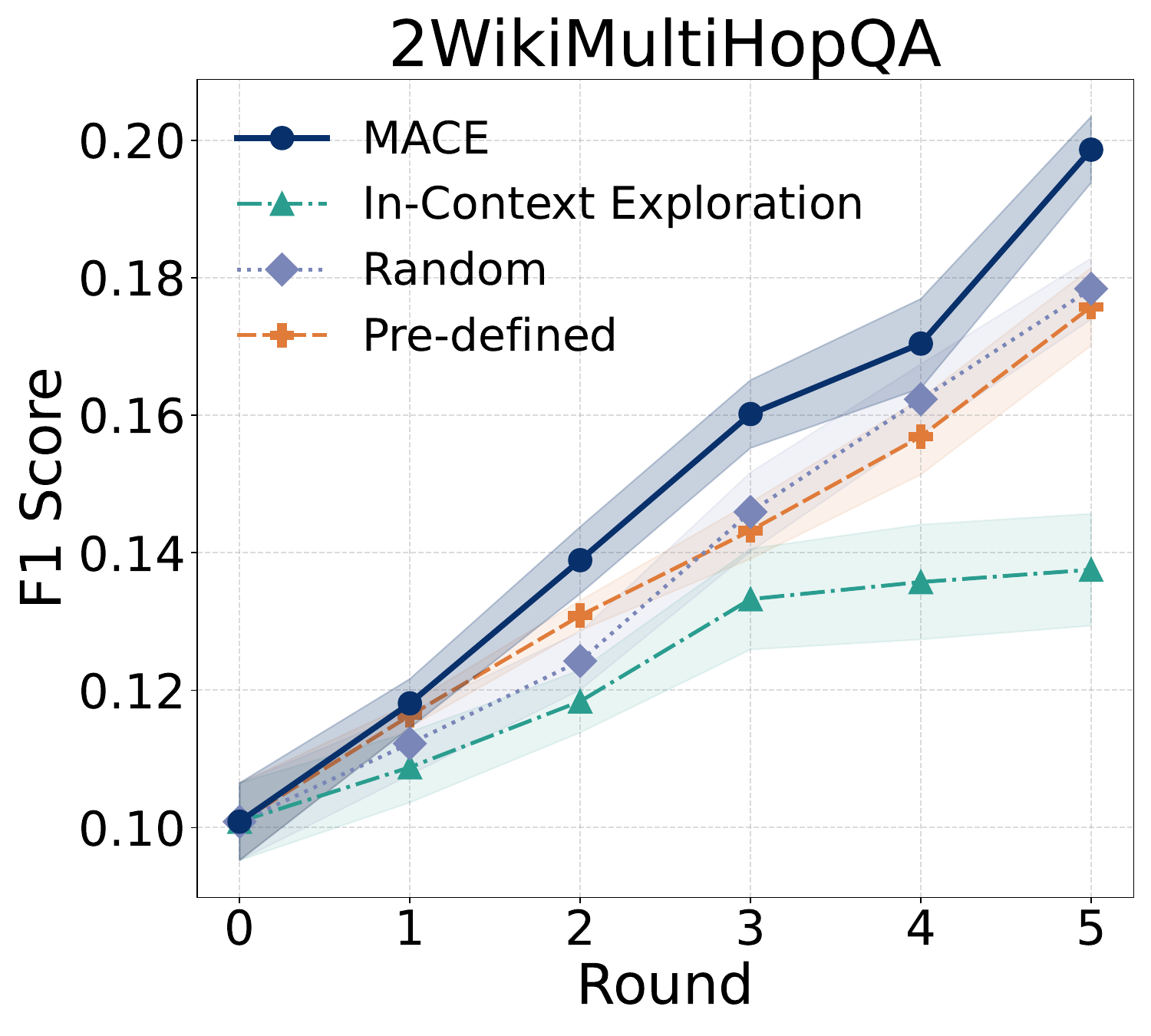}
    \caption{Per-sample interaction round-wise performance}
    \label{fig:transfer_roundwise}
\end{figure}

\paragraph{Round-wise task performance of the parameter transfer experiment.}
Figure~\ref{fig:transfer_roundwise} shows the round-wise performance of parameter transfer to the 2WikiMultihopQA benchmark. 
Across interaction rounds, MACE consistently outperforms the baselines, indicating that the exploration strategy learned on HotpotQA transfers effectively to a different multi-hop QA environment. 
These results support that MACE captures a transferable exploration policy rather than overfitting to the source benchmark.

\paragraph{Full tables for the round-wise performance.}
Here, we provide a full table of the round-wise performances, corresponding to the results shown in Figure~\ref{fig:roundwise_perf}.
The tables also include the performance of each individual agent.

\begin{itemize}[leftmargin=*]
    \item Table~\ref{tab1} -- \ref{tab4}: Contextual diversity experiment results on HotpotQA for MACE, In-Context Exploration, Pre-defined, and Random.
    \item Table~\ref{tab9} -- \ref{tab16}: Parametric diversity experiment results on GPQA and Math500, for MACE, In-Context Exploration, Pre-defined, and Random.
    \item Table~\ref{tab5} -- \ref{tab8}: Parameter Transfer experiment results on 2WikiMultihopQA for MACE, In-Context Exploration, Pre-defined, and Random.
\end{itemize}

\vspace{2cm}

\begin{table}[h!]
\centering
\caption{Round-wise HotpotQA (contextual diversity) performance of \textbf{MACE} across exploration and exploitation phases. 
We report Exact Match (EM) and token-level F1 for each agent over interaction rounds, where round 0 corresponds to the initial response before peer interaction. 
The highlighted Mean column reports the average performance across all 10 agents~(all Qwen2.5-7B-Instruct models).}
\label{tab1}
\resizebox{\textwidth}{!}{
\begin{tabular}{lccccccccccM}
\toprule
\multicolumn{12}{l}{\textbf{(a) Exploration Phase: Exact Match (EM)}} \\
\midrule
Round & Agent 1 & Agent 2 & Agent 3 & Agent 4 & Agent 5 & Agent 6 & Agent 7 & Agent 8 & Agent 9 & Agent 10 & Mean \\
\midrule
0 (init) & 0.1633 & 0.1433 & 0.1533 & 0.1667 & 0.1200 & 0.1567 & 0.1600 & 0.1533 & 0.1300 & 0.1500 & 0.1497 \\
1 & 0.2267 & 0.2333 & 0.2033 & 0.2100 & 0.1667 & 0.2333 & 0.2067 & 0.2233 & 0.1733 & 0.2100 & 0.2087 \\
2 & 0.2300 & 0.2333 & 0.2667 & 0.2500 & 0.2467 & 0.2600 & 0.2667 & 0.2400 & 0.2133 & 0.2433 & 0.2450 \\
3 & 0.2700 & 0.2833 & 0.2767 & 0.2667 & 0.2733 & 0.2867 & 0.2767 & 0.2433 & 0.2533 & 0.2500 & 0.2680 \\
4 & 0.3233 & 0.3067 & 0.3167 & 0.2900 & 0.2800 & 0.2967 & 0.2933 & 0.2800 & 0.2700 & 0.2867 & 0.2943 \\
5 & 0.3433 & 0.3000 & 0.3167 & 0.2900 & 0.3133 & 0.3267 & 0.3267 & 0.2900 & 0.3267 & 0.3000 & 0.3133 \\

\midrule
\multicolumn{12}{l}{\textbf{(b) Exploration Phase: Token-level F1}} \\
\midrule
Round & Agent 1 & Agent 2 & Agent 3 & Agent 4 & Agent 5 & Agent 6 & Agent 7 & Agent 8 & Agent 9 & Agent 10 & Mean \\
\midrule
0 (init) & 0.2032 & 0.1929 & 0.2094 & 0.2035 & 0.1588 & 0.1917 & 0.2039 & 0.1923 & 0.1718 & 0.1955 & 0.1923 \\
1 & 0.2823 & 0.3037 & 0.2641 & 0.2744 & 0.2027 & 0.2957 & 0.2583 & 0.2654 & 0.2327 & 0.2735 & 0.2653 \\
2 & 0.2901 & 0.3074 & 0.3357 & 0.3253 & 0.2979 & 0.3137 & 0.3378 & 0.3032 & 0.2753 & 0.3078 & 0.3094 \\
3 & 0.3302 & 0.3540 & 0.3626 & 0.3542 & 0.3330 & 0.3466 & 0.3561 & 0.3182 & 0.3264 & 0.3250 & 0.3406 \\
4 & 0.3995 & 0.3940 & 0.4054 & 0.3762 & 0.3470 & 0.3619 & 0.3737 & 0.3622 & 0.3491 & 0.3642 & 0.3733 \\
5 & 0.4223 & 0.3888 & 0.3958 & 0.3792 & 0.3827 & 0.4143 & 0.4117 & 0.3678 & 0.3987 & 0.3797 & 0.3941 \\

\midrule
\multicolumn{12}{l}{\textbf{(c) Exploitation Phase: Exact Match (EM)}} \\
\midrule
Round & Agent 1 & Agent 2 & Agent 3 & Agent 4 & Agent 5 & Agent 6 & Agent 7 & Agent 8 & Agent 9 & Agent 10 & Mean \\
\midrule
0 (init) & 0.1533 & 0.1267 & 0.1600 & 0.1833 & 0.1600 & 0.1300 & 0.1500 & 0.1767 & 0.1467 & 0.1567 & 0.1543 \\
1 & 0.2100 & 0.2067 & 0.2300 & 0.2167 & 0.1767 & 0.1867 & 0.1733 & 0.2367 & 0.1867 & 0.2100 & 0.2033 \\
2 & 0.2100 & 0.2133 & 0.2300 & 0.2400 & 0.1967 & 0.2400 & 0.1967 & 0.2233 & 0.2200 & 0.2100 & 0.2180 \\
3 & 0.2333 & 0.2300 & 0.2733 & 0.2500 & 0.2267 & 0.2200 & 0.2333 & 0.2400 & 0.2467 & 0.2333 & 0.2387 \\
4 & 0.2467 & 0.2600 & 0.2500 & 0.2567 & 0.2367 & 0.2533 & 0.2433 & 0.2467 & 0.2467 & 0.2667 & 0.2507 \\
5 & 0.2767 & 0.2600 & 0.2633 & 0.2833 & 0.2767 & 0.2733 & 0.2500 & 0.2800 & 0.2600 & 0.2633 & 0.2687 \\

\midrule
\multicolumn{12}{l}{\textbf{(d) Exploitation Phase: Token-level F1}} \\
\midrule
Round & Agent 1 & Agent 2 & Agent 3 & Agent 4 & Agent 5 & Agent 6 & Agent 7 & Agent 8 & Agent 9 & Agent 10 & Mean \\
\midrule
0 (init) & 0.1901 & 0.1656 & 0.2029 & 0.2213 & 0.1783 & 0.1699 & 0.1876 & 0.2093 & 0.1959 & 0.1971 & 0.1918 \\
1 & 0.2721 & 0.2633 & 0.2904 & 0.2758 & 0.2263 & 0.2336 & 0.2248 & 0.2742 & 0.2508 & 0.2632 & 0.2575 \\
2 & 0.2723 & 0.2726 & 0.2890 & 0.3101 & 0.2413 & 0.3001 & 0.2442 & 0.2829 & 0.2881 & 0.2729 & 0.2774 \\
3 & 0.2966 & 0.2969 & 0.3356 & 0.3203 & 0.2920 & 0.2933 & 0.2973 & 0.2983 & 0.3070 & 0.3157 & 0.3053 \\
4 & 0.3174 & 0.3299 & 0.3235 & 0.3350 & 0.3106 & 0.3240 & 0.3154 & 0.3104 & 0.3121 & 0.3401 & 0.3218 \\
5 & 0.3487 & 0.3329 & 0.3401 & 0.3571 & 0.3351 & 0.3494 & 0.3274 & 0.3547 & 0.3280 & 0.3418 & 0.3415 \\
\bottomrule
\end{tabular}
}
\end{table}
\begin{table}[t]
\centering
\caption{Round-wise HotpotQA (contextual diversity) performance of \textbf{In-Context Exploration} across exploration and exploitation phases. 
We report Exact Match (EM) and token-level F1 for each agent over interaction rounds, where round 0 corresponds to the initial response before peer interaction. 
The highlighted Mean column reports the average performance across all 10 agents.}
\label{tab2}
\resizebox{\textwidth}{!}{
\begin{tabular}{lccccccccccM}
\toprule
\multicolumn{12}{l}{\textbf{(a) Exploration Phase: Exact Match (EM)}} \\
\midrule
Round & Agent 1 & Agent 2 & Agent 3 & Agent 4 & Agent 5 & Agent 6 & Agent 7 & Agent 8 & Agent 9 & Agent 10 & Mean \\
\midrule
0 (init) & 0.1633 & 0.1433 & 0.1533 & 0.1667 & 0.1200 & 0.1567 & 0.1600 & 0.1533 & 0.1300 & 0.1500 & 0.1497 \\
1 & 0.2167 & 0.1800 & 0.2000 & 0.1767 & 0.1500 & 0.1967 & 0.1800 & 0.1500 & 0.1967 & 0.1767 & 0.1823 \\
2 & 0.2133 & 0.2167 & 0.2133 & 0.2133 & 0.1933 & 0.1900 & 0.2067 & 0.1667 & 0.1967 & 0.1933 & 0.2003 \\
3 & 0.2200 & 0.2267 & 0.2200 & 0.2133 & 0.1900 & 0.2133 & 0.2033 & 0.1933 & 0.2233 & 0.1833 & 0.2087 \\
4 & 0.2333 & 0.2433 & 0.2133 & 0.2033 & 0.2267 & 0.2133 & 0.1867 & 0.2067 & 0.2067 & 0.2200 & 0.2153 \\
5 & 0.2433 & 0.2533 & 0.2367 & 0.2133 & 0.2167 & 0.2167 & 0.2233 & 0.2233 & 0.1867 & 0.2433 & 0.2257 \\

\midrule
\multicolumn{12}{l}{\textbf{(b) Exploration Phase: Token-level F1}} \\
\midrule
Round & Agent 1 & Agent 2 & Agent 3 & Agent 4 & Agent 5 & Agent 6 & Agent 7 & Agent 8 & Agent 9 & Agent 10 & Mean \\
\midrule
0 (init) & 0.2032 & 0.1929 & 0.2094 & 0.2035 & 0.1588 & 0.1917 & 0.2039 & 0.1923 & 0.1718 & 0.1955 & 0.1923 \\
1 & 0.2659 & 0.2320 & 0.2439 & 0.2201 & 0.1910 & 0.2448 & 0.2267 & 0.1955 & 0.2441 & 0.2194 & 0.2284 \\
2 & 0.2655 & 0.2739 & 0.2560 & 0.2621 & 0.2453 & 0.2364 & 0.2573 & 0.2096 & 0.2567 & 0.2432 & 0.2506 \\
3 & 0.2731 & 0.2767 & 0.2668 & 0.2715 & 0.2457 & 0.2638 & 0.2548 & 0.2449 & 0.2793 & 0.2433 & 0.2620 \\
4 & 0.2848 & 0.2951 & 0.2768 & 0.2592 & 0.2829 & 0.2721 & 0.2370 & 0.2528 & 0.2628 & 0.2742 & 0.2698 \\
5 & 0.3065 & 0.3081 & 0.2948 & 0.2745 & 0.2663 & 0.2716 & 0.2735 & 0.2815 & 0.2399 & 0.2931 & 0.2810 \\

\midrule
\multicolumn{12}{l}{\textbf{(c) Exploitation Phase: Exact Match (EM)}} \\
\midrule
Round & Agent 1 & Agent 2 & Agent 3 & Agent 4 & Agent 5 & Agent 6 & Agent 7 & Agent 8 & Agent 9 & Agent 10 & Mean \\
\midrule
0 (init) & 0.1533 & 0.1267 & 0.1600 & 0.1833 & 0.1600 & 0.1300 & 0.1500 & 0.1767 & 0.1467 & 0.1567 & 0.1543 \\
1 & 0.2067 & 0.1800 & 0.2200 & 0.1700 & 0.1733 & 0.1633 & 0.1733 & 0.2067 & 0.1867 & 0.2033 & 0.1883 \\
2 & 0.2267 & 0.2133 & 0.2500 & 0.2133 & 0.2067 & 0.1833 & 0.1733 & 0.2100 & 0.1967 & 0.2267 & 0.2100 \\
3 & 0.2333 & 0.2133 & 0.2633 & 0.2333 & 0.2133 & 0.1867 & 0.1967 & 0.2333 & 0.2033 & 0.2167 & 0.2193 \\
4 & 0.2467 & 0.2467 & 0.2767 & 0.2500 & 0.2067 & 0.2067 & 0.2200 & 0.2400 & 0.2100 & 0.2333 & 0.2337 \\
5 & 0.2567 & 0.2600 & 0.2800 & 0.2467 & 0.2067 & 0.2267 & 0.2167 & 0.2500 & 0.2433 & 0.2467 & 0.2433 \\

\midrule
\multicolumn{12}{l}{\textbf{(d) Exploitation Phase: Token-level F1}} \\
\midrule
Round & Agent 1 & Agent 2 & Agent 3 & Agent 4 & Agent 5 & Agent 6 & Agent 7 & Agent 8 & Agent 9 & Agent 10 & Mean \\
\midrule
0 (init) & 0.1901 & 0.1656 & 0.2029 & 0.2213 & 0.1783 & 0.1699 & 0.1876 & 0.2093 & 0.1959 & 0.1971 & 0.1918 \\
1 & 0.2478 & 0.2242 & 0.2658 & 0.2220 & 0.2246 & 0.2093 & 0.2206 & 0.2416 & 0.2337 & 0.2644 & 0.2354 \\
2 & 0.2699 & 0.2555 & 0.3021 & 0.2615 & 0.2499 & 0.2316 & 0.2146 & 0.2595 & 0.2473 & 0.2843 & 0.2576 \\
3 & 0.2903 & 0.2558 & 0.3161 & 0.2849 & 0.2693 & 0.2317 & 0.2428 & 0.2755 & 0.2633 & 0.2651 & 0.2695 \\
4 & 0.3002 & 0.2974 & 0.3292 & 0.3065 & 0.2639 & 0.2637 & 0.2677 & 0.2835 & 0.2743 & 0.2823 & 0.2869 \\
5 & 0.3024 & 0.3256 & 0.3402 & 0.3062 & 0.2693 & 0.2871 & 0.2717 & 0.2930 & 0.3078 & 0.2974 & 0.3001 \\
\bottomrule
\end{tabular}
}
\end{table}
\begin{table}[t]
\centering
\caption{Round-wise HotpotQA (contextual diversity) performance of the \textbf{Pre-defined} baseline across exploration and exploitation phases. 
We report Exact Match (EM) and token-level F1 for each agent over interaction rounds, where round 0 corresponds to the initial response before peer interaction. 
The highlighted Mean column reports the average performance across all 10 agents.}
\label{tab3}
\resizebox{\textwidth}{!}{
\begin{tabular}{lccccccccccM}
\toprule
\multicolumn{12}{l}{\textbf{(a) Exploration Phase: Exact Match (EM)}} \\
\midrule
Round & Agent 1 & Agent 2 & Agent 3 & Agent 4 & Agent 5 & Agent 6 & Agent 7 & Agent 8 & Agent 9 & Agent 10 & Mean \\
\midrule
0 (init) & 0.1633 & 0.1433 & 0.1533 & 0.1667 & 0.1200 & 0.1567 & 0.1600 & 0.1533 & 0.1300 & 0.1500 & 0.1497 \\
1 & 0.2133 & 0.1867 & 0.2100 & 0.1833 & 0.1633 & 0.1933 & 0.1733 & 0.1533 & 0.1600 & 0.1900 & 0.1827 \\
2 & 0.2267 & 0.2367 & 0.2133 & 0.1833 & 0.1933 & 0.2033 & 0.2067 & 0.1867 & 0.1833 & 0.2267 & 0.2060 \\
3 & 0.2633 & 0.2500 & 0.2267 & 0.2100 & 0.2033 & 0.2333 & 0.2267 & 0.2133 & 0.2300 & 0.2400 & 0.2297 \\
4 & 0.2733 & 0.2700 & 0.2400 & 0.2200 & 0.2267 & 0.2533 & 0.2367 & 0.2500 & 0.2533 & 0.2700 & 0.2493 \\
5 & 0.2900 & 0.2667 & 0.2700 & 0.2533 & 0.2567 & 0.2700 & 0.2500 & 0.2767 & 0.2800 & 0.2867 & 0.2700 \\

\midrule
\multicolumn{12}{l}{\textbf{(b) Exploration Phase: Token-level F1}} \\
\midrule
Round & Agent 1 & Agent 2 & Agent 3 & Agent 4 & Agent 5 & Agent 6 & Agent 7 & Agent 8 & Agent 9 & Agent 10 & Mean \\
\midrule
0 (init) & 0.2032 & 0.1929 & 0.2094 & 0.2035 & 0.1588 & 0.1917 & 0.2039 & 0.1923 & 0.1718 & 0.1955 & 0.1923 \\
1 & 0.2696 & 0.2470 & 0.2623 & 0.2310 & 0.2044 & 0.2474 & 0.2154 & 0.1937 & 0.2108 & 0.2410 & 0.2323 \\
2 & 0.2922 & 0.2969 & 0.2696 & 0.2385 & 0.2479 & 0.2535 & 0.2401 & 0.2382 & 0.2296 & 0.2832 & 0.2590 \\
3 & 0.3267 & 0.3179 & 0.2881 & 0.2720 & 0.2600 & 0.2782 & 0.2709 & 0.2616 & 0.2875 & 0.3008 & 0.2864 \\
4 & 0.3401 & 0.3459 & 0.3108 & 0.2899 & 0.2784 & 0.3046 & 0.2929 & 0.3128 & 0.3111 & 0.3326 & 0.3119 \\
5 & 0.3686 & 0.3468 & 0.3506 & 0.3102 & 0.3113 & 0.3298 & 0.3119 & 0.3359 & 0.3442 & 0.3609 & 0.3370 \\

\midrule
\multicolumn{12}{l}{\textbf{(c) Exploitation Phase: Exact Match (EM)}} \\
\midrule
Round & Agent 1 & Agent 2 & Agent 3 & Agent 4 & Agent 5 & Agent 6 & Agent 7 & Agent 8 & Agent 9 & Agent 10 & Mean \\
\midrule
0 (init) & 0.1533 & 0.1267 & 0.1600 & 0.1833 & 0.1600 & 0.1300 & 0.1500 & 0.1767 & 0.1467 & 0.1567 & 0.1543 \\
1 & 0.1767 & 0.1667 & 0.2033 & 0.1933 & 0.1700 & 0.1667 & 0.2033 & 0.1900 & 0.1767 & 0.1767 & 0.1823 \\
2 & 0.1967 & 0.2033 & 0.2400 & 0.1900 & 0.1767 & 0.2033 & 0.2267 & 0.1933 & 0.1867 & 0.1900 & 0.2007 \\
3 & 0.2200 & 0.2267 & 0.2267 & 0.2067 & 0.2200 & 0.2400 & 0.2400 & 0.1933 & 0.2200 & 0.2100 & 0.2203 \\
4 & 0.2300 & 0.2233 & 0.2267 & 0.2267 & 0.2300 & 0.2667 & 0.2300 & 0.2100 & 0.2233 & 0.2200 & 0.2287 \\
5 & 0.2433 & 0.2433 & 0.2500 & 0.2467 & 0.2700 & 0.2633 & 0.2300 & 0.2500 & 0.2367 & 0.2367 & 0.2470 \\

\midrule
\multicolumn{12}{l}{\textbf{(d) Exploitation Phase: Token-level F1}} \\
\midrule
Round & Agent 1 & Agent 2 & Agent 3 & Agent 4 & Agent 5 & Agent 6 & Agent 7 & Agent 8 & Agent 9 & Agent 10 & Mean \\
\midrule
0 (init) & 0.1901 & 0.1656 & 0.2029 & 0.2213 & 0.1783 & 0.1699 & 0.1876 & 0.2093 & 0.1959 & 0.1971 & 0.1918 \\
1 & 0.2189 & 0.2200 & 0.2575 & 0.2334 & 0.2097 & 0.2110 & 0.2484 & 0.2332 & 0.2444 & 0.2185 & 0.2295 \\
2 & 0.2556 & 0.2599 & 0.2985 & 0.2499 & 0.2245 & 0.2562 & 0.2701 & 0.2562 & 0.2526 & 0.2382 & 0.2562 \\
3 & 0.2863 & 0.2874 & 0.2913 & 0.2741 & 0.2678 & 0.2954 & 0.2962 & 0.2631 & 0.2833 & 0.2725 & 0.2817 \\
4 & 0.2941 & 0.2972 & 0.2985 & 0.2884 & 0.2814 & 0.3380 & 0.2914 & 0.2779 & 0.2887 & 0.2945 & 0.2950 \\
5 & 0.3149 & 0.3110 & 0.3188 & 0.3094 & 0.3377 & 0.3391 & 0.2925 & 0.3267 & 0.3179 & 0.3065 & 0.3174 \\
\bottomrule
\end{tabular}
}
\end{table}
\begin{table}[t]
\centering
\caption{Round-wise HotpotQA (contextual diversity) performance of the \textbf{Random} baseline across exploration and exploitation phases. 
We report Exact Match (EM) and token-level F1 for each agent over interaction rounds, where round 0 corresponds to the initial response before peer interaction. 
The highlighted Mean column reports the average performance across all 10 agents.}
\label{tab4}
\resizebox{\textwidth}{!}{
\begin{tabular}{lccccccccccM}
\toprule
\multicolumn{12}{l}{\textbf{(a) Exploration Phase: Exact Match (EM)}} \\
\midrule
Round & Agent 1 & Agent 2 & Agent 3 & Agent 4 & Agent 5 & Agent 6 & Agent 7 & Agent 8 & Agent 9 & Agent 10 & Mean \\
\midrule
0 (init) & 0.1633 & 0.1433 & 0.1533 & 0.1667 & 0.1200 & 0.1567 & 0.1600 & 0.1533 & 0.1300 & 0.1500 & 0.1497 \\
1 & 0.2067 & 0.1700 & 0.1900 & 0.1733 & 0.1900 & 0.2067 & 0.1767 & 0.1600 & 0.1800 & 0.1700 & 0.1823 \\
2 & 0.2300 & 0.1933 & 0.1900 & 0.2067 & 0.1967 & 0.2233 & 0.1967 & 0.2233 & 0.2233 & 0.2033 & 0.2087 \\
3 & 0.2600 & 0.2167 & 0.2133 & 0.2533 & 0.2133 & 0.2400 & 0.2233 & 0.2367 & 0.2467 & 0.2367 & 0.2340 \\
4 & 0.2900 & 0.2667 & 0.2300 & 0.2600 & 0.2500 & 0.2733 & 0.2267 & 0.2433 & 0.2600 & 0.2467 & 0.2547 \\
5 & 0.3000 & 0.2933 & 0.2600 & 0.2567 & 0.2633 & 0.2600 & 0.2633 & 0.2967 & 0.2700 & 0.2700 & 0.2733 \\

\midrule
\multicolumn{12}{l}{\textbf{(b) Exploration Phase: Token-level F1}} \\
\midrule
Round & Agent 1 & Agent 2 & Agent 3 & Agent 4 & Agent 5 & Agent 6 & Agent 7 & Agent 8 & Agent 9 & Agent 10 & Mean \\
\midrule
0 (init) & 0.2032 & 0.1929 & 0.2094 & 0.2035 & 0.1588 & 0.1917 & 0.2039 & 0.1923 & 0.1718 & 0.1955 & 0.1923 \\
1 & 0.2531 & 0.2237 & 0.2469 & 0.2217 & 0.2431 & 0.2434 & 0.2189 & 0.1967 & 0.2296 & 0.2249 & 0.2302 \\
2 & 0.2785 & 0.2532 & 0.2413 & 0.2657 & 0.2565 & 0.2707 & 0.2470 & 0.2690 & 0.2742 & 0.2606 & 0.2617 \\
3 & 0.3168 & 0.2792 & 0.2666 & 0.3166 & 0.2605 & 0.2856 & 0.2771 & 0.2979 & 0.2978 & 0.2918 & 0.2890 \\
4 & 0.3399 & 0.3381 & 0.2842 & 0.3195 & 0.3038 & 0.3212 & 0.2722 & 0.3138 & 0.3120 & 0.3033 & 0.3108 \\
5 & 0.3564 & 0.3561 & 0.3080 & 0.3241 & 0.3290 & 0.3157 & 0.3281 & 0.3560 & 0.3311 & 0.3339 & 0.3338 \\

\midrule
\multicolumn{12}{l}{\textbf{(c) Exploitation Phase: Exact Match (EM)}} \\
\midrule
Round & Agent 1 & Agent 2 & Agent 3 & Agent 4 & Agent 5 & Agent 6 & Agent 7 & Agent 8 & Agent 9 & Agent 10 & Mean \\
\midrule
0 (init) & 0.1533 & 0.1267 & 0.1600 & 0.1833 & 0.1600 & 0.1300 & 0.1500 & 0.1767 & 0.1467 & 0.1567 & 0.1543 \\
1 & 0.1667 & 0.1633 & 0.2100 & 0.1700 & 0.1833 & 0.1600 & 0.1633 & 0.1900 & 0.1700 & 0.1800 & 0.1757 \\
2 & 0.2000 & 0.1867 & 0.1833 & 0.1900 & 0.2000 & 0.1967 & 0.2000 & 0.1733 & 0.2200 & 0.2000 & 0.1950 \\
3 & 0.2033 & 0.1933 & 0.2167 & 0.2233 & 0.2033 & 0.2267 & 0.2100 & 0.2233 & 0.2167 & 0.2167 & 0.2133 \\
4 & 0.2233 & 0.2133 & 0.2533 & 0.2400 & 0.2267 & 0.2467 & 0.2033 & 0.2200 & 0.2233 & 0.2133 & 0.2263 \\
5 & 0.2300 & 0.2267 & 0.2267 & 0.2400 & 0.2367 & 0.2533 & 0.2233 & 0.2467 & 0.2733 & 0.2267 & 0.2383 \\

\midrule
\multicolumn{12}{l}{\textbf{(d) Exploitation Phase: Token-level F1}} \\
\midrule
Round & Agent 1 & Agent 2 & Agent 3 & Agent 4 & Agent 5 & Agent 6 & Agent 7 & Agent 8 & Agent 9 & Agent 10 & Mean \\
\midrule
0 (init) & 0.1901 & 0.1656 & 0.2029 & 0.2213 & 0.1783 & 0.1699 & 0.1876 & 0.2093 & 0.1959 & 0.1971 & 0.1918 \\
1 & 0.2074 & 0.2072 & 0.2516 & 0.2191 & 0.2310 & 0.2130 & 0.2146 & 0.2415 & 0.2200 & 0.2218 & 0.2227 \\
2 & 0.2554 & 0.2387 & 0.2317 & 0.2360 & 0.2442 & 0.2655 & 0.2582 & 0.2166 & 0.2873 & 0.2476 & 0.2481 \\
3 & 0.2599 & 0.2619 & 0.2673 & 0.2900 & 0.2527 & 0.2809 & 0.2657 & 0.2709 & 0.2776 & 0.2765 & 0.2703 \\
4 & 0.2860 & 0.2905 & 0.3071 & 0.3011 & 0.2779 & 0.3162 & 0.2592 & 0.2820 & 0.2897 & 0.2835 & 0.2893 \\
5 & 0.3001 & 0.2952 & 0.2844 & 0.3126 & 0.2979 & 0.3233 & 0.2798 & 0.3069 & 0.3412 & 0.3117 & 0.3053 \\
\bottomrule
\end{tabular}
}
\end{table}
\begin{table}[t]
\centering
\caption{Round-wise performance on 2WikiMultihopQA~(parameter transfer experiment) of \textbf{MACE} measured by Exact Match (EM) and token-level F1. 
We report performance for each agent over interaction rounds, where round 0 corresponds to the initial response before peer interaction. 
The highlighted Mean column reports the average performance across all 10 agents.}
\label{tab5}
\resizebox{\textwidth}{!}{
\begin{tabular}{lccccccccccM}
\toprule
\multicolumn{12}{l}{\textbf{(a) Exact Match (EM)}} \\
\midrule
Round & Agent 1 & Agent 2 & Agent 3 & Agent 4 & Agent 5 & Agent 6 & Agent 7 & Agent 8 & Agent 9 & Agent 10 & Mean \\
\midrule
0 (init) & 0.0767 & 0.1000 & 0.0967 & 0.0867 & 0.1267 & 0.0800 & 0.0800 & 0.0667 & 0.0800 & 0.0867 & 0.0880 \\
1 & 0.1033 & 0.1100 & 0.1267 & 0.1000 & 0.1267 & 0.0933 & 0.0867 & 0.1000 & 0.1000 & 0.0933 & 0.1040 \\
2 & 0.1467 & 0.1100 & 0.1300 & 0.1133 & 0.1167 & 0.1300 & 0.1033 & 0.1067 & 0.1433 & 0.1233 & 0.1223 \\
3 & 0.1500 & 0.1567 & 0.1467 & 0.1433 & 0.1367 & 0.1367 & 0.1100 & 0.1567 & 0.1533 & 0.1167 & 0.1407 \\
4 & 0.1500 & 0.1567 & 0.1867 & 0.1533 & 0.1100 & 0.1600 & 0.1200 & 0.1600 & 0.1600 & 0.1700 & 0.1527 \\
5 & 0.1700 & 0.1800 & 0.1900 & 0.1633 & 0.1767 & 0.2000 & 0.1667 & 0.1967 & 0.1833 & 0.1933 & 0.1820 \\

\midrule
\multicolumn{12}{l}{\textbf{(b) Token-level F1}} \\
\midrule
Round & Agent 1 & Agent 2 & Agent 3 & Agent 4 & Agent 5 & Agent 6 & Agent 7 & Agent 8 & Agent 9 & Agent 10 & Mean \\
\midrule
0 (init) & 0.0854 & 0.1161 & 0.1030 & 0.0993 & 0.1453 & 0.0849 & 0.1026 & 0.0808 & 0.0963 & 0.0949 & 0.1009 \\
1 & 0.1095 & 0.1275 & 0.1347 & 0.1155 & 0.1400 & 0.1114 & 0.1063 & 0.1093 & 0.1175 & 0.1091 & 0.1181 \\
2 & 0.1617 & 0.1277 & 0.1453 & 0.1295 & 0.1400 & 0.1521 & 0.1208 & 0.1147 & 0.1612 & 0.1362 & 0.1389 \\
3 & 0.1619 & 0.1763 & 0.1613 & 0.1629 & 0.1588 & 0.1538 & 0.1310 & 0.1811 & 0.1778 & 0.1370 & 0.1602 \\
4 & 0.1654 & 0.1726 & 0.2035 & 0.1739 & 0.1353 & 0.1788 & 0.1328 & 0.1763 & 0.1873 & 0.1785 & 0.1705 \\
5 & 0.1823 & 0.1978 & 0.2017 & 0.1729 & 0.1939 & 0.2164 & 0.1829 & 0.2232 & 0.2049 & 0.2107 & 0.1987 \\
\bottomrule
\end{tabular}
}
\end{table}
\begin{table}[t]
\centering
\caption{Round-wise performance on 2WikiMultihopQA~(parameter transfer experiment) of \textbf{In-Context Exploration} measured by Exact Match (EM) and token-level F1. 
We report performance for each agent over interaction rounds, where round 0 corresponds to the initial response before peer interaction. 
The highlighted Mean column reports the average performance across all 10 agents.}
\label{tab6}
\resizebox{\textwidth}{!}{
\begin{tabular}{lccccccccccM}
\toprule
\multicolumn{12}{l}{\textbf{(a) Exact Match (EM)}} \\
\midrule
Round & Agent 1 & Agent 2 & Agent 3 & Agent 4 & Agent 5 & Agent 6 & Agent 7 & Agent 8 & Agent 9 & Agent 10 & Mean \\
\midrule
0 (init) & 0.0767 & 0.1000 & 0.0967 & 0.0867 & 0.1267 & 0.0800 & 0.0800 & 0.0667 & 0.0800 & 0.0867 & 0.0880 \\
1 & 0.0833 & 0.1100 & 0.0933 & 0.0767 & 0.1167 & 0.1000 & 0.0933 & 0.0967 & 0.0667 & 0.1133 & 0.0950 \\
2 & 0.1133 & 0.1033 & 0.1000 & 0.0833 & 0.1100 & 0.1233 & 0.1067 & 0.1133 & 0.0767 & 0.0967 & 0.1027 \\
3 & 0.1367 & 0.1167 & 0.0967 & 0.1133 & 0.1067 & 0.1467 & 0.1100 & 0.1133 & 0.0767 & 0.1500 & 0.1167 \\
4 & 0.1300 & 0.1167 & 0.1067 & 0.0967 & 0.1167 & 0.1533 & 0.1333 & 0.1167 & 0.0733 & 0.1633 & 0.1207 \\
5 & 0.1333 & 0.1100 & 0.1033 & 0.0967 & 0.1133 & 0.1633 & 0.1300 & 0.1433 & 0.0833 & 0.1500 & 0.1227 \\

\midrule
\multicolumn{12}{l}{\textbf{(b) Token-level F1}} \\
\midrule
Round & Agent 1 & Agent 2 & Agent 3 & Agent 4 & Agent 5 & Agent 6 & Agent 7 & Agent 8 & Agent 9 & Agent 10 & Mean \\
\midrule
0 (init) & 0.0854 & 0.1161 & 0.1030 & 0.0993 & 0.1453 & 0.0849 & 0.1026 & 0.0808 & 0.0963 & 0.0949 & 0.1009 \\
1 & 0.0890 & 0.1303 & 0.1036 & 0.0928 & 0.1335 & 0.1130 & 0.1089 & 0.1105 & 0.0830 & 0.1232 & 0.1088 \\
2 & 0.1205 & 0.1243 & 0.1097 & 0.0994 & 0.1250 & 0.1387 & 0.1262 & 0.1361 & 0.0917 & 0.1121 & 0.1184 \\
3 & 0.1489 & 0.1386 & 0.1053 & 0.1310 & 0.1221 & 0.1670 & 0.1251 & 0.1325 & 0.0924 & 0.1696 & 0.1333 \\
4 & 0.1455 & 0.1307 & 0.1127 & 0.1136 & 0.1328 & 0.1721 & 0.1506 & 0.1350 & 0.0860 & 0.1784 & 0.1357 \\
5 & 0.1467 & 0.1259 & 0.1094 & 0.1121 & 0.1256 & 0.1795 & 0.1516 & 0.1532 & 0.0994 & 0.1719 & 0.1375 \\
\bottomrule
\end{tabular}
}
\end{table}
\begin{table}[t]
\centering
\caption{Round-wise performance on 2WikiMultihopQA~(parameter transfer experiment) of the \textbf{Pre-defined} baseline measured by Exact Match (EM) and token-level F1. 
We report performance for each agent over interaction rounds, where round 0 corresponds to the initial response before peer interaction. 
The highlighted Mean column reports the average performance across all 10 agents.}
\label{tab7}
\resizebox{\textwidth}{!}{
\begin{tabular}{lccccccccccM}
\toprule
\multicolumn{12}{l}{\textbf{(a) Exact Match (EM)}} \\
\midrule
Round & Agent 1 & Agent 2 & Agent 3 & Agent 4 & Agent 5 & Agent 6 & Agent 7 & Agent 8 & Agent 9 & Agent 10 & Mean \\
\midrule
0 (init) & 0.0767 & 0.1000 & 0.0967 & 0.0867 & 0.1267 & 0.0800 & 0.0800 & 0.0667 & 0.0800 & 0.0867 & 0.0880 \\
1 & 0.1000 & 0.1067 & 0.1067 & 0.1033 & 0.1133 & 0.1000 & 0.0900 & 0.0967 & 0.1000 & 0.1067 & 0.1023 \\
2 & 0.1267 & 0.1100 & 0.1233 & 0.1167 & 0.1067 & 0.1167 & 0.1167 & 0.1167 & 0.1033 & 0.1300 & 0.1167 \\
3 & 0.1300 & 0.1200 & 0.1233 & 0.1133 & 0.1133 & 0.1333 & 0.1300 & 0.1200 & 0.1400 & 0.1600 & 0.1283 \\
4 & 0.1433 & 0.1433 & 0.1200 & 0.1167 & 0.1300 & 0.1567 & 0.1233 & 0.1300 & 0.1633 & 0.1667 & 0.1393 \\
5 & 0.1533 & 0.1567 & 0.1500 & 0.1233 & 0.1767 & 0.1533 & 0.1433 & 0.1633 & 0.1833 & 0.1700 & 0.1573 \\

\midrule
\multicolumn{12}{l}{\textbf{(b) Token-level F1}} \\
\midrule
Round & Agent 1 & Agent 2 & Agent 3 & Agent 4 & Agent 5 & Agent 6 & Agent 7 & Agent 8 & Agent 9 & Agent 10 & Mean \\
\midrule
0 (init) & 0.0854 & 0.1161 & 0.1030 & 0.0993 & 0.1453 & 0.0849 & 0.1026 & 0.0808 & 0.0963 & 0.0949 & 0.1009 \\
1 & 0.1087 & 0.1199 & 0.1162 & 0.1226 & 0.1244 & 0.1137 & 0.1070 & 0.1150 & 0.1154 & 0.1203 & 0.1163 \\
2 & 0.1392 & 0.1229 & 0.1386 & 0.1345 & 0.1230 & 0.1284 & 0.1313 & 0.1275 & 0.1219 & 0.1413 & 0.1309 \\
3 & 0.1470 & 0.1351 & 0.1346 & 0.1295 & 0.1307 & 0.1413 & 0.1426 & 0.1385 & 0.1599 & 0.1730 & 0.1432 \\
4 & 0.1596 & 0.1634 & 0.1323 & 0.1359 & 0.1431 & 0.1723 & 0.1415 & 0.1564 & 0.1807 & 0.1839 & 0.1569 \\
5 & 0.1744 & 0.1754 & 0.1668 & 0.1347 & 0.1920 & 0.1682 & 0.1705 & 0.1858 & 0.2058 & 0.1845 & 0.1758 \\
\bottomrule
\end{tabular}
}
\end{table}
\begin{table}[t]
\centering
\caption{Round-wise performance on 2WikiMultihopQA~(parameter transfer experiment) of the \textbf{Random} baseline measured by Exact Match (EM) and token-level F1. 
We report performance for each agent over interaction rounds, where round 0 corresponds to the initial response before peer interaction. 
The highlighted Mean column reports the average performance across all 10 agents.}
\label{tab8}
\resizebox{\textwidth}{!}{
\begin{tabular}{lccccccccccM}
\toprule
\multicolumn{12}{l}{\textbf{(a) Exact Match (EM)}} \\
\midrule
Round & Agent 1 & Agent 2 & Agent 3 & Agent 4 & Agent 5 & Agent 6 & Agent 7 & Agent 8 & Agent 9 & Agent 10 & Mean \\
\midrule
0 (init) & 0.0767 & 0.1000 & 0.0967 & 0.0867 & 0.1267 & 0.0800 & 0.0800 & 0.0667 & 0.0800 & 0.0867 & 0.0880 \\
1 & 0.1033 & 0.1067 & 0.1100 & 0.1133 & 0.1233 & 0.0767 & 0.0833 & 0.1033 & 0.0800 & 0.1000 & 0.1000 \\
2 & 0.1400 & 0.1133 & 0.0967 & 0.1100 & 0.1233 & 0.1067 & 0.1033 & 0.0900 & 0.1000 & 0.1200 & 0.1103 \\
3 & 0.1433 & 0.1433 & 0.0933 & 0.1233 & 0.1600 & 0.1300 & 0.1167 & 0.1300 & 0.1333 & 0.1400 & 0.1313 \\
4 & 0.1433 & 0.1600 & 0.1233 & 0.1267 & 0.1767 & 0.1433 & 0.1433 & 0.1333 & 0.1567 & 0.1433 & 0.1450 \\
5 & 0.1433 & 0.1600 & 0.1600 & 0.1633 & 0.1700 & 0.1567 & 0.1700 & 0.1500 & 0.1633 & 0.1867 & 0.1623 \\

\midrule
\multicolumn{12}{l}{\textbf{(b) Token-level F1}} \\
\midrule
Round & Agent 1 & Agent 2 & Agent 3 & Agent 4 & Agent 5 & Agent 6 & Agent 7 & Agent 8 & Agent 9 & Agent 10 & Mean \\
\midrule
0 (init) & 0.0854 & 0.1161 & 0.1030 & 0.0993 & 0.1453 & 0.0849 & 0.1026 & 0.0808 & 0.0963 & 0.0949 & 0.1009 \\
1 & 0.1094 & 0.1245 & 0.1193 & 0.1221 & 0.1390 & 0.0857 & 0.1032 & 0.1143 & 0.0961 & 0.1087 & 0.1122 \\
2 & 0.1468 & 0.1295 & 0.1133 & 0.1226 & 0.1393 & 0.1165 & 0.1187 & 0.1011 & 0.1163 & 0.1384 & 0.1243 \\
3 & 0.1607 & 0.1551 & 0.1053 & 0.1350 & 0.1761 & 0.1410 & 0.1373 & 0.1384 & 0.1530 & 0.1575 & 0.1459 \\
4 & 0.1573 & 0.1775 & 0.1424 & 0.1416 & 0.1912 & 0.1547 & 0.1627 & 0.1494 & 0.1838 & 0.1628 & 0.1623 \\
5 & 0.1566 & 0.1828 & 0.1749 & 0.1730 & 0.1871 & 0.1663 & 0.1946 & 0.1603 & 0.1866 & 0.2020 & 0.1784 \\
\bottomrule
\end{tabular}
}
\end{table}
\begin{table}[t]
\centering
\caption{Round-wise Math500~(parametric diversity) performance of \textbf{MACE} across exploration and exploitation phases. 
We report accuracy for each agent and the mean accuracy across agents over interaction rounds. 
Round 0 corresponds to the initial response before peer interaction.
Agent 1 through 4 correspond to GPT-5, Qwen-7B, Llama-8B, Mistral-7B, respectively.}
\label{tab9}

\textbf{(a) Exploration Phase}
\vspace{0.4em}

\resizebox{0.55\textwidth}{!}{
\begin{tabular}{lccccM}
\toprule
Round & Agent 1 & Agent 2 & Agent 3 & Agent 4 & Mean \\
\midrule
0 (init) & 0.8626 & 0.3969 & 0.3206 & 0.0305 & 0.4027 \\
1        & 0.8779 & 0.4504 & 0.6718 & 0.3740 & 0.5935 \\
2        & 0.8702 & 0.5038 & 0.6412 & 0.5038 & 0.6298 \\
3        & 0.8931 & 0.5191 & 0.5573 & 0.6107 & 0.6450 \\
\bottomrule
\end{tabular}
}

\vspace{1.2em}

\textbf{(b) Exploitation Phase}
\vspace{0.4em}

\resizebox{0.55\textwidth}{!}{
\begin{tabular}{lccccM}
\toprule
Round & Agent 1 & Agent 2 & Agent 3 & Agent 4 & Mean \\
\midrule
0 (init) & 0.8397 & 0.4733 & 0.3206 & 0.0611 & 0.4237 \\
1        & 0.8550 & 0.4962 & 0.7252 & 0.3893 & 0.6164 \\
2        & 0.8779 & 0.4809 & 0.7252 & 0.5649 & 0.6622 \\
3        & 0.8779 & 0.5267 & 0.6947 & 0.6336 & 0.6832 \\
\bottomrule
\end{tabular}
}
\end{table}

\begin{table}[t]
\centering
\caption{Round-wise GPQA~(parametric diversity) performance of \textbf{MACE} across exploration and exploitation phases. 
We report accuracy for each agent and the mean accuracy across agents over interaction rounds. 
Round 0 corresponds to the initial response before peer interaction.
Agent 1 through 4 correspond to GPT-5, Qwen-7B, Llama-8B, Mistral-7B, respectively.}
\label{tab10}

\textbf{(a) Exploration Phase}
\vspace{0.4em}

\resizebox{0.55\textwidth}{!}{
\begin{tabular}{lccccM}
\toprule
Round & Agent 1 & Agent 2 & Agent 3 & Agent 4 & Mean \\
\midrule
0 (init) & 0.6061 & 0.3535 & 0.2828 & 0.1010 & 0.3359 \\
1        & 0.6465 & 0.5354 & 0.3939 & 0.3737 & 0.4874 \\
2        & 0.6465 & 0.5859 & 0.4545 & 0.4545 & 0.5354 \\
3        & 0.6667 & 0.5758 & 0.4949 & 0.4242 & 0.5404 \\
\bottomrule
\end{tabular}
}

\vspace{1.2em}

\textbf{(b) Exploitation Phase}
\vspace{0.4em}

\resizebox{0.55\textwidth}{!}{
\begin{tabular}{lccccM}
\toprule
Round & Agent 1 & Agent 2 & Agent 3 & Agent 4 & Mean \\
\midrule
0 (init) & 0.6465 & 0.3535 & 0.1919 & 0.1212 & 0.3283 \\
1        & 0.6364 & 0.5859 & 0.5960 & 0.3838 & 0.5505 \\
2        & 0.6566 & 0.5960 & 0.5758 & 0.4343 & 0.5657 \\
3        & 0.6566 & 0.6162 & 0.5758 & 0.4545 & 0.5758 \\
\bottomrule
\end{tabular}
}
\end{table}
\begin{table}[t]
\centering
\caption{Round-wise Math500~(parametric diversity) performance of \textbf{In-Context Exploration} across exploration and exploitation phases. 
We report accuracy for each agent and the mean accuracy across agents over interaction rounds. 
Round 0 corresponds to the initial response before peer interaction.
Agent 1 through 4 correspond to GPT-5, Qwen-7B, Llama-8B, Mistral-7B, respectively.}
\label{tab11}

\textbf{(a) Exploration Phase}
\vspace{0.4em}

\resizebox{0.55\textwidth}{!}{
\begin{tabular}{lccccM}
\toprule
Round & Agent 1 & Agent 2 & Agent 3 & Agent 4 & Mean \\
\midrule
0 (init) & 0.8855 & 0.3969 & 0.3206 & 0.0305 & 0.4084 \\
1        & 0.8931 & 0.4046 & 0.4504 & 0.4122 & 0.5401 \\
2        & 0.8931 & 0.3511 & 0.4809 & 0.5496 & 0.5687 \\
3        & 0.8779 & 0.3893 & 0.6031 & 0.6031 & 0.6183 \\
\bottomrule
\end{tabular}
}

\vspace{1.2em}

\textbf{(b) Exploitation Phase}
\vspace{0.4em}

\resizebox{0.55\textwidth}{!}{
\begin{tabular}{lccccM}
\toprule
Round & Agent 1 & Agent 2 & Agent 3 & Agent 4 & Mean \\
\midrule
0 (init) & 0.8473 & 0.4733 & 0.3206 & 0.0611 & 0.4256 \\
1        & 0.8626 & 0.4580 & 0.4580 & 0.4427 & 0.5553 \\
2        & 0.8702 & 0.4427 & 0.5267 & 0.5344 & 0.5935 \\
3        & 0.8702 & 0.4198 & 0.5344 & 0.5878 & 0.6031 \\
\bottomrule
\end{tabular}
}
\end{table}

\begin{table}[t]
\centering
\caption{Round-wise GPQA~(parametric diversity) performance of \textbf{In-Context Exploration} across exploration and exploitation phases. 
We report accuracy for each agent and the mean accuracy across agents over interaction rounds. 
Round 0 corresponds to the initial response before peer interaction.
Agent 1 through 4 correspond to GPT-5, Qwen-7B, Llama-8B, Mistral-7B, respectively.}
\label{tab12}

\textbf{(a) Exploration Phase}
\vspace{0.4em}

\resizebox{0.55\textwidth}{!}{
\begin{tabular}{lccccM}
\toprule
Round & Agent 1 & Agent 2 & Agent 3 & Agent 4 & Mean \\
\midrule
0 (init) & 0.6364 & 0.3535 & 0.2828 & 0.1010 & 0.3434 \\
1        & 0.6869 & 0.2424 & 0.2626 & 0.4646 & 0.4141 \\
2        & 0.6566 & 0.3434 & 0.2929 & 0.4848 & 0.4444 \\
3        & 0.6566 & 0.3333 & 0.2727 & 0.4747 & 0.4343 \\
\bottomrule
\end{tabular}
}

\vspace{1.2em}

\textbf{(b) Exploitation Phase}
\vspace{0.4em}

\resizebox{0.55\textwidth}{!}{
\begin{tabular}{lccccM}
\toprule
Round & Agent 1 & Agent 2 & Agent 3 & Agent 4 & Mean \\
\midrule
0 (init) & 0.6263 & 0.3535 & 0.1919 & 0.1212 & 0.3232 \\
1        & 0.6465 & 0.2626 & 0.2626 & 0.3030 & 0.3687 \\
2        & 0.6667 & 0.2828 & 0.3434 & 0.3838 & 0.4192 \\
3        & 0.6566 & 0.3838 & 0.3939 & 0.4040 & 0.4596 \\
\bottomrule
\end{tabular}
}
\end{table}
\begin{table}[t]
\centering
\caption{Round-wise Math500~(parametric diversity) performance of the \textbf{Pre-defined} baseline across exploration and exploitation phases. 
We report accuracy for each agent and the mean accuracy across agents over interaction rounds. 
Round 0 corresponds to the initial response before peer interaction.
Agent 1 through 4 correspond to GPT-5, Qwen-7B, Llama-8B, Mistral-7B, respectively.}
\label{tab13}

\textbf{(a) Exploration Phase}
\vspace{0.4em}

\resizebox{0.55\textwidth}{!}{
\begin{tabular}{lccccM}
\toprule
Round & Agent 1 & Agent 2 & Agent 3 & Agent 4 & Mean \\
\midrule
0 (init) & 0.8779 & 0.3969 & 0.3206 & 0.0305 & 0.4065 \\
1        & 0.8702 & 0.4809 & 0.1756 & 0.4122 & 0.4847 \\
2        & 0.8702 & 0.4427 & 0.3282 & 0.5344 & 0.5439 \\
3        & 0.8779 & 0.4351 & 0.4427 & 0.6412 & 0.5992 \\
\bottomrule
\end{tabular}
}

\vspace{1.2em}

\textbf{(b) Exploitation Phase}
\vspace{0.4em}

\resizebox{0.55\textwidth}{!}{
\begin{tabular}{lccccM}
\toprule
Round & Agent 1 & Agent 2 & Agent 3 & Agent 4 & Mean \\
\midrule
0 (init) & 0.8473 & 0.4733 & 0.3206 & 0.0611 & 0.4256 \\
1        & 0.8702 & 0.3740 & 0.1756 & 0.4122 & 0.4580 \\
2        & 0.8779 & 0.3664 & 0.3740 & 0.5420 & 0.5401 \\
3        & 0.8626 & 0.3206 & 0.3969 & 0.5878 & 0.5420 \\
\bottomrule
\end{tabular}
}
\end{table}

\begin{table}[t]
\centering
\caption{Round-wise GPQA~(parametric diversity) performance of the \textbf{Pre-defined} baseline across exploration and exploitation phases. 
We report accuracy for each agent and the mean accuracy across agents over interaction rounds. 
Round 0 corresponds to the initial response before peer interaction.
Agent 1 through 4 correspond to GPT-5, Qwen-7B, Llama-8B, Mistral-7B, respectively.}
\label{tab14}

\textbf{(a) Exploration Phase}
\vspace{0.4em}

\resizebox{0.55\textwidth}{!}{
\begin{tabular}{lccccM}
\toprule
Round & Agent 1 & Agent 2 & Agent 3 & Agent 4 & Mean \\
\midrule
0 (init) & 0.6263 & 0.3535 & 0.2828 & 0.1010 & 0.3409 \\
1        & 0.6566 & 0.3030 & 0.1717 & 0.3535 & 0.3712 \\
2        & 0.6566 & 0.3131 & 0.3232 & 0.4040 & 0.4242 \\
3        & 0.6566 & 0.3131 & 0.4040 & 0.4646 & 0.4596 \\
\bottomrule
\end{tabular}
}

\vspace{1.2em}

\textbf{(b) Exploitation Phase}
\vspace{0.4em}

\resizebox{0.55\textwidth}{!}{
\begin{tabular}{lccccM}
\toprule
Round & Agent 1 & Agent 2 & Agent 3 & Agent 4 & Mean \\
\midrule
0 (init) & 0.6061 & 0.3535 & 0.1919 & 0.1212 & 0.3182 \\
1        & 0.6465 & 0.3030 & 0.1818 & 0.3737 & 0.3763 \\
2        & 0.6667 & 0.2323 & 0.2929 & 0.4444 & 0.4091 \\
3        & 0.6566 & 0.2828 & 0.4040 & 0.4646 & 0.4520 \\
\bottomrule
\end{tabular}
}
\end{table}
\begin{table}[t]
\centering
\caption{Round-wise Math500~(parametric diversity) performance of the \textbf{Random} baseline across exploration and exploitation phases. 
We report accuracy for each agent and the mean accuracy across agents over interaction rounds. 
Round 0 corresponds to the initial response before peer interaction.
Agent 1 through 4 correspond to GPT-5, Qwen-7B, Llama-8B, Mistral-7B, respectively.}
\label{tab15}

\textbf{(a) Exploration Phase}
\vspace{0.4em}

\resizebox{0.55\textwidth}{!}{
\begin{tabular}{lccccM}
\toprule
Round & Agent 1 & Agent 2 & Agent 3 & Agent 4 & Mean \\
\midrule
0 (init) & 0.8702 & 0.3969 & 0.3206 & 0.0305 & 0.4046 \\
1        & 0.8473 & 0.4733 & 0.3893 & 0.2748 & 0.4962 \\
2        & 0.8473 & 0.4656 & 0.4656 & 0.3664 & 0.5363 \\
3        & 0.8855 & 0.4809 & 0.5115 & 0.3664 & 0.5611 \\
\bottomrule
\end{tabular}
}

\vspace{1.2em}

\textbf{(b) Exploitation Phase}
\vspace{0.4em}

\resizebox{0.55\textwidth}{!}{
\begin{tabular}{lccccM}
\toprule
Round & Agent 1 & Agent 2 & Agent 3 & Agent 4 & Mean \\
\midrule
0 (init) & 0.8321 & 0.4733 & 0.3206 & 0.0611 & 0.4218 \\
1        & 0.8702 & 0.4427 & 0.3130 & 0.2443 & 0.4676 \\
2        & 0.8855 & 0.4275 & 0.4427 & 0.3817 & 0.5344 \\
3        & 0.8855 & 0.4885 & 0.5191 & 0.4122 & 0.5763 \\
\bottomrule
\end{tabular}
}
\end{table}

\begin{table}[t]
\centering
\caption{Round-wise GPQA~(parametric diversity) performance of the \textbf{Random} baseline across exploration and exploitation phases. 
We report accuracy for each agent and the mean accuracy across agents over interaction rounds. 
Round 0 corresponds to the initial response before peer interaction.
Agent 1 through 4 correspond to GPT-5, Qwen-7B, Llama-8B, Mistral-7B, respectively.}
\label{tab16}

\textbf{(a) Exploration Phase}
\vspace{0.4em}

\resizebox{0.55\textwidth}{!}{
\begin{tabular}{lccccM}
\toprule
Round & Agent 1 & Agent 2 & Agent 3 & Agent 4 & Mean \\
\midrule
0 (init) & 0.5758 & 0.3535 & 0.2828 & 0.1010 & 0.3283 \\
1        & 0.6465 & 0.3434 & 0.2424 & 0.2121 & 0.3611 \\
2        & 0.6768 & 0.3939 & 0.3333 & 0.2323 & 0.4091 \\
3        & 0.7071 & 0.4444 & 0.4040 & 0.3434 & 0.4747 \\
\bottomrule
\end{tabular}
}

\vspace{1.2em}

\textbf{(b) Exploitation Phase}
\vspace{0.4em}

\resizebox{0.55\textwidth}{!}{
\begin{tabular}{lccccM}
\toprule
Round & Agent 1 & Agent 2 & Agent 3 & Agent 4 & Mean \\
\midrule
0 (init) & 0.6162 & 0.3535 & 0.1919 & 0.1212 & 0.3207 \\
1        & 0.6263 & 0.3737 & 0.3131 & 0.1818 & 0.3737 \\
2        & 0.6566 & 0.3939 & 0.3636 & 0.2525 & 0.4167 \\
3        & 0.6566 & 0.4747 & 0.3535 & 0.2727 & 0.4394 \\
\bottomrule
\end{tabular}
}
\end{table}

\clearpage

\subsection{Sensitivity Analysis of Exploration Coefficient $\alpha$}
\label{apdx:alpha_sensitivity}

\begin{figure}[h!]
    \centering
    \includegraphics[width=0.75\linewidth]{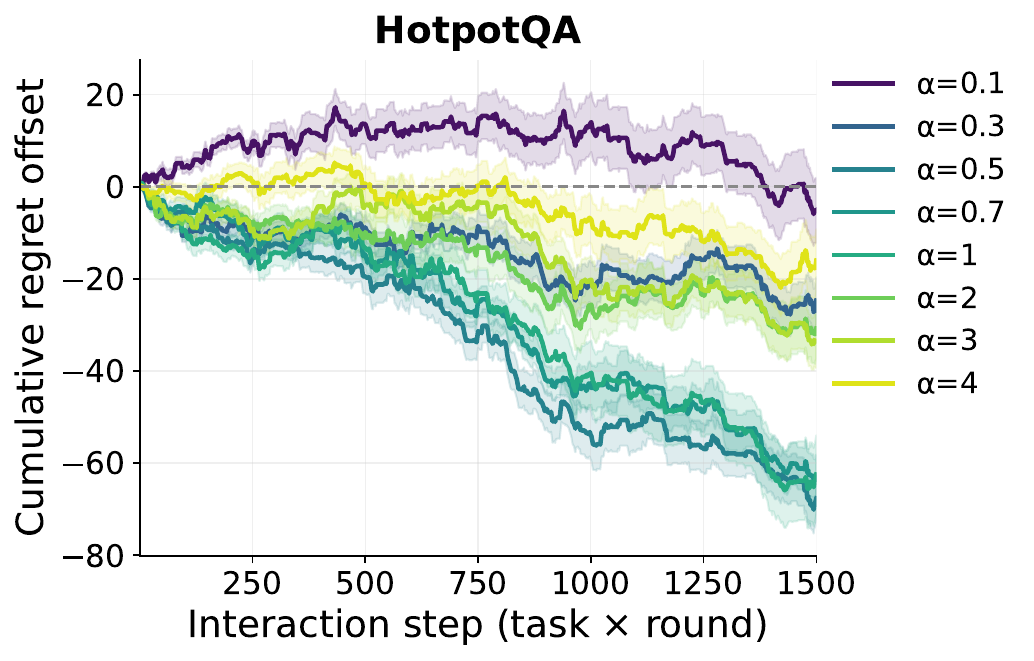}
    \caption{Sensitivity of $\alpha$ on the HotpotQA benchmark.}
    \label{fig:alpha_sensitivity}
\end{figure}

We analyze the sensitivity of MACE to the exploration coefficient $\alpha$ on the HotpotQA benchmark~(contextual diversity), as shown in Figure~\ref{fig:alpha_sensitivity}. 
The coefficient $\alpha$ controls the strength of the uncertainty bonus in the LinUCB selection rule~(\eqref{eq:ucb}), thereby determining how aggressively agents explore under-tested peers. 

Overall, we find that MACE is sensitive to this trade-off: both insufficient and excessive exploration lead to worse regret behavior.
When $\alpha$ is too small, such as $\alpha=0.1$, the uncertainty bonus becomes negligible, and the policy effectively reduces to a greedy peer-selection strategy.
In this regime, agents are more likely to prematurely commit to peers based on noisy early observations, leading to poor exploration and regret even worse than the random baseline, indicated by the dashed horizontal line in the figure. 
This is consistent with our broader observation that non-exploring or weakly exploring policies can fail severely in heterogeneous multi-agent environments.
On the other hand, overly large values of $\alpha$ also degrade exploratory performance. Although a higher $\alpha$ encourages broader peer exploration, excessive exploration slows down the decrease in cumulative regret because agents continue to prioritize uncertain peers even after sufficient evidence has been collected. 
In other words, the policy overweights exploration relative to exploitation, delaying convergence toward effective collaborators. 
Empirically, intermediate values around $\alpha=0.5$ to $\alpha=1.0$ achieve the strongest regret reduction, suggesting that MACE benefits from a moderate exploration bonus that is large enough to prevent premature commitment but not so large that it prevents efficient exploitation.

\subsection{Experiments on Stronger LLMs}
\label{apdx:frontier}
To further examine the role of model capabilities, we revisit the parametric diversity setting using a set of stronger frontier models: GPT-5, GPT-4, GPT-5.4-mini, and GPT-5.4-nano. 
Although these models vary in scale, they are all sufficiently capable of making meaningful progress on GPQA, enforcing more homogeneity in the model capabilities.

\begin{figure}[t]
    \centering
    \vspace{-3mm}
    \begin{subfigure}[t]{0.45\textwidth}
        \centering
        \includegraphics[width=\linewidth]{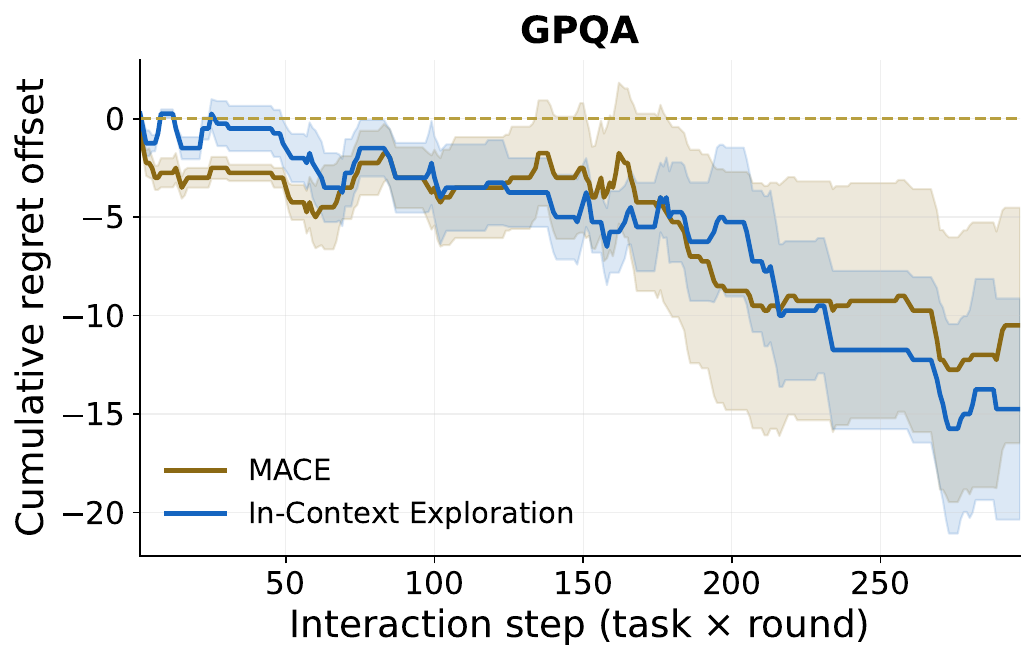}
        \caption{Trial-and-Error Phase}
        \label{fig:exploration_f1}
    \end{subfigure}
    \hfill
    \begin{subfigure}[t]{0.45\textwidth}
        \centering
        \includegraphics[width=\linewidth]{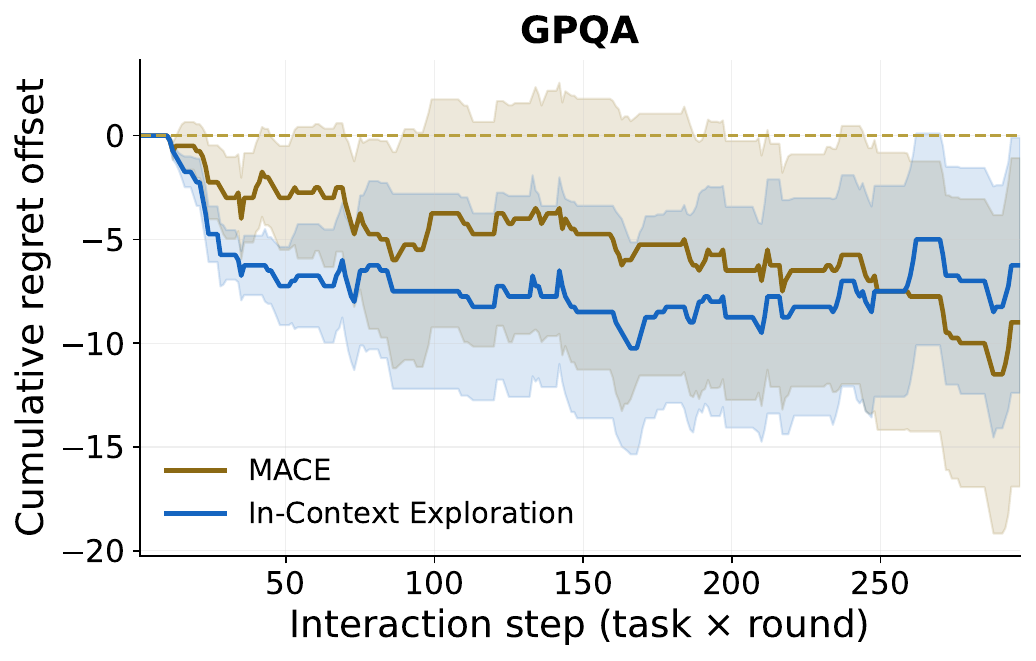}
        \caption{Exploitation Phase}
        \label{fig:exploration_f1b}
    \end{subfigure}
    \caption{\textbf{Cumulative regret offset on a set of stronger LLMs.} The average cumulative regret offset~(Baseline -- Random) across interaction steps are shown~(lower the better). The horizontal dashed line at Offset $=0$ is the Random baseline, and the shaded areas indicate the standard error across participating agents.}
    \vspace{-4mm}
    \label{fig:frontier}
\end{figure}

Figure~\ref{fig:frontier} compares the cumulative regret offset of MACE and In-Context Exploration in this setting. 
The two methods exhibit broadly similar regret trends in both the Trial-and-Error and Exploitation phases. 
We conjecture that this is because most agents are comparably capable, and identifying the ``best'' peer becomes less consequential, and exploration contributes less to cumulative regret reduction.
This is consistent with our theoretical intuition from Definition~\ref{def:diversity}: when the capability diversity $\delta$ is small, the penalty for selecting one peer over another is reduced, and the benefit of explicit exploration correspondingly diminishes. 

Interestingly, however, the round-wise task performance in Table~\ref{tab:frontier} reveals a more nuanced picture. 
Despite the similar regret trajectories, MACE still yields stronger downstream accuracy than In-Context Exploration, with the advantage becoming especially clear in the Exploitation phase. 
MACE achieves the best final-round mean accuracy in both phases, reaching 0.7475 during Trial-and-Error and 0.7652 during Exploitation, compared to 0.7273 and 0.7222 for In-Context Exploration, respectively. 
This suggests that even when capability diversity is not large enough to produce a pronounced regret gap, structured exploration can still improve the quality of information exchange and lead to better final predictions. 
Overall, these results support the view that the regret benefit of exploration depends strongly on diversity, while the performance benefit can persist even in relatively homogeneous pools of strong agents.
In light of these observations, we further provide experimental results on a fully homogeneous setting in Appendix~\ref{apdx:homogeneous}, where agents have identical capabilities and contexts, all instructed to solve shared tasks.
\begin{table*}[h!]
\centering
\caption{Round-wise GPQA performance across Trial-and-Error and Exploitation phases, using a set of stronger models.
We report accuracy for each agent and the mean accuracy across agents over interaction rounds.
Round 0 corresponds to the initial response before peer interaction.
Agent 1 through 4 correspond to GPT-5, GPT-4, GPT-5.4-mini, GPT-5.4-nano respectively.
Best final-round performance is underlined.}
\label{tab:frontier}

\scriptsize
\setlength{\tabcolsep}{3pt}

\begin{tabular}{cc}
\toprule
\textbf{Trial-and-Error Phase} & \textbf{Exploitation Phase} \\
\midrule

\begin{minipage}{0.48\textwidth}
\centering
\textbf{(a) MACE}\\[0.3em]
\resizebox{\linewidth}{!}{
\begin{tabular}{lccccM}
\toprule
Round & Agent 1 & Agent 2 & Agent 3 & Agent 4 & Mean \\
\midrule
0 (init) & 0.6263 & 0.7172 & 0.6263 & 0.5556 & 0.6313 \\
1        & 0.6667 & 0.7172 & 0.6869 & 0.6061 & 0.6692 \\
2        & 0.6566 & 0.7475 & 0.7374 & 0.7273 & 0.7172 \\
3        & 0.6970 & 0.7576 & 0.7879 & 0.7475 & \underline{0.7475} \\
\bottomrule
\end{tabular}
}
\end{minipage}
&
\begin{minipage}{0.48\textwidth}
\centering
\textbf{(b) MACE}\\[0.3em]
\resizebox{\linewidth}{!}{
\begin{tabular}{lccccM}
\toprule
Round & Agent 1 & Agent 2 & Agent 3 & Agent 4 & Mean \\
\midrule
0 (init) & 0.6263 & 0.7374 & 0.6667 & 0.4545 & 0.6212 \\
1        & 0.6566 & 0.7475 & 0.7778 & 0.6465 & 0.7071 \\
2        & 0.6667 & 0.7677 & 0.7980 & 0.7879 & 0.7551 \\
3        & 0.6768 & 0.7677 & 0.8081 & 0.8081 & \underline{0.7652} \\
\bottomrule
\end{tabular}
}
\end{minipage}
\\[1.4em]
\begin{minipage}{0.48\textwidth}
\centering
\vspace{0.5em}
\textbf{(c) In-Context Exploration}\\[0.3em]
\resizebox{\linewidth}{!}{
\begin{tabular}{lccccM}
\toprule
Round & Agent 1 & Agent 2 & Agent 3 & Agent 4 & Mean \\
\midrule
0 (init) & 0.6162 & 0.7677 & 0.6465 & 0.4747 & 0.6263 \\
1        & 0.6869 & 0.7677 & 0.6768 & 0.6566 & 0.6970 \\
2        & 0.6869 & 0.7475 & 0.7374 & 0.7172 & 0.7222 \\
3        & 0.6970 & 0.7576 & 0.7273 & 0.7273 & 0.7273 \\
\bottomrule
\end{tabular}
}
\end{minipage}
&
\begin{minipage}{0.48\textwidth}
\centering
\vspace{0.5em}
\textbf{(d) In-Context Exploration}\\[0.3em]
\resizebox{\linewidth}{!}{
\begin{tabular}{lccccM}
\toprule
Round & Agent 1 & Agent 2 & Agent 3 & Agent 4 & Mean \\
\midrule
0 (init) & 0.5758 & 0.7273 & 0.6364 & 0.5556 & 0.6237 \\
1        & 0.6465 & 0.7172 & 0.7374 & 0.6566 & 0.6894 \\
2        & 0.6869 & 0.7576 & 0.7374 & 0.7273 & 0.7273 \\
3        & 0.6768 & 0.7374 & 0.7071 & 0.7677 & 0.7222 \\
\bottomrule
\end{tabular}
}
\end{minipage}
\\[1.4em]

\end{tabular}
\end{table*}

\subsection{Fully Homogeneous Agent Setup}
\label{apdx:homogeneous}

\begin{figure}[t]
    \centering
    \vspace{-3mm}
    \begin{subfigure}[t]{0.45\textwidth}
        \centering
        \includegraphics[width=\linewidth]{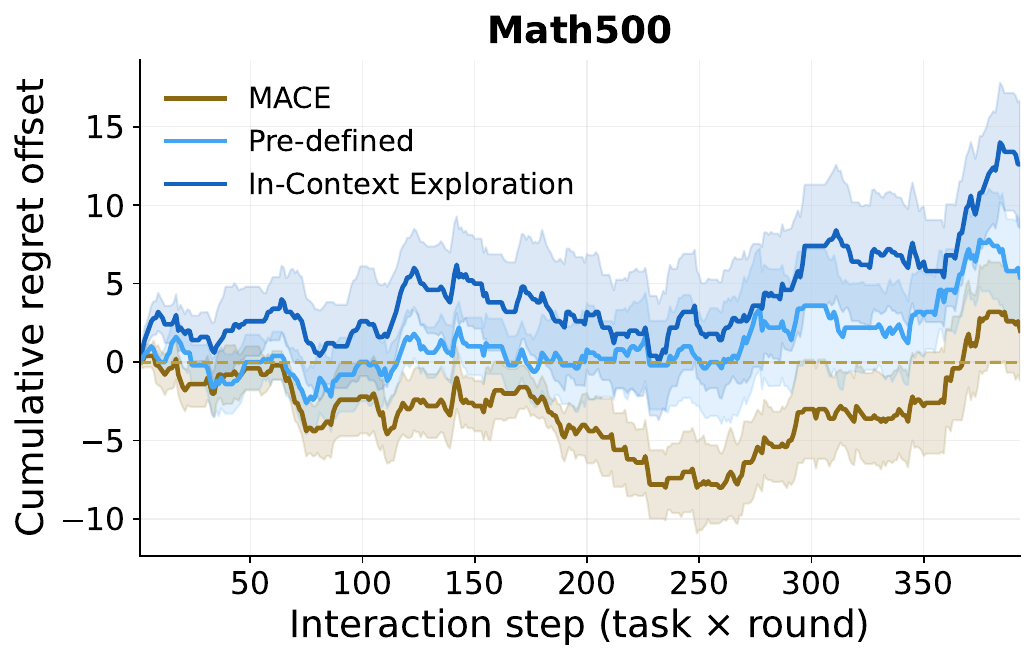}
        \caption{Trial-and-Error Phase}
        \label{fig:exploration_f1}
    \end{subfigure}
    \hfill
    \begin{subfigure}[t]{0.45\textwidth}
        \centering
        \includegraphics[width=\linewidth]{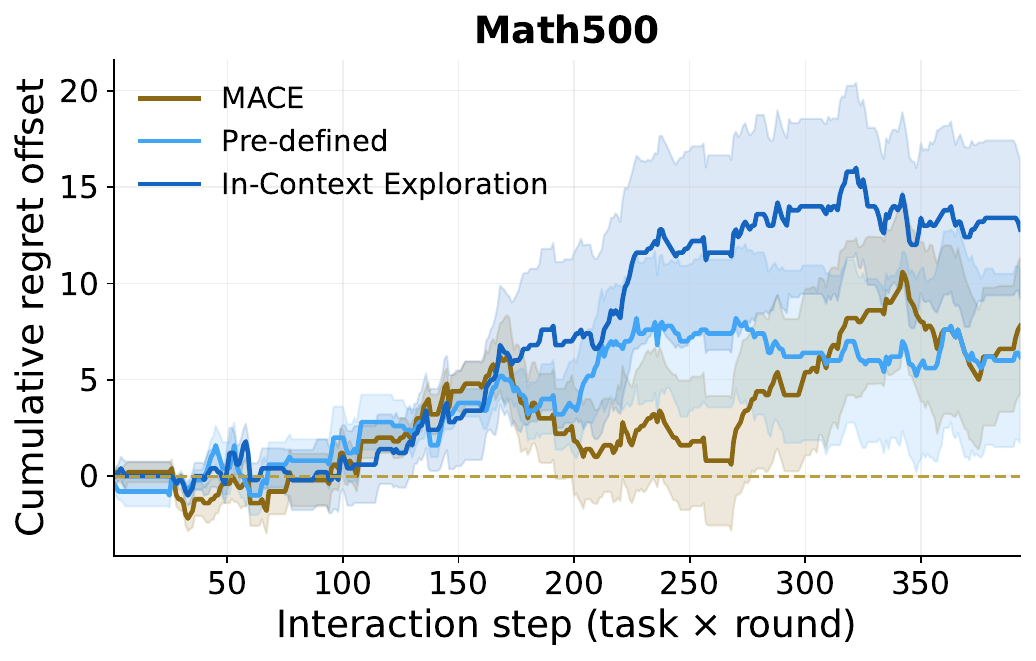}
        \caption{Exploitation Phase}
        \label{fig:exploration_f1}
    \end{subfigure}
    \caption{\textbf{Cumulative regret offset on a set of homogeneous LLMs.} The average cumulative regret offset~(Baseline -- Random) across interaction steps are shown~(lower the better). The horizontal dashed line at Offset $=0$ is the Random baseline, and the shaded areas indicate the standard error across participating agents.}
    \vspace{-4mm}
    \label{fig:homogeneous}
\end{figure}

To isolate the role of capability diversity, we consider a strictly homogeneous setting on Math500 using five copies of Qwen2.5-7B-Instruct. 
In this setup, all agents share the same backbone model, receive the same input content, and solve the same task distribution, so the capability diversity $\delta$ in Definition~\ref{def:diversity} is effectively zero. 
As shown in Figure~\ref{fig:homogeneous}, the cumulative regret trends are indeed much less profound than in the heterogeneous settings. 
MACE generally maintains a lower regret offset than In-Context Exploration, especially during the Trial-and-Error phase, but the advantage is modest and often close to the dashed random-baseline line. 
In the Exploitation phase, the regret offset of MACE also rises above zero in later rounds, indicating that explicit exploration offers little regret advantage once agents become nearly interchangeable. 
This mirrors, and further amplifies, the low-diversity trend observed with the stronger frontier-model pool in Appendix~\ref{apdx:frontier}: when peers have similar capabilities, identifying a particular collaborator matters much less, so exploration has limited room to reduce regret.

That said, Table~\ref{tab:homogeneous} shows that MACE still yields small but consistent improvements in round-wise task performance. 
Although the gains are subtle, MACE attains the best final-round mean accuracy in both phases: in Trial-and-Error, it reaches $0.4779$, compared to $0.4687$ for In-Context Exploration and $0.4611$ for Pre-defined; in Exploitation, it reaches $0.5053$, compared to $0.4962$ and $0.4641$, respectively.
Thus, even when explicit exploration does not substantially lower cumulative regret, structured peer selection can still improve the quality of interaction and lead to slightly better final predictions. 
Overall, these results are consistent with our theory: as $\delta \to 0$, the regret benefit of exploration largely vanishes, but modest performance gains can still remain through better coordination over otherwise similar agents.

\begin{table*}[h!]
\centering
\caption{Round-wise Math500 performance across Trial-and-Error and Exploitation phases, using a set of stronger models.
We report accuracy for each agent and the mean accuracy across agents over interaction rounds.
Round 0 corresponds to the initial response before peer interaction.
Agent 1 through 5 correspond to the five agents in the homogeneous-model setup, where each agent is an instance of the Qwen2.5-7B-Instruct model.
Best final-round performance is underlined.}
\label{tab:homogeneous}

\scriptsize
\setlength{\tabcolsep}{3pt}

\begin{tabular}{cc}
\toprule
\textbf{Trial-and-Error Phase} & \textbf{Exploitation Phase} \\
\midrule

\begin{minipage}{0.48\textwidth}
\centering
\textbf{(a) MACE}\\[0.3em]
\resizebox{\linewidth}{!}{
\begin{tabular}{lcccccM}
\toprule
Round & Agent 1 & Agent 2 & Agent 3 & Agent 4 & Agent 5 & Mean \\
\midrule
0 (init) & 0.4809 & 0.4351 & 0.4504 & 0.4122 & 0.4122 & 0.4382 \\
1        & 0.4656 & 0.4122 & 0.4275 & 0.4504 & 0.4504 & 0.4412 \\
2        & 0.4351 & 0.4275 & 0.4809 & 0.4962 & 0.4427 & 0.4565 \\
3        & 0.4885 & 0.4809 & 0.4885 & 0.5267 & 0.4046 & \underline{0.4779} \\
\bottomrule
\end{tabular}
}
\end{minipage}
&
\begin{minipage}{0.48\textwidth}
\centering
\textbf{(b) MACE}\\[0.3em]
\resizebox{\linewidth}{!}{
\begin{tabular}{lcccccM}
\toprule
Round & Agent 1 & Agent 2 & Agent 3 & Agent 4 & Agent 5 & Mean \\
\midrule
0 (init) & 0.4580 & 0.4733 & 0.4122 & 0.4351 & 0.5267 & 0.4611 \\
1        & 0.4733 & 0.4198 & 0.5038 & 0.4351 & 0.4656 & 0.4595 \\
2        & 0.5038 & 0.4962 & 0.4809 & 0.4733 & 0.4275 & 0.4763 \\
3        & 0.5038 & 0.4885 & 0.5344 & 0.5267 & 0.4733 & \underline{0.5053} \\
\bottomrule
\end{tabular}
}
\end{minipage}
\\[1.4em]

\begin{minipage}{0.48\textwidth}
\centering
\vspace{0.5em}
\textbf{(c) In-Context Exploration}\\[0.3em]
\resizebox{\linewidth}{!}{
\begin{tabular}{lcccccM}
\toprule
Round & Agent 1 & Agent 2 & Agent 3 & Agent 4 & Agent 5 & Mean \\
\midrule
0 (init) & 0.4809 & 0.4351 & 0.4504 & 0.4122 & 0.4122 & 0.4382 \\
1        & 0.4580 & 0.4580 & 0.4275 & 0.4351 & 0.4351 & 0.4427 \\
2        & 0.4733 & 0.4885 & 0.4580 & 0.4580 & 0.4962 & 0.4748 \\
3        & 0.4656 & 0.5573 & 0.4427 & 0.4275 & 0.4504 & 0.4687 \\
\bottomrule
\end{tabular}
}
\end{minipage}
&
\begin{minipage}{0.48\textwidth}
\centering
\vspace{0.5em}
\textbf{(d) In-Context Exploration}\\[0.3em]
\resizebox{\linewidth}{!}{
\begin{tabular}{lcccccM}
\toprule
Round & Agent 1 & Agent 2 & Agent 3 & Agent 4 & Agent 5 & Mean \\
\midrule
0 (init) & 0.4580 & 0.4733 & 0.4122 & 0.4351 & 0.5267 & 0.4611 \\
1        & 0.4733 & 0.4656 & 0.4656 & 0.4351 & 0.4046 & 0.4489 \\
2        & 0.4885 & 0.4656 & 0.4351 & 0.4351 & 0.4733 & 0.4595 \\
3        & 0.5038 & 0.4580 & 0.4046 & 0.4504 & 0.5038 & 0.4641 \\
\bottomrule
\end{tabular}
}
\end{minipage}
\\[1.4em]

\begin{minipage}{0.48\textwidth}
\centering
\vspace{0.5em}
\textbf{(e) Pre-defined}\\[0.3em]
\resizebox{\linewidth}{!}{
\begin{tabular}{lcccccM}
\toprule
Round & Agent 1 & Agent 2 & Agent 3 & Agent 4 & Agent 5 & Mean \\
\midrule
0 (init) & 0.4809 & 0.4351 & 0.4504 & 0.4122 & 0.4122 & 0.4382 \\
1        & 0.4962 & 0.4504 & 0.4122 & 0.4122 & 0.4198 & 0.4382 \\
2        & 0.4504 & 0.4351 & 0.4809 & 0.4046 & 0.4427 & 0.4427 \\
3        & 0.5038 & 0.5038 & 0.3969 & 0.4504 & 0.4504 & 0.4611 \\
\bottomrule
\end{tabular}
}
\end{minipage}
&
\begin{minipage}{0.48\textwidth}
\centering
\vspace{0.5em}
\textbf{(f) Pre-defined}\\[0.3em]
\resizebox{\linewidth}{!}{
\begin{tabular}{lcccccM}
\toprule
Round & Agent 1 & Agent 2 & Agent 3 & Agent 4 & Agent 5 & Mean \\
\midrule
0 (init) & 0.4580 & 0.4733 & 0.4122 & 0.4351 & 0.5267 & 0.4611 \\
1        & 0.4656 & 0.4580 & 0.4427 & 0.4580 & 0.4504 & 0.4550 \\
2        & 0.4504 & 0.4656 & 0.5191 & 0.4656 & 0.4962 & 0.4794 \\
3        & 0.4580 & 0.4885 & 0.5267 & 0.5115 & 0.4962 & 0.4962 \\
\bottomrule
\end{tabular}
}
\end{minipage}
\\[1.4em]


\end{tabular}
\end{table*}

\subsection{MACE-TD: A Temporal-Difference Extension}
\label{apdx:mace_td}

\paragraph{MACE-TD.}
The main MACE formulation treats each interaction as a one-step contextual bandit decision. 
However, if agents engage in multiple rounds of discussion for each task question~(like the setting in our work), an early peer query may improve not only the current response but also the agent's future information state. 
To capture this delayed effect, we naturally consider a temporal-difference~\citep{sutton1988learning} extension, MACE-TD, which models the $R$ interaction rounds for each task as a finite-horizon sequential decision process.

Let $h_{i,r}$ denote agent $i$'s interaction history before interaction round $r \in [0,R]$, and let $\boldsymbol{\mathbf{x}}_{i,a,r}$ be the relational feature vector for querying peer $a$.
We approximate the action-value function by a linear model
\[
Q_i(h_{i,r},a) \approx \boldsymbol{\mathbf{x}}_{i,a,r}^{\top}\boldsymbol{\theta}_{i,a}.
\]
After agent $i$ selects peer $a_{i,r}$, observes reward $\rho_{i,r}$, and transitions to history $h_{i,r+1}$, we form the temporal-difference target
\[
y_{i,r}=\rho_{i,r}+\gamma \max_{a'\in[N]} Q_i(h_{i,r+1},a'),
\]
where $\rho_{i,r}$ is the reward agent $i$ received in round $r$, $\gamma$ is set to 0.9.
The linear parameters are then updated by replacing the immediate reward target in MACE with the temporal-difference target:
\[
\mathbf{A}_{i,a_{i,r}} \leftarrow \mathbf{A}_{i,a_{i,r}}+
\boldsymbol{\mathbf{x}}_{i,a_{i,r},r}\boldsymbol{\mathbf{x}}_{i,a_{i,r},r}^{\top},
\qquad
\mathbf{b}_{i,a_{i,r}} \leftarrow \mathbf{b}_{i,a_{i,r}}+
y_{i,r}\boldsymbol{\mathbf{x}}_{i,a_{i,r},r}.
\]
Action selection follows the same optimistic rule as MACE, using the estimated temporal-difference target together with an uncertainty bonus. 
This extension allows peer selection to account for delayed benefits of interaction, such as early exploration that improves information propagation or enables better decisions in later rounds.

\paragraph{Selection distribution comparison.}
To begin with, we compare the overall peer selection distribution of MACE-TD to the original MACE method.
Interestingly, MACE-TD exhibits a slightly sharper distribution compared to MACE. 
We conjecture that this arises from its ability to internalize multi-turn interaction effects, potentially promoting emergent cooperative structures~(see the last paragraph of this section). 
In particular, agents learn to favor peers that not only yield immediate gains but also lead to improved downstream outcomes, resulting in role specialization~(i.e., each agent focuses on exploring a small subset of agents to reach maximum coverage across agents) and more stable interaction patterns over rounds. 

\begin{figure}[h!]
    \centering
    \includegraphics[width=\linewidth]{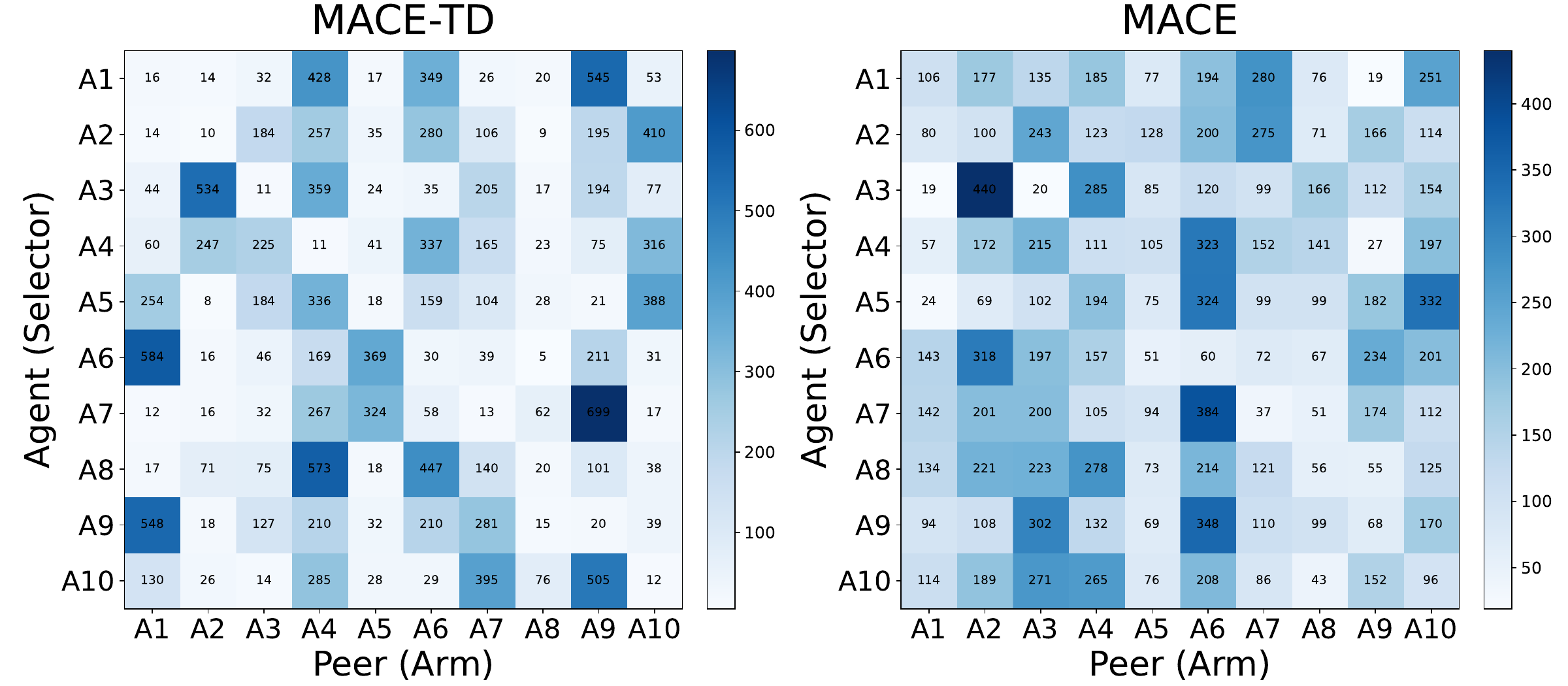}
    \caption{\textbf{Peer Selection distribution comparison: MACE vs MACE-TD}}
    \label{fig:heatmap_TD}
\end{figure}

\begin{figure}[h!]
    \centering
    \vspace{-3mm}
    \begin{subfigure}[t]{0.32\textwidth}
        \centering
        \includegraphics[width=\linewidth]{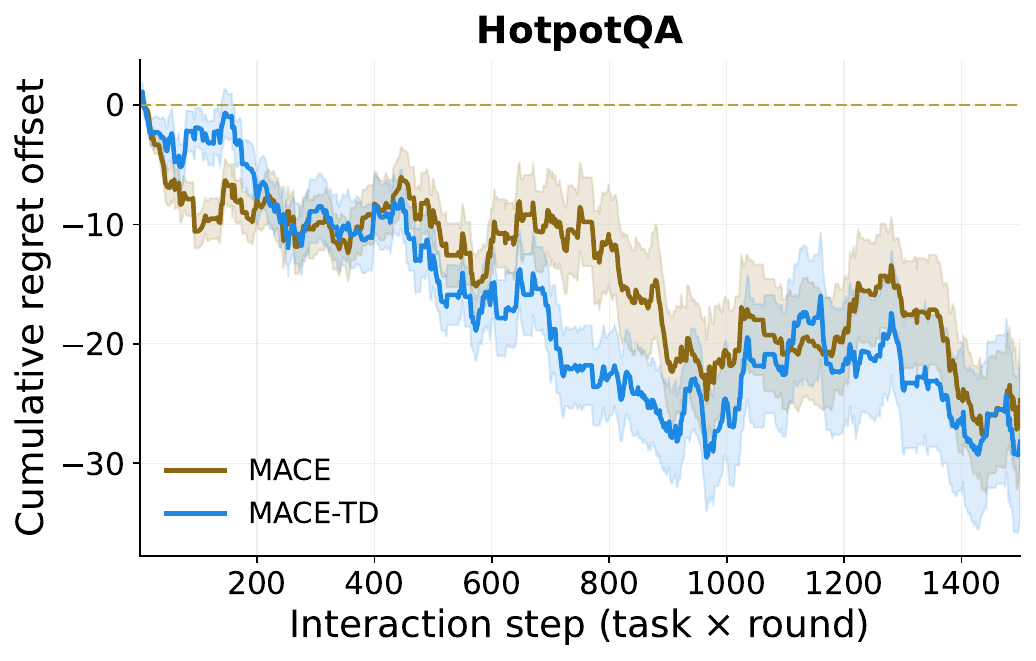}
    \end{subfigure}
    \hfill
    \begin{subfigure}[t]{0.32\textwidth}
        \centering
        \includegraphics[width=\linewidth]{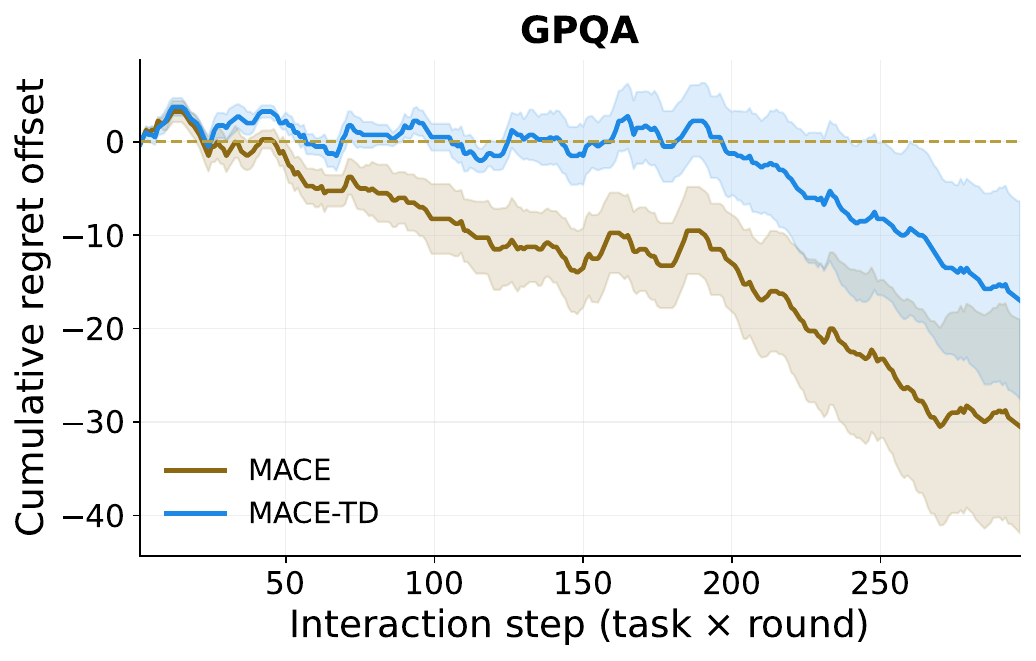}
    \end{subfigure}
    \hfill
    \begin{subfigure}[t]{0.32\textwidth}
        \centering
        \includegraphics[width=\linewidth]{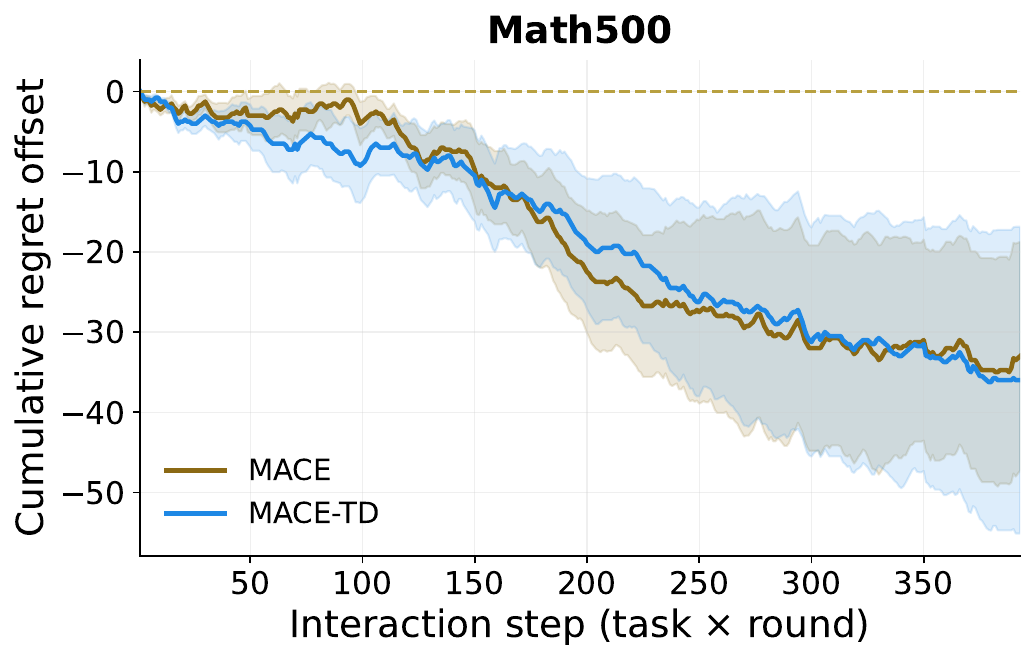}
    \end{subfigure}
    \caption{\textbf{Cumulative regret offset across tasks and phases: MACE-TD vs MACE.} The average cumulative reward offset~(Baseline -- Random) across interaction steps are shown~(lower the better). The horizontal dashed line at Offset $=0$ is the Random baseline, and the shaded areas indicate the standard error across participating agents.}
    \label{fig:macetd_regret}
\end{figure}

\paragraph{Cumulative regrets and round-wise task performance.}
Figure~\ref{fig:macetd_regret} compares MACE-TD against the original MACE formulation in terms of cumulative regret offset and round-wise downstream performance. 
Overall, MACE-TD does not exhibit a clear advantage over MACE in cumulative regret. 
On HotpotQA and Math500, the regret trajectories of the two methods are broadly comparable, while on GPQA, MACE-TD underperforms MACE by accumulating noticeably higher regret, suggesting that temporal-difference updates may introduce instability when the reward signal is sparse or noisy. 
Interestingly, however, this disadvantage in regret does not translate into worse final task performance. 
In Figure~\ref{fig:roundwise_perf}, in GPQA, MACE-TD reaches round-wise performance that is similar to, or even slightly better than, MACE after multiple interaction rounds. 
This suggests that while MACE-TD may be less efficient in minimizing cumulative regret during peer selection, its sequential value estimation can still support effective multi-round information aggregation, leading to comparable downstream performance after sufficient interaction.
On the other hand, HotpotQA presents a contrasting case: although the cumulative regret trend is comparable to MACE, the exploration-phase F1 score remains noticeably lower.

\begin{figure}[h!]
    \centering
    \begin{subfigure}[h!]{0.32\textwidth}
        \centering
        \includegraphics[width=\linewidth]{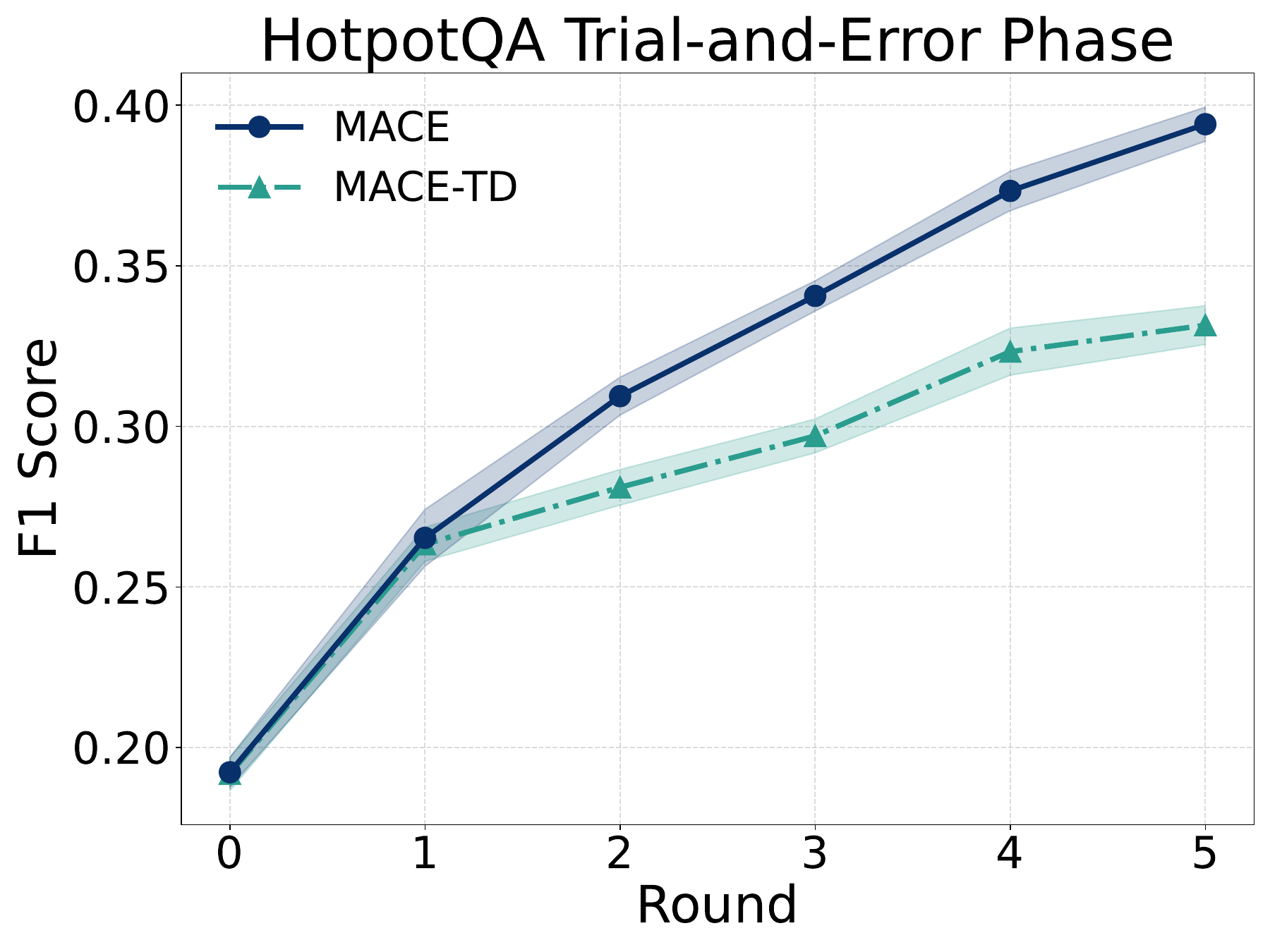}
    \end{subfigure}
    \hfill
    \begin{subfigure}[h!]{0.32\textwidth}
        \centering
        \includegraphics[width=\linewidth]{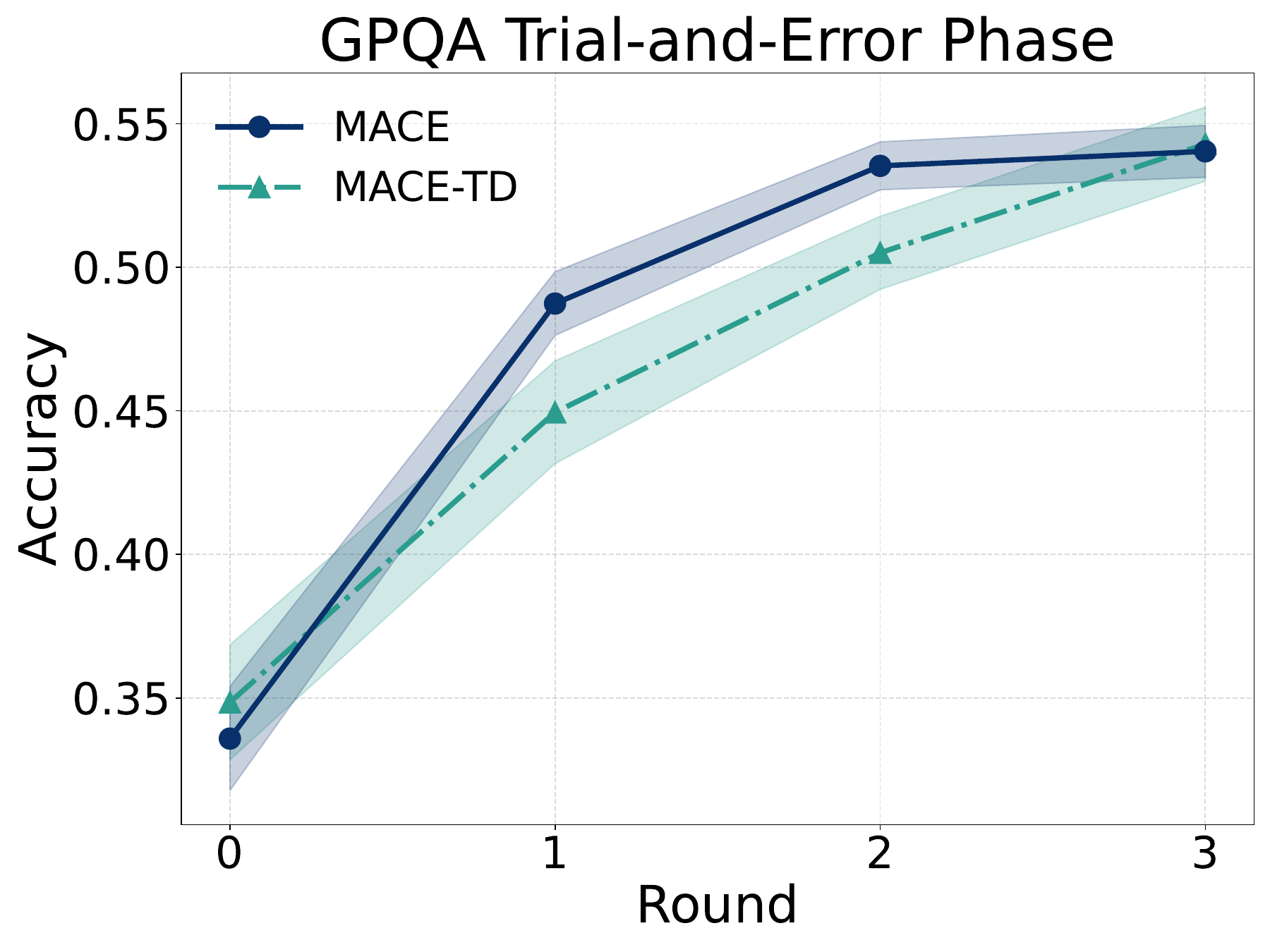}
    \end{subfigure}
    \hfill
    \begin{subfigure}[h!]{0.32\textwidth}
        \centering
        \includegraphics[width=\linewidth]{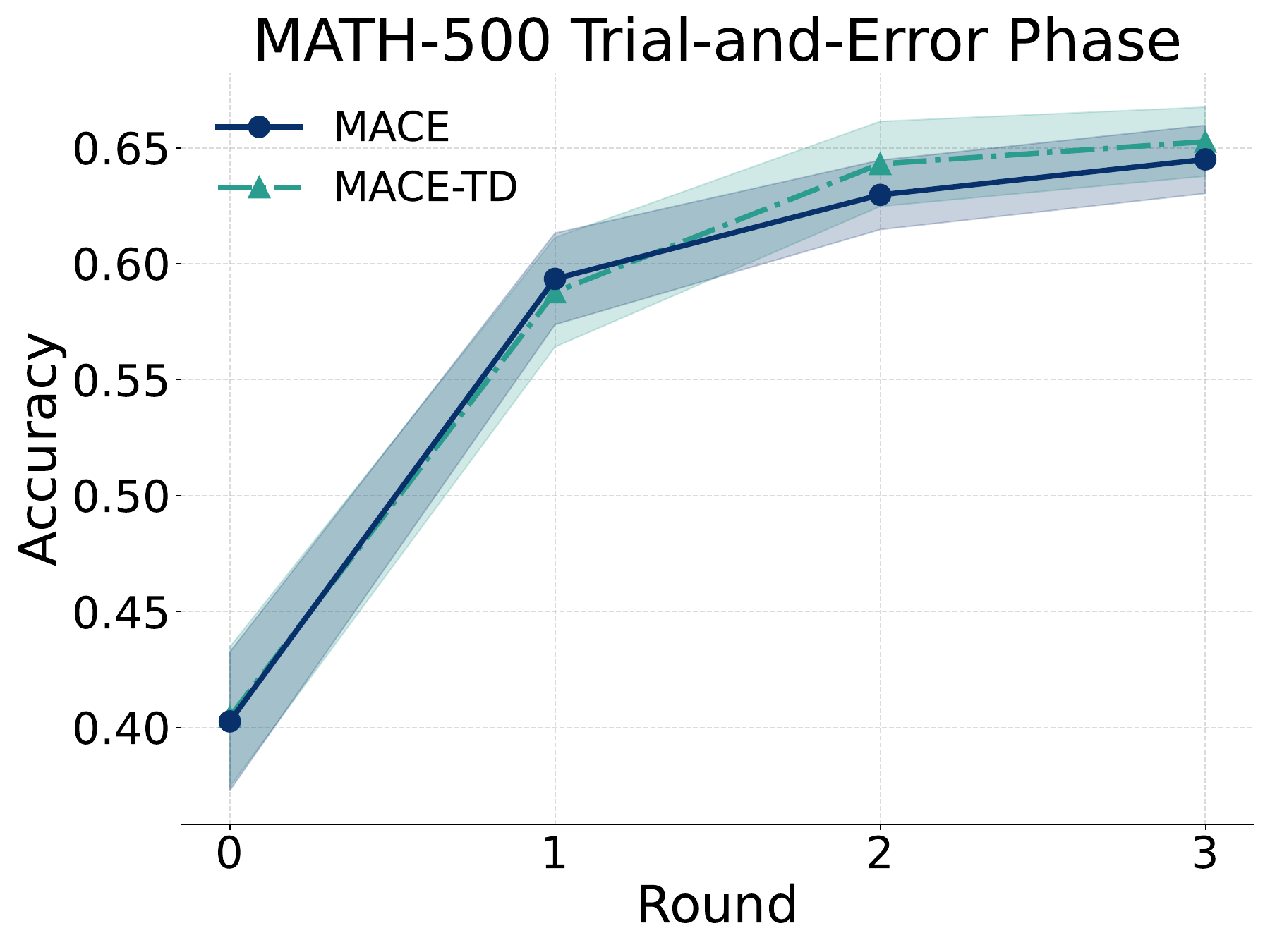}
    \end{subfigure}
    \hfill
    \begin{subfigure}[h!]{0.32\textwidth}
        \centering
        \includegraphics[width=\linewidth]{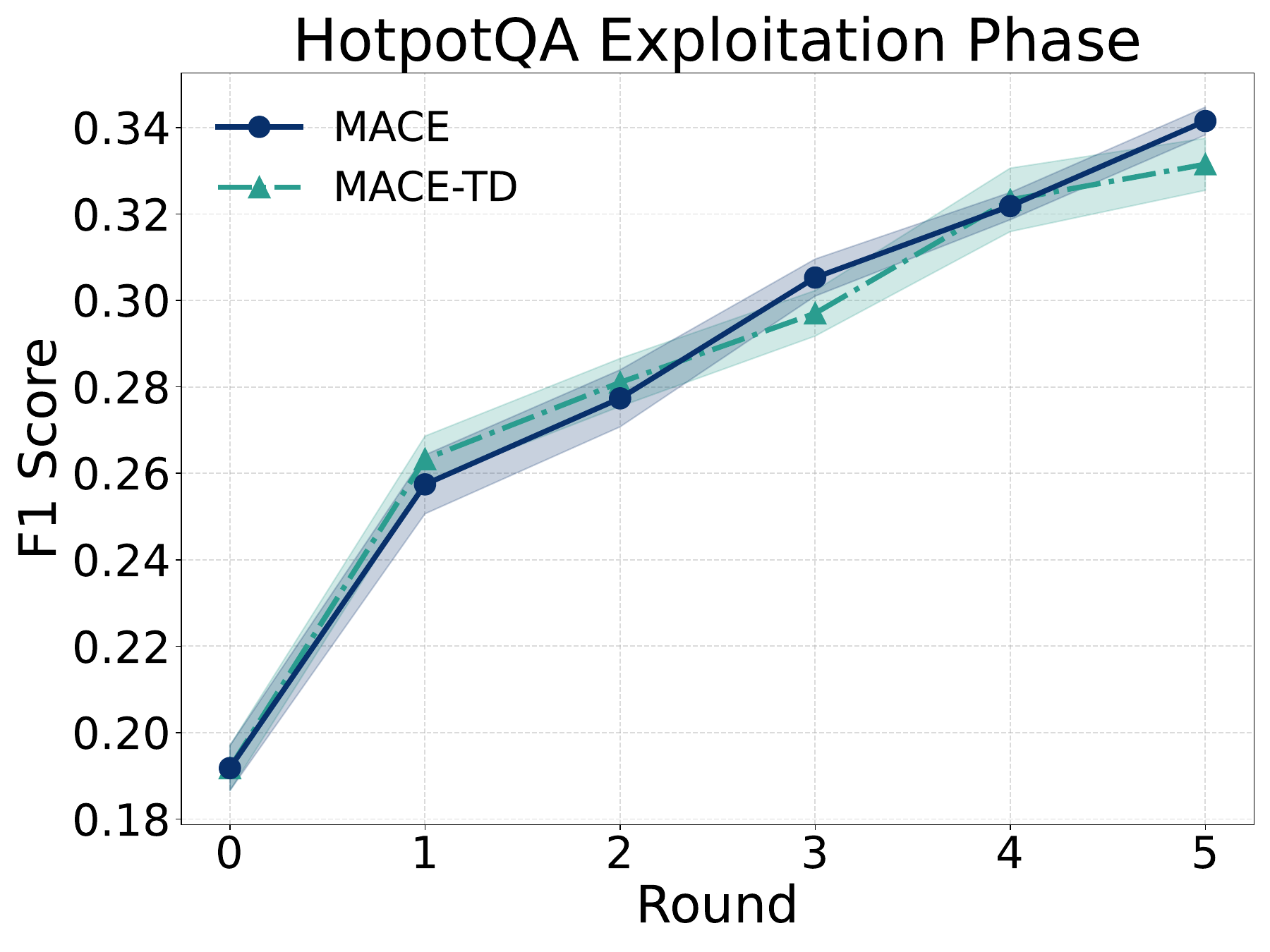}
    \end{subfigure}
    \hfill
    \begin{subfigure}[h!]{0.32\textwidth}
        \centering
        \includegraphics[width=\linewidth]{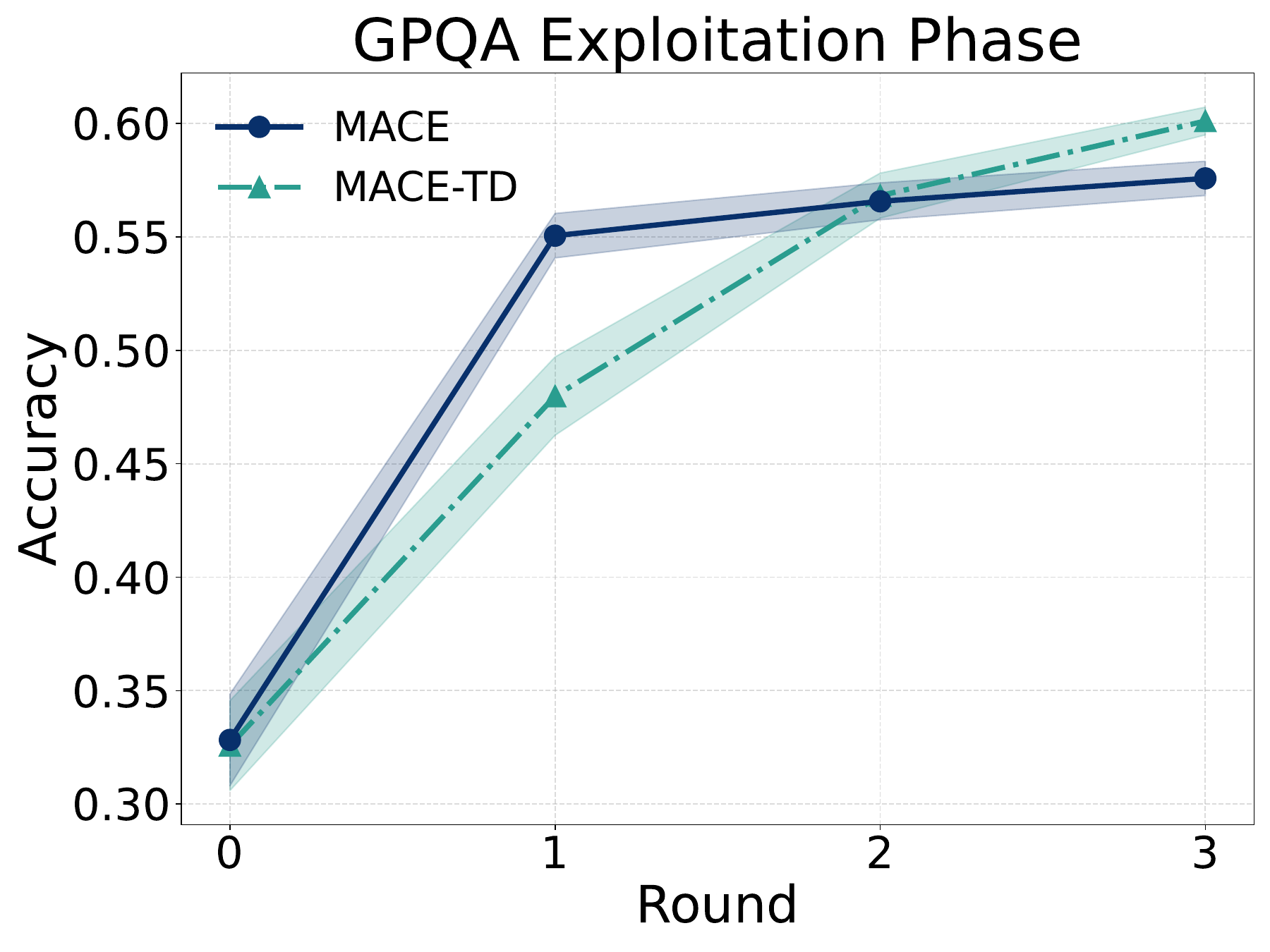}
    \end{subfigure}
    \hfill
    \begin{subfigure}[h!]{0.32\textwidth}
        \centering
        \includegraphics[width=\linewidth]{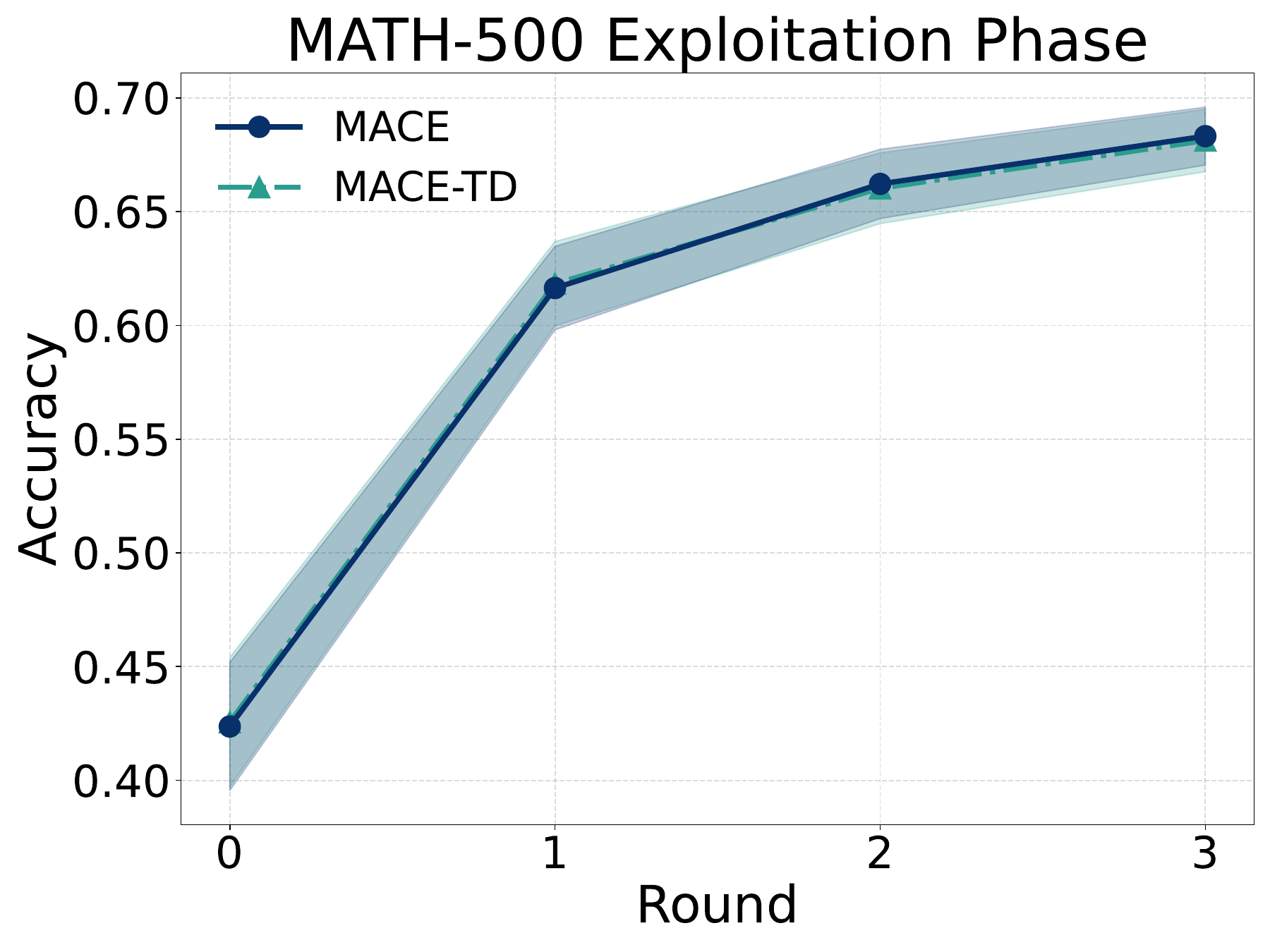}
    \end{subfigure}
    \caption{\textbf{Comparison of exploration and exploitation performance for each interaction round: MACE-TD vs MACE.} Shaded are the standard errors across participating agents; standard error * 0.2 are shown for Math500 and GPQA to avoid visual clutter. Also, note that the variance in the 0-th round is induced by GPT-5. }
    \label{fig:roundwise_perf}
\end{figure}

\paragraph{Potential Emergence of Cooperative Behaviors.}
Although MACE-TD does not yield a consistent advantage over MACE in cumulative regret or final task performance, it reveals an interesting qualitative pattern in how agents distribute their interactions. 
In particular, MACE-TD appears to encourage agents to specialize over different subsets of peers, which can expand the system-level ``receptive field'' of information exchange. 
This behavior is desirable from a cooperative perspective: if all agents repeatedly query the same peer, then only that peer's information is propagated through the system, limiting the diversity of evidence available to the group and making the problem more difficult to solve through interaction. 
In contrast, when different agents query different peers, information from a broader portion of the agent population can be indirectly incorporated through subsequent interactions.

\begin{figure}[h!]
    \centering
    \includegraphics[width=0.7\linewidth]{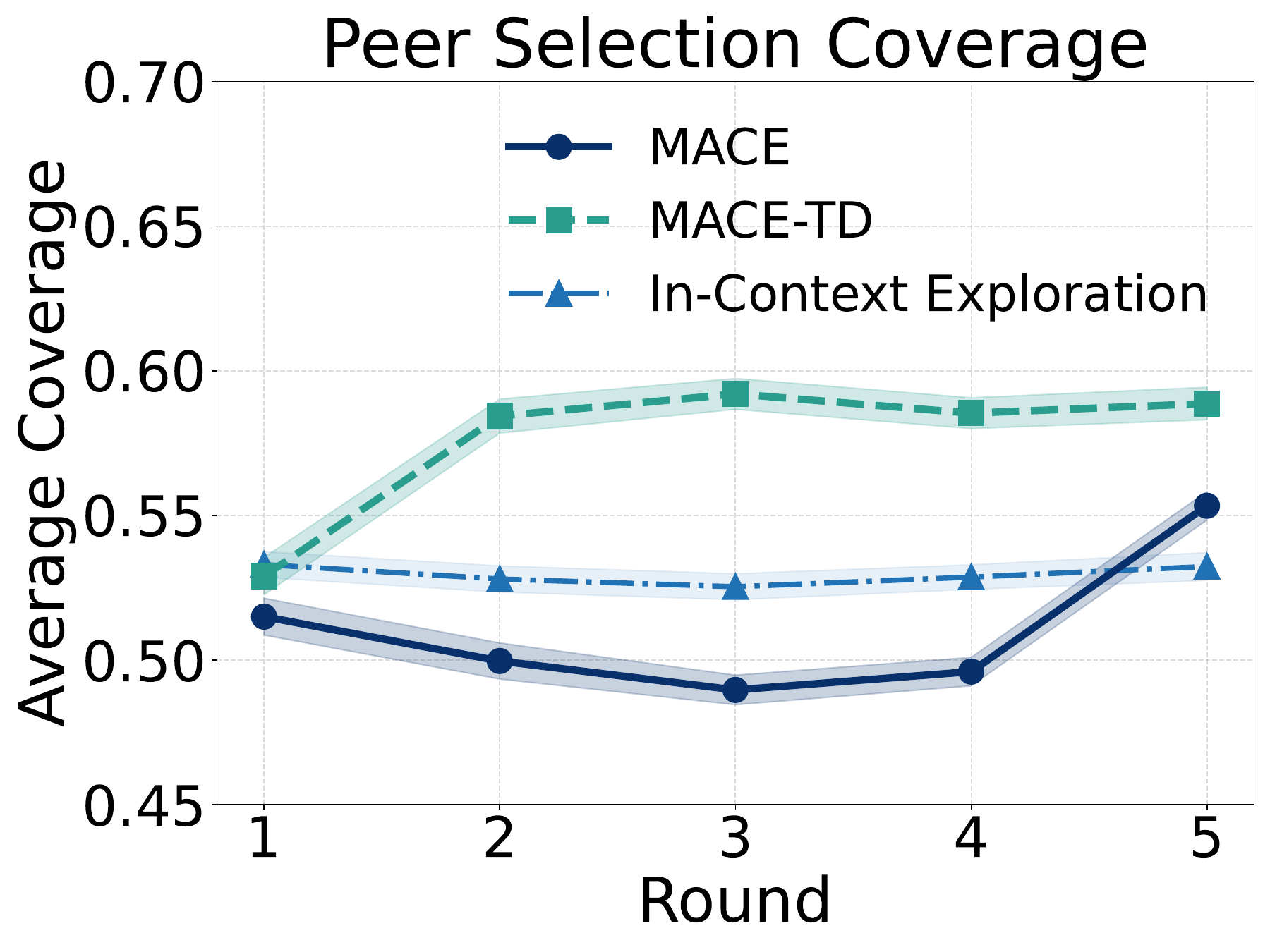}
    \caption{\textbf{The trend of peer coverage throughout interaction rounds}}
    \label{fig:coop}
\end{figure}

To quantify this effect, we measure \emph{peer selection coverage}, defined as the fraction of peers that are selected by at least one agent in a given interaction round. 
We average this quantity across agents and samples, and report its round-wise trend in Figure~\ref{fig:coop}. 
Interestingly, MACE-TD consistently achieves higher peer selection coverage than In-Context Exploration, whereas standard MACE does not exhibit the same trend.
This suggests that temporal-difference learning may induce a more system-aware interaction pattern: by optimizing for delayed rewards rather than only immediate gains, agents may learn that diversifying peer coverage improves downstream information flow. 
While this observation does not by itself establish the emergence of cooperation, it provides preliminary evidence that sequential credit assignment in MACE-TD can encourage interaction structures that are more favorable for collective information sharing.

{\color{black}
\subsection{Explicit Exploration and Relational Features}
\label{sec:hotpotqa-ucb1}

In this section, we examine how explicit exploration and relational features contribute to MACE's performance in the Contextual Diversity setup. 
We compare MACE with UCB1 and Greedy MACE under the same agent configuration, reward feedback, and interaction budget.
UCB1 provides a context-free reference: it first queries unvisited peers and then selects the peer with the highest empirical mean reward plus a visit-count bonus. 
Greedy MACE isolates the effect of explicit exploration by setting the uncertainty bonus to zero during the trial-and-error phase (i.e., $\alpha=0$).

\begin{table}[h!]
\centering
\caption{Final-round exact match (EM) and F1 on HotpotQA in the
trial-and-error and exploitation phases.}
\label{tab:hotpotqa-ucb1}
\begin{tabular}{lcccc}
\toprule
& \multicolumn{2}{c}{Trial-and-Error}
& \multicolumn{2}{c}{Exploitation} \\
\cmidrule(lr){2-3}\cmidrule(lr){4-5}
Method & EM & F1 & EM & F1 \\
\midrule
UCB1                   & 29.33 & 36.18 & 22.70 & 29.18 \\
Greedy MACE            & 21.00 & 26.61 & 19.27 & 24.06 \\
\textsc{Full MACE}          & \textbf{31.33} & \textbf{39.41}
                       & \textbf{26.87} & \textbf{34.15} \\
\bottomrule
\end{tabular}
\end{table}

Table~\ref{tab:hotpotqa-ucb1} shows that removing the exploration
bonus yields the larger performance drop in both phases. 
On held-out questions, Greedy MACE trails full MACE by 7.60 EM points
and 10.09 F1 points, while UCB1 trails by 4.17 EM points and 4.97
F1 points. 
UCB1 also outperforms Greedy MACE in both phases, suggesting that explicit exploration contributes more in this setting than using relational features alone. 
The comparison with Greedy MACE isolates the effect of the exploration bonus, whereas UCB1 also differs in how it estimates rewards and uncertainty.

During exploitation, freezing UCB1's statistics fixes each agent's
peer ranking. 
MACE freezes its learned parameters and recomputes relational features as agents revise their answers, allowing its peer rankings to adapt across questions and rounds.}

\subsection{Feature Analyses}
\label{apdx:feature_importance}
In this section, we analyze the feature weights learned by MACE across agents and experimental settings. 

\subsubsection{Natural-Language versus Numerical Exploration Signals}
\label{apdx:natural-language-exploration}

We investigate whether the performance of In-Context Exploration depends on how peer information is represented. The original baseline receives the same numerical evidence used by MACE, enabling a controlled comparison between LLM-based and algorithmic peer selection. However, numerical features may be a less natural interface for LLMs than textual descriptions.

To examine this possibility, we evaluate a \emph{Natural-Language In-Context Exploration} baseline that converts numerical statistics into qualitative descriptions of each peer's query frequency, response similarity to the selector, response distinctiveness relative to the agent population, and historical helpfulness. The selector is explicitly instructed to balance exploring uncertain peers against consulting historically helpful peers.

As shown in Table~\ref{tab:natural-language-exploration}, the numerical and natural-language variants achieve similar performance in both phases on the HotpotQA benchmark, while MACE consistently outperforms both. These findings suggest that the gap is not primarily attributable to the numerical presentation of evidence. Instead, they support the value of an explicit exploration mechanism beyond providing peer information and prompting the LLM to balance exploration and exploitation.

\begin{table}[h!]
    \centering
    \caption{\textbf{Effect of exploration-signal representation.}
    In-Context Exploration performs similarly with numerical features
    and qualitative natural-language descriptions. MACE achieves the
    highest F1 and exact-match scores in both phases.}
    \label{tab:natural-language-exploration}
    \renewcommand{\arraystretch}{1.15}
    \begin{tabular}{lcccc}
        \toprule
        & \multicolumn{2}{c}{\textbf{Trial-and-Error}}
        & \multicolumn{2}{c}{\textbf{Exploitation}} \\
        \cmidrule(lr){2-3} \cmidrule(lr){4-5}
        \textbf{Method}
        & \textbf{F1} & \textbf{EM}
        & \textbf{F1} & \textbf{EM} \\
        \midrule
        In-Context (Numerical)
        & 0.2810 & 0.2257 & 0.3001 & 0.2433 \\
        In-Context (Natural Language)
        & 0.2773 & 0.2193 & 0.3016 & 0.2427 \\
        MACE
        & \textbf{0.3941} & \textbf{0.3133}
        & \textbf{0.3415} & \textbf{0.2687} \\
        \bottomrule
    \end{tabular}
\end{table}

\subsubsection{Feature Importance}
\begin{figure}[h!]
    \centering
    \includegraphics[width=0.8\linewidth]{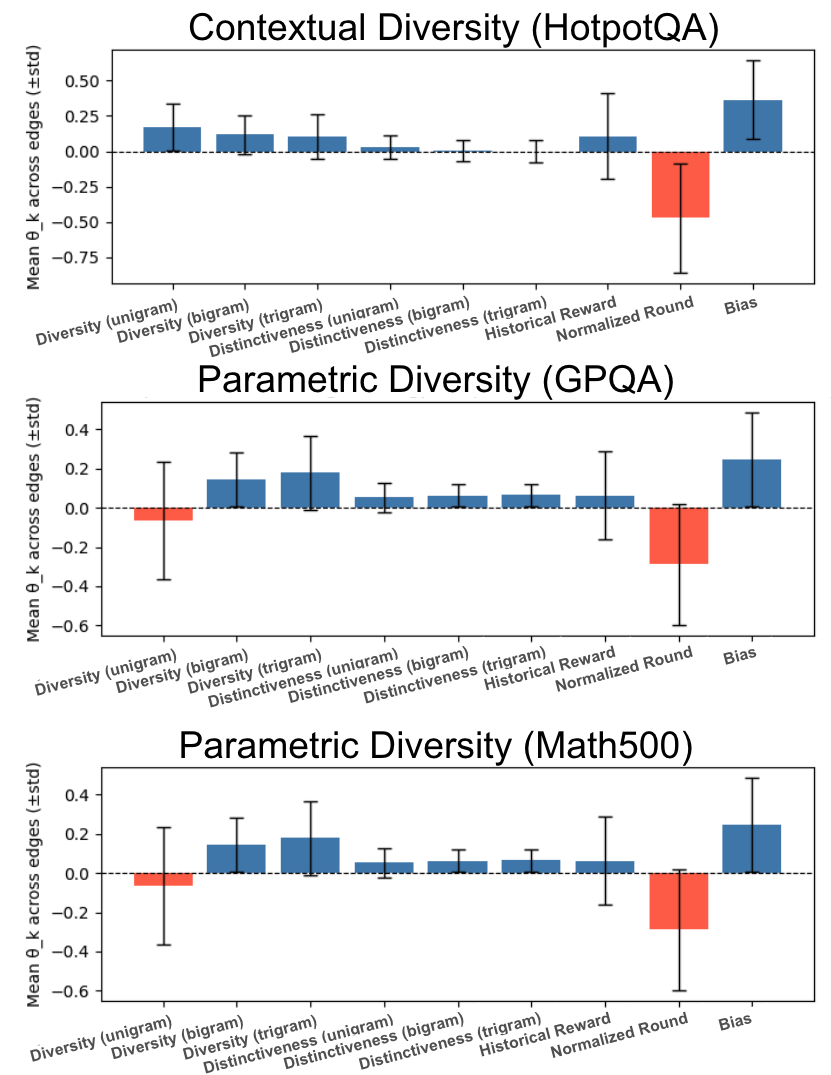}
    \caption{{Learned feature weights averaged across agents.}}
    \label{fig:feat_imp}
\end{figure}

Figure~\ref{fig:feat_imp} reports the mean learned weights $\boldsymbol{\theta}_{i,a}$ for each feature, averaged across agent pairs $(i, a)$. 
Overall, the n-gram-based diversity features largely received positive weights, suggesting that MACE learns to favor peers whose responses differ from the agent's own response. 
This is expected, since disagreement can signal complementary information that may be useful for exploration.

Interestingly, the learned weights are broadly similar across contextual and parametric diversity settings. 
The main exception is that distinctiveness features are close to zero in the contextual diversity setting. This is intuitive: in HotpotQA, agents differ primarily by the passage they observe, rather than by their intrinsic reasoning ability. Thus, merely producing a distinctive response does not necessarily indicate that an agent is reliable or informative.
It may simply reflect access to a different, possibly irrelevant, context. 
In contrast, under parametric diversity, distinctiveness is more useful because differences in responses can reflect differences in model capability or reasoning behavior.

\subsubsection{Feature-Group Ablations}
\label{apdx:feature-ablations}

MACE uses a simple relational representation comprising intuitive signals that may inform peer selection. These features were not extensively curated or optimized; our goal is to examine whether explicit exploration can be effective with readily available peer information. To assess the contribution of each feature group, we conduct leave-one-group-out ablations on HotpotQA.

Table~\ref{tab:feature-ablations} reports performance during the Trial-and-Error phase. Full MACE achieves the highest F1 score and ties for the highest exact-match score. Removing response diversity produces the largest decrease in both metrics, reducing F1 from $0.3941$ to $0.3736$ and exact match from $0.3133$ to $0.2963$. Removing historical performance or interaction round also reduces both metrics, whereas removing peer distinctiveness has a smaller effect.

These results suggest that response diversity provides particularly useful information for learning the peer-selection policy. 
Although this observable signal is distinct from the capability diversity $\delta$ considered in our theoretical analysis, its contribution is qualitatively consistent with the importance of distinguishing peers when guiding exploration.

\begin{table}[t]
    \centering
    \caption{\textbf{Feature-group ablations on HotpotQA.}
    Performance during the Trial-and-Error phase when each feature
    group is individually removed from MACE.}
    \label{tab:feature-ablations}
    \renewcommand{\arraystretch}{1.15}
    \begin{tabular}{lcc}
        \toprule
        \textbf{Method} & \textbf{F1 Score} & \textbf{Exact Match} \\
        \midrule
        Full MACE
        & \textbf{0.3941} & \textbf{0.3133} \\
        \quad w/o Peer distinctiveness
        & 0.3900 & \textbf{0.3133} \\
        \quad w/o Interaction round
        & 0.3801 & 0.2997 \\
        \quad w/o Historical performance
        & 0.3768 & 0.3047 \\
        \quad w/o Response diversity
        & 0.3736 & 0.2963 \\
        \bottomrule
    \end{tabular}
\end{table}

\section{Theoretical Assumptions and Proofs}
\label{apdx:proofs}

\subsection{Assumptions}
\label{app:theory}

Here, we state the assumptions used in the theoretical analysis of MACE. 
For a fixed agent $i$, let $\mathbf{x}_{i,a,t} \in \mathbb{R}^d$ denote the contextual feature vector associated with querying peer $a$ at round $t$. 
MACE estimates the utility of each candidate peer using a linear reward model:
\begin{equation}
    \mathbb{E}\!\left[r_{i,t} \mid \mathcal{F}_{t-1}, a_{i,t}=a\right]
    =
    \mu_{i,a,t}
    =
    \mathbf{x}_{i,a,t}^{\top}\boldsymbol{\theta}_{i,a},
    \label{eq:lin_reward_theory}
\end{equation}
where $\boldsymbol{\theta}_{i,a}\in\mathbb{R}^d$ is the unknown parameter associated with agent $i$'s interaction with peer $a$, and $\mathcal{F}_{t-1}$ denotes the history observed before round $t$. 
Let
\begin{equation}
    a_{i,t}^* \in \underset{a \in \mathcal{N}}{\arg\max} \;\; \mu_{i,a,t}
\end{equation}
denote the optimal peer at round $t$. 
The cumulative regret of agent $i$ is
\begin{equation}
    \mathrm{Regret}_i
    =
    \sum_{t=1}^{T}
    \left(
        \mu_{i,a_{i,t}^*,t}
        -
        \mu_{i,a_{i,t},t}
    \right).
\end{equation}

We use standard assumptions for contextual linear bandits, along with our several mild assumptions.

\paragraph{Assumption 1. Bounded features and parameters.}
For all agents $i$, peers $a$, and rounds $t$,
\begin{equation}
    \|\mathbf{x}_{i,a,t}\|_2 \leq 1,
    \qquad
    \|\boldsymbol{\theta}_{i,a}\|_2 \leq S .
\end{equation}

\paragraph{Assumption 2. Sub-Gaussian reward noise.}
The observed reward satisfies
\begin{equation}
    r_{i,t}
    =
    \mathbf{x}_{i,a_{i,t},t}^{\top}\boldsymbol{\theta}_{i,a_{i,t}}
    +
    \eta_{i,t},
\end{equation}
where the noise $\eta_{i,t}$ is conditionally $R$-sub-Gaussian with respect to $\mathcal{F}_{t-1}$:
\begin{equation}
    \mathbb{E}
    \left[
        \exp(\lambda \eta_{i,t})
        \mid
        \mathcal{F}_{t-1}
    \right]
    \leq
    \exp\!\left(\frac{\lambda^2 R^2}{2}\right),
    \qquad
    \forall \lambda \in \mathbb{R}.
\end{equation}

\subsection{Proof of Theorem~\ref{thm:mace_regret}}
\label{app:proof_mace}
Fix a selector $i$. Assume that the contextual features are
available before peer selection and that
\[
    \mu_{i,a,t}
    =
    \mathbb{E}[r_{i,t}\mid\mathcal F_{t-1},a_{i,t}=a]
    =
    x_{i,a,t}^{\top}\theta_{i,a},
\]
where $\theta_{i,a}$ is fixed over time,
$\|x_{i,a,t}\|_2\leq1$, and $\|\theta_{i,a}\|_2\leq S$.
Here, $\mathcal F_{t-1}$ contains the information available
before selection, including the current contextual features.
The reward noise is conditionally $\sigma$-sub-Gaussian.

For each peer $a$, define
\[
\begin{aligned}
    A_{i,a}(t)
    &=
    \lambda I
    +
    \sum_{\substack{s<t\\a_{i,s}=a}}
    x_{i,a,s}x_{i,a,s}^{\top},\\
    b_{i,a}(t)
    &=
    \sum_{\substack{s<t\\a_{i,s}=a}}
    r_{i,s}x_{i,a,s},\\
    \widehat{\theta}_{i,a}(t)
    &=
    A_{i,a}(t)^{-1}b_{i,a}(t).
\end{aligned}
\]
We choose
\begin{equation}
    \alpha_T
    =
    \sqrt{\lambda}\,S
    +
    \sigma\sqrt{
        d\log\!\left(1+\frac{T}{d\lambda}\right)
        +2\log(NT)
    }.
    \label{eq:confidence_radius}
\end{equation}

\paragraph{Step 1: Simultaneous confidence bounds.}
For each peer, the self-normalized concentration inequality
for linear bandits~\citep{abbasi2011improved} gives, with
failure probability at most $1/(NT)$, simultaneously
over $t\leq T$,
\[
    \left\|
        \widehat{\theta}_{i,a}(t)-\theta_{i,a}
    \right\|_{A_{i,a}(t)}
    \leq
    \sqrt{\lambda}\,S
    +
    \sigma\sqrt{
        \log\frac{\det A_{i,a}(t)}{\det(\lambda I)}
        +2\log(NT)
    }.
\]
The feature norm bound and the determinant--trace
inequality imply
\[
    \log\frac{\det A_{i,a}(t)}{\det(\lambda I)}
    \leq
    d\log\!\left(1+\frac{T}{d\lambda}\right).
\]
A union bound over the $N$ peers therefore yields an
event $\mathcal G$ with $\Pr(\mathcal G)\geq1-1/T$
on which, for every peer and round,
\begin{equation}
    \left|
        x_{i,a,t}^{\top}
        \bigl(\widehat{\theta}_{i,a}(t)-\theta_{i,a}\bigr)
    \right|
    \leq
    \alpha_T
    \|x_{i,a,t}\|_{A_{i,a}(t)^{-1}}.
    \label{eq:mace_confidence}
\end{equation}

\paragraph{Step 2: Instantaneous regret.}
Let $a_{i,t}^{\star}\in\arg\max_a\mu_{i,a,t}$ and define
\[
    U_{i,a,t}
    =
    x_{i,a,t}^{\top}\widehat{\theta}_{i,a}(t)
    +
    \alpha_T\|x_{i,a,t}\|_{A_{i,a}(t)^{-1}}.
\]
MACE selects $a_{i,t}\in\arg\max_a U_{i,a,t}$.
On $\mathcal G$, optimism and the selection rule give
\[
    \mu_{i,a_{i,t}^{\star},t}
    \leq U_{i,a_{i,t}^{\star},t}
    \leq U_{i,a_{i,t},t}.
\]
Applying \eqref{eq:mace_confidence} to the
selected peer yields
\begin{equation}
    \mu_{i,a_{i,t}^{\star},t}-\mu_{i,a_{i,t},t}
    \leq
    2\alpha_T
    \|x_{i,a_{i,t},t}\|_{A_{i,a_{i,t}}(t)^{-1}}.
    \label{eq:mace_instantaneous}
\end{equation}

\paragraph{Step 3: Cumulative regret on the confidence event.}
Write
\[
    w_t
    =
    \|x_{i,a_{i,t},t}\|_{A_{i,a_{i,t}}(t)^{-1}},
\]
and let $T_a$ denote the number of selections of peer $a$
through round $T$. Since $\lambda\geq1$ and the features
have norm at most one, $w_t^2\leq1$.

The matrix determinant lemma and the inequality
$u\leq2\log(1+u)$ for $u\in[0,1]$ imply
\[
\begin{aligned}
    \sum_{t=1}^{T}w_t^2
    &\leq
    2\sum_{a=1}^{N}
    \log\frac{\det A_{i,a}(T+1)}{\det(\lambda I)}\\
    &\leq
    2d\sum_{a=1}^{N}
    \log\!\left(1+\frac{T_a}{d\lambda}\right)\\
    &\leq
    2Nd\log\!\left(1+\frac{T}{d\lambda}\right).
\end{aligned}
\]
Summing \eqref{eq:mace_instantaneous} and applying
Cauchy--Schwarz gives, on $\mathcal G$,
\[
\begin{aligned}
    \mathrm{Regret}_i^{\mathrm{MACE}}
    &\leq
    2\alpha_T\sum_{t=1}^{T}w_t\\
    &\leq
    2\alpha_T\sqrt{T\sum_{t=1}^{T}w_t^2}\\
    &\leq
    2\alpha_T
    \sqrt{
        2TNd\log\!\left(1+\frac{T}{d\lambda}\right)
    }
    =:B_T.
\end{aligned}
\]

\paragraph{Step 4: Expected regret.}
Rewards lie in $[-1/2,1]$, as does the reward
$r_{i,t}=s_i^{(t)}-\tfrac12s_i^{(t-1)}$ used in our
experiments. Hence, conditional expected rewards also
lie in this interval, and cumulative pseudo-regret
is bounded by $3T/2$ on every trajectory.
Splitting the expectation according to $\mathcal G$,
\[
\begin{aligned}
    \mathbb{E}\!\left[
        \mathrm{Regret}_i^{\mathrm{MACE}}
    \right]
    &\leq
    B_T\Pr(\mathcal G)
    +
    \frac{3T}{2}\Pr(\mathcal G^c)\\
    &\leq
    2\alpha_T
    \sqrt{
        2TNd\log\!\left(1+\frac{T}{d\lambda}\right)
    }
    +\frac{3}{2}.
\end{aligned}
\]
This establishes the claimed bound. For fixed
$N,d,\lambda,S,\sigma$, the choice in
\eqref{eq:confidence_radius} satisfies
$\alpha_T=O(\sqrt{\log T})$, giving
$O(\sqrt{T}\log T)$ expected regret.
$\hfill\square$

\subsection{Proof of Theorem~\ref{thm:greedy_regret}}
\label{app:proof_greedy}

Fix $N\geq2$. Consider $N$ peers with independent Bernoulli
rewards and means
\[
    \mu_1=\frac{3}{4},
    \qquad
    \mu_a=\frac{1}{4},
    \quad a=2,\ldots,N.
\]
Peer $1$ is optimal, and every other peer has the same
suboptimality gap $\Delta=1/2$.

Set $c_a=\mu_a$. The mean capability is
\[
    \bar c
    =
    \frac{1}{4}+\frac{1}{2N}.
\]
Consequently, Definition~\ref{def:diversity} gives
\[
\begin{aligned}
    \delta
    &=
    \frac{1}{N}
    \left(
        \frac{N-1}{2N}
        +(N-1)\frac{1}{2N}
    \right)\\
    &=
    \frac{N-1}{N^2}.
\end{aligned}
\]

Let $X_{a,1}$ denote the reward from the initialization
query to peer $a$. Consider the event
\[
    \mathcal E
    =
    \left\{
        X_{1,1}=0,\;
        \max_{2\leq a\leq N}X_{a,1}=1
    \right\}.
\]
By independence,
\[
    \Pr(\mathcal E)
    =
    \frac{1}{4}
    \left[
        1-\left(\frac{3}{4}\right)^{N-1}
    \right]
    \geq
    \frac{1}{16}.
\]

On $\mathcal E$, the optimal peer's empirical mean is
zero, while at least one inferior peer has an initial
reward of one. Fix such an inferior peer $b$. After any
finite number $n\geq1$ of queries to this peer,
\[
    \widehat{\mu}_{i,b}
    =
    \frac{1+\sum_{k=2}^{n}X_{b,k}}{n}
    \geq
    \frac{1}{n}
    >0.
\]
Thus, at every subsequent round, at least one inferior
peer has a strictly positive empirical mean. As long
as peer $1$ is not queried again, its empirical mean
remains zero. By induction, greedy selection never
returns to peer $1$.

Therefore, on $\mathcal E$, the optimal peer is queried
exactly once during initialization, and all other
$T-1$ queries select inferior peers. Hence,
\[
    \mathrm{Regret}_i^{\mathrm{non\text{-}exploring}}
    =
    \Delta(T-1)
    =
    \frac{T-1}{2}.
\]
Using nonnegativity of pseudo-regret,
\[
    \mathbb{E}\!\left[
        \mathrm{Regret}_i^{\mathrm{non\text{-}exploring}}
    \right]
    \geq
    \Pr(\mathcal E)\frac{T-1}{2}
    \geq
    \frac{T-1}{32}.
\]
Finally,
\[
    \delta=\frac{N-1}{N^2}\leq\frac{1}{4},
\]
so
\[
    \frac{T-1}{32}
    \geq
    \frac{\delta}{8}(T-1).
\]
This proves the theorem.
$\hfill\square$

\subsection{Discussion}

Theorem~\ref{thm:mace_regret} shows that MACE achieves sublinear regret in the number of interaction rounds, scaling as
\begin{equation}
    \widetilde{O}\!\left(\sqrt{T \log T}\right).
\end{equation}
Thus, the average regret per round vanishes as $T$ grows. 
In contrast, Theorem~\ref{thm:greedy_regret} shows that a greedy non-exploring policy can suffer regret that grows linearly in $T$ once it prematurely commits to a suboptimal peer.

The role of capability diversity is captured by the factor $\delta$. 
When $\delta \approx 0$, peers are nearly interchangeable, and the cost of selecting the wrong peer is small. 
In this regime, explicit exploration may provide only limited benefit. 
When $\delta$ is large, however, agents are specialized, and peer selection becomes consequential. 
The regret gap in Corollary~\ref{cor:gap} therefore grows as
\begin{equation}
    \Omega(\delta T)
    -
    \widetilde{O}(\sqrt{T \log T}),
\end{equation}
showing that the value of exploration increases directly with the diversity of the agent pool.
\section{Appendix for Related Works}
\label{apdx:related_works}

Here, we provide an extended version of the Related Works section of the manuscript. 
We give a broad overview of the landscape of multi-agent LLM systems, and discuss approaches that specifically target autonomous multi-agent systems~(Section~\ref{apdx:relwork1}).
We further discuss works that conducted in-depth experiments and analysis on the exploratory ability of single LLM agents~(Section~\ref{apdx:relwork2}).

\subsection{Autonomous Multi-Agent LLM Systems}
\label{apdx:relwork1}

\paragraph{Autonomous multi-agent Overview.}
Recent autonomous LLM systems have shifted from isolated, single-agent prompting toward coordinated groups of agents that divide roles, exchange information, and jointly plan or act. 
Recent overviews~\citep{guo2024large,li2024survey,ferrag2025llm,tran2025multi} collectively organize the space around agent profiling, communication structure, collaboration protocol, and system evolution.

\paragraph{Frameworks for multi-agent coordination.}
A major line of work builds general-purpose infrastructures for multi-agent coordination. 
CAMEL~\citep{li2023camel} introduced role-playing communicative agents for autonomous cooperation, while AutoGen~\citep{wu2024autogen} cast multi-agent interaction as programmable conversation among agents, humans, and tools. 
ChatDev~\citep{qian2024chatdev}, MetaGPT~\citep{hong2023metagpt}, and AgentScope~\citep{gao2024agentscope} further instantiated multi-agent orchestration in which specialized agents collaborate through structured workflows.
Beyond these, DyLAN~\citep{liu2024dynamic} studied learnable designs for dynamic agent networks, while recent work on self-resource allocation~\citep{amayuelas2025self} examines how LLM-based planners and orchestrators allocate tasks across worker agents under cost, efficiency, and capability constraints.
AgentVerse~\citep{chen2023agentverse} and MegaAgent~\citep{wang2025megaagent} further proposed 
collaborative groups that are autonomously designed by LLM agents.
In parallel, benchmark-oriented work has begun to isolate the coordination abilities required by such systems: LLM-Coordination~\citep{agashe2025llm} evaluates LLM agents in pure coordination games and shows that, while LLMs can coordinate well when decisions depend on observable environmental state, they remain limited in theory-of-mind reasoning and joint planning.
Self-Consistency~\citep{wang2023self} and ModeX~\citep{choi2026modex} instead devise multi-agent response aggregation methods based on voting and consensus.
Moreover, Multi-Agent Debate~(MAD) frameworks have been commonly explored as a hallmark of collective intelligence~\citep{du2024improving,chan2024chateval,liu2024groupdebate,li2024improving,smit2023should,liang2024encouraging,becker2025mallm}, while several recent works report failure modes and limitations of the MAD protocol~\citep{smit2023should,huang2023large,estornell2024multi,liu2024groupdebate,cemri2025multi,choi2025debate,kaesberg2025voting,choi2026identity}.
Subsequent works have also proposed methods to enhance self-improvement through group-evolution~\citep{weng2026group}.
Together, these works establish the engineering foundations of autonomous multi-agent LLM systems, although most still rely on externally specified roles, orchestration rules, or known worker capabilities rather than learning whom to interact with and how to coordinate from experience.

\paragraph{Granting Autonomy to Multi-Agent LLM Systems.}
Several works have focused on the autonomy aspect of multi-agent LLM systems.
A work on iAgents~\citep{liu2024autonomous} discussed human-computer interaction paradigms for autonomous communication between agents, while a body of works studied approaches to autonomously design and assign roles to LLM agents.
Specifically, HALO~\citep{hou2025halo} introduced a system of hierarchical orchestration with high-level planning, role design, and workflow search.
Puppeteer~\citep{dang2025multi} proposes a reinforcement learning approach to train a central agent that orchestrates sub-agents dynamically and autonomously.
MegaAgent~\citep{wang2025megaagent} and AgentVerse~\citep{chen2023agentverse}, on the other hand, autonomously design the role specifications in recruiting sub-agents to solve the given task.
Other works enforced autonomy in solving domain-specific tasks~\citep{sanwal2024autonomous}, while \citep{handler2023balancing} attempted to assess the current state and taxonomy of autonomy in multi-agent LLM architectures.

\subsection{LLM Exploration Capabilities}
\label{apdx:relwork2}

\paragraph{LLM exploration capabilities.}
A recent line of work asks whether a \emph{single} LLM can explore effectively when placed in sequential decision-making problems. 
\citep{krishnamurthy2024can} studied in-context bandits and found that robust exploration is rare in LLMs.
\citep{nie2025evolve} deepened this evaluation perspective by introducing a bandit benchmark suite for measuring LLM exploration and showing that explicit algorithmic support, distillation, and fine-tuning can improve exploratory behavior. 
Beyond bandits, \citep{pan2025large} study open-ended exploration and report that most LLMs underperform humans because they rely too heavily on uncertainty and make premature decisions, while \citep{zhang2025comparing} compare LLMs and humans on standard multi-armed bandits and show that thinking-enabled models become more human-like in simple stationary settings but still struggle with directed exploration in more complex environments. 
\citep{arumugam2025toward} sharpens this picture by showing that explicitly instantiating a classical exploration algorithm such as PSRL inside an LLM can be substantially more effective than expecting exploration to emerge implicitly from prompting alone.
Most recently, \citep{park2026exploration} showed that exploration-exploitation errors are quantifiable in LLMs and revealed that there are systematic inefficiencies in how they balance exploration and exploitation.
{\color{black}Moreover, there have been works that also use contextual bandits to route evolving user queries to LLMs~\citep{poon2026online} or subtasks to agents~\citep{guan2026symphony}.
Compared to the routing setup, our setting is a partially observable stochastic game~(POSG) in which multiple agents choose whom to consult and revise their own answers over successive rounds.
Because each agent sees only part of the evolving debate, and its queries can change the answers available to others in later rounds, the usefulness of a peer changes over time and feedback about that peer is noisy.}

\paragraph{Delegation, trust, and adjacent capabilities for exploration.}
A closely related strand studies the capabilities a single LLM agent needs in order to decide \emph{whom} to rely on under uncertainty, which is adjacent to exploration in delegation-style settings. 
\citep{tomavsev2026intelligent} frame intelligent AI delegation as an adaptive sequence of decisions over task allocation, transfer of authority, and trust, emphasizing that autonomous delegation requires judging when another agent is worth relying on. 
\citep{buyl2025building} show that explicit self-reported trust between LLMs can be weakly aligned, or even negatively aligned, with implicit behavioral measures such as persuasion susceptibility and collaborative willingness, suggesting that naive verbalized confidence is a poor proxy for reliable partner selection. 
Relatedly, \citep{debnath2025can} examine whether LLMs can reason about and actively foster trust in dyadic interactions. 
Although these papers are not exploration papers in the bandit sense, they are highly relevant to our setting because effective exploration in multi-agent delegation depends not only on trying alternatives, but also on assessing partner reliability, calibrating trust, and avoiding over-commitment to noisy early impressions.






\section{Broader Impact}

Most discussions of AI reliability focus on single models acting in isolation. 
However, many future deployments may consist of populations of interacting agents that communicate, delegate, negotiate, and coordinate with one another while operating alongside humans. Despite this shift, the question of how multi-agent AI systems can integrate seamlessly into broader social and organizational environments remains comparatively underexplored. 
We view this as the challenge of building \emph{socially reliable multi-agent AI}.
This work contributes to this agenda by identifying a fundamental obstacle to reliable multi-agent autonomy: insufficient exploration of peers and interaction strategies. 

A positive implication is that socially reliable multi-agent systems could better support domains such as scientific collaboration, digital workforces, education, healthcare coordination, and public-service decision support, where success depends on combining distributed expertise rather than relying on a single model. 
By improving how agents discover useful partners, adapt to changing environments, and allocate communication efficiently, methods such as MACE may help AI systems participate more naturally in human institutions and mixed human-AI teams.
At the same time, stronger coordination among autonomous agents may introduce risks, including scalable manipulation, collusion, disinformation campaigns, or opaque collective behavior that becomes difficult to monitor. 
As multi-agent capabilities improve, governance and safety mechanisms become increasingly important. 
Future research should therefore pair advances in coordination with controllable communication protocols, fairness-aware objectives, and human oversight in high-stakes settings.

Overall, we hope this work helps broaden the conversation from reliable \emph{models} to reliable \emph{societies of models}. 
Enabling seamless and trustworthy integration of multi-agent AI systems into real-world social environments may become one of the central challenges of the next generation of AI.

\end{document}